\documentclass[final,authoryear,11pt]{elsarticle}
\UseRawInputEncoding
\usepackage{natbib}
\usepackage{amssymb, amsmath}
\usepackage{amsfonts}
\usepackage{amsthm}
\usepackage{graphics}
\usepackage{bm}
\usepackage{pdflscape}
\usepackage{float}
\usepackage{breqn}
\usepackage{amssymb}
\usepackage[mathscr]{eucal}
\usepackage{appendix}
\usepackage{comment}
\usepackage[multiple]{footmisc}
\usepackage{verbatim}
\usepackage{pdflscape}
\usepackage{tikz} 
\usepackage{setspace}
\usepackage{rotating}
\usepackage{breqn}
\usepackage{rotating}
\usepackage{bbm}
\usepackage{bm}
\usepackage{resizegather}
\usepackage{subcaption}
\usepackage[colorlinks=true,linkcolor=blue,citecolor=blue]{hyperref}
\usepackage{enumerate}

\usepackage[top=2.5cm, bottom=2.5cm, left=3.0cm, right=2.5cm]{geometry}
\input{ee.sty}
\newtheorem{theorem}{Theorem}
\newtheorem{lemma}{Lemma}
\newtheorem{proposition}{Proposition}

\newtheorem{assumption}{Assumption}[section]
\journal{arXiv}
\theoremstyle{definition}
\newtheorem{definition}{Definition}
\newtheorem{remark}{Remark}

\begin{document}

\bibpunct{(}{)}{,}{a}{}{,}
\bibliographystyle{ecta}

\begin{frontmatter}

% add here more packages or macros if needed

\title{On the use of U-statistics for linear dyadic interaction models}

\tnotetext[label1]{I would like to thank my advisors Frank Kleibergen and Art\={u}ras Juodis for all comments and suggestions. Moreover, I grateful for comments from Bo Honor\'{e} and Timo Schenk. I also thank Andrea Titton for his advice on optimizing the codes for the MC implementation and participants at the University of Amsterdam PhD Seminar Lunch for their helpful discussions.}

\author{Gabriela M. Miyazato Szini\corref{cor1}}
\ead{g.m.m.szini@uva.nl.}

\address{Amsterdam School of Economics, University of Amsterdam}
\address{Tinbergen Institute}

\begin{abstract}
    Even though dyadic regressions are widely used in empirical applications, the (asymptotic) properties of estimation methods only began to be studied recently in the literature. This paper aims to provide in a step-by-step manner how U-statistics tools \citep{serfling2009approximation} can be applied to obtain the asymptotic properties of pairwise differences estimators for a two-way fixed effects model of dyadic interactions. More specifically, we first propose an estimator for the model that relies on pairwise differencing such that the fixed effects are differenced out. As a result, the summands of the influence function will not be independent anymore, showing dependence on the individual level and translating to the fact that the usual law of large numbers and central limit theorems do not straightforwardly apply. To overcome such obstacles, we show how to generalize tools of U-statistics for single-index variables to the double-indices context of dyadic datasets. A key result is that there can be different ways of defining the H\'{a}jek projection for a directed dyadic structure, which will lead to distinct, but equivalent, consistent estimators for the asymptotic variances. The results presented in this paper are easily extended to non-linear models, as in \cite{graham2017econometric} and \cite{jochmans2018semiparametric}. 
\end{abstract}

\end{frontmatter}
\clearpage

\section{Introduction} \label{introduction}

Dyadic regression analysis is a common practice in several applications for network models. It is used, for instance, in the estimation of gravity models for international trade flows since its establishment by \cite{tinbergen1962shaping}. As defined by \cite{graham2020dyadic}, a dyadic dataset corresponds to a situation where the outcome of interest reflects a pairwise interaction among the sample units. Therefore, it is natural that datasets on trade flows are characterized by such a dyadic structure, as the value of imports and exports are determined by both the importer and the exporter countries. Other examples of applications of dyadic settings are, for instance, the estimation of models of migration, equity, international financial flows, and information flows \citep{jackson2013diffusion}.

Even though dyadic regressions are widely used in empirical applications, the (asymptotic) properties of estimation methods only began to be studied recently in the literature. One key feature that all studies mentioned above contain is the presence of two-way unit-specific effects, one for each individual in the dyadic interaction, and the fact that the outcome variable (and, in most cases, explanatory variables) is double indexed. For linear models with the two-way unobserved heterogeneity and the idiosyncratic error term entering additively in the specification it is possible to estimate the model consistently and (asymptotically) unbiasedly using the two-way fixed effects estimator (as long as the model is correcly specified; see \cite{juodis2020shock}). .

However, many economic models are non-linear, more specifically, in the context of network datasets, network formation models have the structure of discrete choice models (where the outcome of interest is binary), and outcomes of interest that are bounded below at zero (such as trade flows) can be approximated by a gravity equation in its multiplicative form. Naturally, one approach be to estimate such models with a probit/logit, and a poisson-pseudo maximum likelihood estimator \citep{silva2006log}, respectively. The challenge is that, while it is desirable to treat the unit-specific effects as parameters to be estimated (i.e., fixed effects, such that the conditional distribution of the unobserved heterogeneity given the covariates is left unrestricted), as \cite{fernandez2016individual} show, even if both dimensions of the (pseudo-)panel dataset grow with the sample size, these estimators suffer from the incidental parameter problem \citep{neyman1948consistent} in the presence of two-way fixed effects. This problem occurs since, in non-linear models, the estimates of the coefficients of the covariates depend on the estimates of the fixed-effects, and the latter converges at a slower rate than the first, resulting in an asymptotic bias in the estimates (and, therefore, invalid inference).

To address the incidental parameter problem in estimates for non-linear models with two-way fixed effects, such as logit, probit and poisson-pseudo maximum likelihood estimators, \cite{fernandez2016individual} proposed analytical and jackknife bias correction methods. However, for network formation models (discrete choice models), others propose a conditional maximum-likelihood approach under the logistic specification, such as in \cite{charbonneau2017multiple}, \cite{jochmans2018semiparametric} and \cite{graham2019network}. The conditioning sets in this approach translates to a model where pairwise differences of the outcomes (and covariates) are taken such that the two-way fixed effects are differenced out from the objective function, eliminating the incidental parameter problem. 

The advantage of the conditional maximum-likelihood approach as opposed to the bias-correction methods is that it accomodates sparse networks in the case of a network formation model \citep{jochmans2018semiparametric}. However, when taking such differences in the model, in general, the summands of the influence functions will not be independent anymore, showing some dependence on the unit level and translating to the fact that usual law of large numbers and central limit theorems do not straightforwardly apply.

To overcome such obstacle, U-statistics tools are generally applied such that one can show the asymptotic properties of those estimators. Although the U-statistics properties are well-known for single-index variables (see \citet[Chapter 5]{serfling2009approximation}, and \citet[Chapters 11 and 12]{van2000asymptotic}), those of double-indexed variables are generally not treated (up to my knowledge) in textbooks, with few exceptions related to applications for dyadic contexts, such as \citet[Chapter 4]{graham2019network}.

The main purpose of this paper is to illustrate, step-by-step in a comprehensive way, how to obtain the asymptotic properties of pairwise differences estimators (such that the fixed effects are cancelled out) for models of dyadic interactions using tools from the literature on U-statistics. More specifically, we show how to accomodate such tools to double-indexed variables, as the outcome is and the covariates are indexed by both individuals in the interaction. Even though, as mentioned earlier, the classical two-way fixed effects estimators delivers consistent and (asymptotically) unbiased estimates in the linear model (such that the pairwise differencing approach is not needed), for simplicity, we consider a linear two-way fixed effects, but the arguments can be generalized (with additional regularity conditions) to non-linear models or models with multiplicative individual heterogeneity and errors structure (as showed in \cite{jochmans2017two}).

A U-statistic for single-indexed variables is formally defined as an unbiased estimator that is of the form of an average of a function (kernel) of i.i.d. random variables. The main idea to determine the asymptotic properties of this estimator is to define a projection of the U-statistic, the so-called H\'{a}jek projection, that is asymptotically equivalent to the U-statistic itself. This projection consists on the average of the conditional expected value of the U-statistics, where each summand is the expected value of the U-statistic conditional on each index. Thus, by conditional independence arguments the H\'{a}jek projection becomes a simple average of i.i.d. random variables, to which laws of large numbers and central limit theorems can be applied. This concept will become more clear in the following sections.

A key result provided in this paper, is that, for a directed dyadic structure, there can be different ways of defining the H\'{a}jek projection, which will lead to distinct, but equivalent, consistent estimators for the asymptotic variance of the proposed estimator. More specifically, we provide two possible projections depending on which random variables one conditions the expected value of the U-statistics. Central to both possibilities is the fact that the summands of the influence function have a conditional independence structure once conditioned on dyad level attributes and the individual heterogeneity of both individuals forming a dyad (a fundamental difference with respect to single-index contexts). This is intuitive to dyadic settings, where the dependence across the dyads arises only through the individual fixed effects and the possible correlated observations for the same individual in different pairwise interactions and thus, outcomes and covariates. 

The organization of this paper is as follows: in Section 2 we define the linear model of directed dyadic interactions, in Section 3 we propose a pairwise differences estimator, in Section 4 we explain how some tools of U-statistics can be employed in this context and extended to dyadic settings, in Section 5 we discuss the asymptotic properties of the estimator and possible consistent estimates of its asymptotic variance and in Section 6 we demonstrate a Monte Carlo simulation exercise to investigate the finite sample properties of the estimator.\\
\\
\noindent \textbf{Notation}

Random variables are denoted by capital letters, specific realizations thereof by lower case, and their support by blackboard bold capital letters. That is, $Y$, $y$, and $\mathbb{Y}$ respectively denote a generic draw of, a specific value of, and the support of $Y$. 

Calligraphic letters denote sets. For instance, denote by $\mathcal{N} = \{1,2,...N\}$ the set of indices for $N$ individuals (or nodes). Denote by $\mathcal{C}(\mathcal{N},4)$ the multiset containing all sets of combinations of four individuals from the sampled $N$ observations. Moreover, denote $\rvert \mathcal{C}(\mathcal{N},4) \rvert = {N \choose 4}$ the number of obtained combinations, and denote by $\mathcal{C}$ an unordered set formed by a given combination, say $\mathcal{C} = \{i,j,k,l\}$.

Set $\mathcal{P}(\mathcal{C},4)$ to be the multiset containing all sets of permutations containing four elements of a given combination $\mathcal{C}$. Also, let $\rvert \mathcal{P}(\mathcal{C},4) \rvert = 4!$ to be the number of possible permutations, and $\pi$ to be the ordered set formed by a given permutation, where $\pi_1$, $\pi_2$, $\pi_3$ and $\pi_4$ denotes its first, second, third and fourth elements. For instance, given a permutation $\pi = \{k,l,j,i\}$, we have that $\pi_1 = k$, $\pi_2 = l$, $\pi_3 = j$ and $\pi_4 = i$. 

\section{A linear model of dyadic interactions}

We consider a linear model of dyadic interactions between $N$ agents, in which we assume that all variables of all pairwise interactions are observed (therefore, there is no sample selection). Let $(y_{ij}, x_{ij})$ denote the realizations of the random vector of outcomes and covariates $(Y_{ij}, X_{ij})$ for the dyad $(i,j)$, i.e., related to the interaction between agents $i$ and $j$. Importantly, $Y_{ij}$ is an outcome variable generated by the interaction of the individuals and it is continuous in this setting. We allow for directed interactions, such that $(Y_{ij}, X_{ij})$ need not be equal to $(Y_{ji}, X_{ji})$, and we do not include self links. Therefore, for a set $\mathcal{N} = \{1,2,... N\}$ of $N$ agents, we have $N(N-1)$ observed dyads. Following the notation of \cite{graham2020dyadic}, we denote that the first subscript on $Y_{ij}$ or $X_{ij}$ to be the \textit{ego}, or sending agent, and the second to be the \textit{alter}, or receiving agent.

Consider the following linear model of dyadic direct interactions taking into account two-way fixed effects:
\begin{align} \label{eq:model}
    Y_{ij} = \beta_1 X_{ij} + \theta_i + \xi_j + U_{ij}.
\end{align}
For simplicity, we consider for now only one regressor $X_{ij}$ (which can easily be relaxed to a vector). We assume that an agent-level attribute $A_i$ (which also can be relaxed to be a vector), and an attribute $B_j$ are observed, such that $X_{ij} = f(A_i, B_j)$ is a constructed dyad-level attribute. On the other hand, the sequences of individual-level heterogeneity $\{\theta_i\}_{i=1}^N$ and $\{\xi_i\}_{i=1}^N$ (for the \textit{ego} and for the \textit{alter}, respectively) are unobserved and we treat them as fixed-effects. In particular, there are no restrictions on correlations between $\theta_i$, $\xi_j$ and $X_{ij}$. In other words, the joint distribution between the observed and unobserved agent-level characteristics, $\{A_i, B_i, \theta_i, \xi_i \}$ is left unrestricted, such that the model is semiparametric. Finally, $U_{ij}$ is an idiosyncratic component that is also not necessarily equal to $U_{ji}$. 

Taking for instance the classical gravity model for international trade flows given by \cite{anderson2003gravity}, and usually applied in the empirical literature, the variable $Y_{ij}$ would refer to the log of the value of exports from country $i$ to country $j$, $X_{ij}$ refers to characteristics of the dyad $(i,j)$, for instance, the distance between the two countries, and $\theta_i$ and $\xi_j$ refers to the so-called unobserved \textit{multilateral resistance} terms. The latter terms refer, for example, to unmodeled export orientation of an economy, undervalued currencies and consumption taste.

We impose the following assumptions on this model: 
\begin{assumption} \label{ass:1}
    The error term $U_{ij}$ is i.i.d., independent of the sequence $\{A_i, B_i, \theta_i, \xi_i\}_{i=1}^N$ for any $i$ and $j$, and satisfies: 
    $$\mathbbm{E}[U_{ij}] = 0 $$
    $$ \mathbbm{E} [U_{ij} U_{lk}] = \begin{cases}
    \sigma_u^2 \quad \text{if } i=l,j=k\\
    0 \quad \text{otherwise.}
\end{cases}$$ 
\end{assumption}

\begin{assumption} \label{ass:2}
    The dyad-level observed variable $X_{ij}$ is given by:
    $$ X_{ij} = f(A_i, B_j)$$
    where $f$ is a measurable function, $A_i$ and $B_j$ are observed individual-level characteristics of the \textit{ego} and the \textit{alter}, respectively. Moreover, $A_i$, $B_i$ are i.i.d. and the sequences $\{A_i, B_i\}_{i=1}^N$ are mutually independent.
\end{assumption}

\begin{assumption} \label{ass:3}
    (Analogous to \cite{graham2017econometric}) Random sampling: Let $i=1,...N$ index a random sample of agents from a population satisfying Assumption 1. It is observed $(Y_{ij}, X_{ij})$ for $i=1,...N$, $j \neq i$ (i.e., all sampled dyads).
\end{assumption} 

Given the presence of the two-way fixed effects, namely $\theta_i$ and $\xi_j$, we have that conditional independence between the outputs of different dyads given the sequences of $A_i$ and $B_j$ (or, given the covariates $X_{ij}$) is unlikely to hold. Even conditioning on the sequence of covariates, outcomes that share the same \textit{ego} or \textit{alter} indices are likely to not be independent. For instance, the outcomes $Y_{12}$ and $Y_{34}$ are independent of each other, but the outcomes $Y_{12}$ and $Y_{13}$ are likely to be dependent, even after conditioning on $X_{12}$ and $X_{13}$. As pointed out by \cite{graham2020dyadic}, in the international trade example, this translates to the fact that exports from Japan to Korea will likely covary with exports from Japan to the United States, even after controlling for covariates, due to the Japan exporter effect. \cite{graham2020dyadic} denotes these patterns as dyadic dependence.

However, after conditioning also on the fixed-effects, that is, conditional on $\{ X_{12}, X_{13}, \theta_1, \xi_2, \xi_3 \}$, or, equivalently from Assumption \ref{ass:2}, conditional on $\{A_1, B_2, B_3, \theta_1, \xi_2, \xi_3\}$, the outcomes $Y_{12}$ and $Y_{13}$ are independent. This result follows from Assumption \ref{ass:1}. This conditional independence structure will be essential for the asymptotic properties of the estimator proposed in the following Section, mainly because this structure is well-suited for applying the tools of U-statistics. \cite{shalizi2016cid} denotes such models of dyadic interactions with such independency structure as conditionally independent dyad models (CID). 

\section{A pairwise differences estimator}

Even though the model given by Equation \eqref{eq:model} and under Assumptions \ref{ass:1} and \ref{ass:3} could be consistently and (asymptotically) unbiasedly estimated with a two-way fixed-effects estimator, we propose an estimator that differences out the fixed effects through pairwise differences. This estimator builds up on differencing arguments for a similar model, however, non-linear, introduced by \cite{charbonneau2017multiple}. She considers a model of network formation, where the outcome variable $Y_{ij}$ is binary, indicating whether an individual $i$ forms a \textit{directed} link with individual $j$.

As mentioned before, \cite{fernandez2016individual} shows that maximum likelihood estimators for nonlinear models with two-way fixed effects, such as probit/logit, suffer from the incidental parameter problem even if both dimensions of the (pseudo-)panel dataset tend to infinity. This is due to the fact that the dimensions of the vectors of nuisance parameters (how the fixed effects are treated in both \cite{charbonneau2017multiple} and in this paper) grows with the number of observations. At this point, it is important to notice that datasets of dyadic interactions can be seen as a pseudo panel data, where both dimensions of the panel tend to infinity as the number of individuals grow. \cite{fernandez2016individual} proposes analytical bias corrections to reduce the incidental parameter (asymptotic) bias, which was implemented by \cite{dzemski2019empirical} to a network formation context. However, as explained by \cite{jochmans2018semiparametric}, the problem with the bias correction approach is that for sparse networks the individual-specific parameters (the fixed effects) may not be consistently estimable or may be estimable only at a very slow rate. 

The approach proposed by \cite{charbonneau2017multiple} becomes very attractive for sparse networks, since, through a conditional maximum likelihood approach for logistic models, it delivers an estimator that differences out the fixed-effects. The estimator essentially is based on a set of conditions that translates to a transformation of the dependent and covariates, where pairwise differences are taken, such that the fixed-effects in the model are cancelled out. Even though a classic logit estimation can be used to obtain the estimates of the coefficients of the covariates, inference does not follow the textbook usual procedures, since agent-level dependencies arise when taking such pairwise differences. The asymptotic properties of this estimator are studied by \cite{jochmans2018semiparametric} and are obtained by employing tools of U-statistics. He shows that this estimator is consistent, asymptotically unbiased and the estimated variances deliver correct sizes for the t-test.

The estimator that we introduce in this Section is based on a similar pairwise differences methodology for transforming the dependent variable and covariates to difference out fixed-effects as presented by \cite{charbonneau2017multiple}, however, for a linear model. Our purpose when introducing this estimator is to provide a better understanding, through a simpler and linear model, on how to apply the tools of U-statistics to derive the asymptotic properties of estimators based on such pairwise differences for dyadic data.

First, we define the following notation for the random variable obtained by taking the specified pairwise differences among different dyads' outputs:
\begin{align} \label{eq:pairwisediff}
    \Tilde{Y}_{ijkl} \equiv (Y_{ij} - Y_{ik}) - (Y_{lj} - Y_{lk}),
\end{align}
\noindent and analogously for $\Tilde{X}_{ijkl}$ and $\Tilde{U}_{ijkl}$.

If we substitute the expressions for each of the outcomes $Y_{ij}$, $Y_{ik}$, $Y_{lj}$ and $Y_{lk}$ given by the model in Equation \eqref{eq:model} to the expression for $\Tilde{Y}_{ijkl}$ in Equation \eqref{eq:pairwisediff}, we obtain:
\begin{align} \label{eq:modeldiff}
    \Tilde{Y}_{ijkl} = \beta_1 \Tilde{X}_{ijkl} + \Tilde{U}_{ijkl},
\end{align}
\noindent where the fixed effects are differenced out. The equation above is simply a linear regression with the transformed variables obtained by taking the pairwise differences between the dyads $(i,j)$, $(i,k)$ and $(l,j)$, $(l,k)$. 

This form of differencing out the fixed effects depends heavily on the fact that the individual-specific heterogeneity parameters (i.e., the fixed effects themselves) enter the model additively. For more general specifications, this transformation fails to difference out the fixed effects. However, other studies, such as \cite{chen2021nonlinear} and \cite{jochmans2017two}, study cases with interactive fixed-effects. The former proposes an analytical bias correction estimator and the latter provides also an argument for differencing out the individual-specific parameters. 

Inspired by the same methodology of \cite{charbonneau2017multiple} that is further studied by \cite{jochmans2018semiparametric}, we can then estimate $\beta_1$ with an ordinary least squares estimator by taking into account the transformed variables $\Tilde{Y}_{ijkl}$ and $\Tilde{X}_{ijkl}$. Notice that the model given by Equation \eqref{eq:modeldiff}, where the fixed effects are differenced out, holds for all combinations of quadruples of indices from the set $\mathcal{N} = \{1,\dots,N\}$ and its permutations. Therefore, we can write the pairwise differences OLS estimator as:
\begin{align} \label{eq:OLSestimator}
\hat{\beta}_{1, PD} &= \left[\sum_{i=1}^N \sum_{j \neq i} \sum_{k \neq i,j} \sum_{l \neq i,j,k} \Tilde{X}_{ijkl} \Tilde{X}_{ijkl}' \right]^{-1}  \left[\sum_{i=1}^N \sum_{j \neq i} \sum_{k \neq i,j} \sum_{l \neq i,j,k} \Tilde{X}_{ijkl} \Tilde{Y}_{ijkl} \right] \\
&= \left[ \frac{1}{{N \choose 4}} \sum_{\mathcal{C} \in \mathcal{C}(\mathcal{N},4)}\frac{1}{4!} \sum_{\pi \in \mathcal{P}(\mathcal{C},4)} \Tilde{X}_{\pi_1\pi_2\pi_3\pi_4} \Tilde{X}_{\pi_1\pi_2\pi_3\pi_4}' \right]^{-1} \left[ \frac{1}{{N \choose 4}} \sum_{\mathcal{C} \in \mathcal{C}(\mathcal{N},4)}\frac{1}{4!} \sum_{\pi \in \mathcal{P}(\mathcal{C},4)} \Tilde{X}_{\pi_1\pi_2\pi_3\pi_4} \Tilde{Y}_{\pi_1\pi_2\pi_3\pi_4} \right], \nonumber 
\end{align}
\noindent where, in the second line, we use the fact that summing over all possible permutations of quadruples is equivalent to summing over all possible combinations of quadruples and then all permutations of such combinations. Therefore, say we look at a specific combination given by $ \mathcal{C}$, then the multiset denoted by $\mathcal{P}(\mathcal{C},4)$ corresponds to all permutations of those indices. Then, given a permutation, $\pi = \{i,j,k,l\}$, we have that $\pi_1 = i$, $\pi_2 = j$, $\pi_3 = k$ and $\pi_4 = l$, such that $\pi_1$ refers to the index occuping the first position in the permutation set, and analogously for $\pi_2$, $\pi_3$ and $\pi_4$.

To obtain the properties of the estimator $\hat{\beta}_{1,PD}$, it is useful, as shown in the regular textbook case for OLS estimators to rewrite the previous expression in terms of its influence function:
\begin{align} \label{eq:influenceOLS}
    \hat{\beta}_{1, PD} = \beta_1 + &\left[ \frac{1}{{N \choose 4}} \sum_{\mathcal{C} \in \mathcal{C}(\mathcal{N},4)} \frac{1}{4!} \sum_{\pi \in \mathcal{P}(\mathcal{C},4)} \Tilde{X}_{\pi_1\pi_2\pi_3\pi_4} \Tilde{X}_{\pi_1\pi_2\pi_3\pi_4}' \right]^{-1}  \left[ \frac{1}{{N \choose 4}} \sum_{\mathcal{C} \in \mathcal{C}(\mathcal{N},4)} \frac{1}{4!} \sum_{\pi \in \mathcal{P}(\mathcal{C},4)} \Tilde{X}_{\pi_1\pi_2\pi_3\pi_4} \Tilde{U}_{\pi_1\pi_2\pi_3\pi_4} \right]. 
\end{align}
In order to derive the asymptotic properties of this estimator, it is necessary to first derive the asymptotic properties of the last term in the equation above, namely: $$\left[ \frac{1}{{N \choose 4}} \sum_{\mathcal{C} \in \mathcal{C}(\mathcal{N},4)} \frac{1}{4!} \sum_{\pi \in \mathcal{P}(\mathcal{C},4)} \Tilde{X}_{\pi_1\pi_2\pi_3\pi_4} \Tilde{U}_{\pi_1\pi_2\pi_3\pi_4} \right]. $$ 
Notice that the transformed error terms $\Tilde{U}_{\pi_1\pi_2\pi_3\pi_4} \equiv (U_{\pi_1\pi_2} - U_{\pi_1\pi_3}) - (U_{\pi_4\pi_2} - U_{\pi_4\pi_3})$ are not independent over the dataset obtained when applying the transformation over all possible combinations and its permutations of quadruples, since the same dyads will appear in different terms, leading to a correlation amongst the terms. Therefore, the traditional application of LLNs and CLTs does not hold straightforwardly. 

From now on we will denote the last term in Equation \eqref{eq:influenceOLS} by:
\begin{align} \label{eq:UN}
    U_N &= \frac{1}{{N \choose 4}} \sum_{\mathcal{C} \in \mathcal{C}(\mathcal{N},4)} \frac{1}{4!} \sum_{\pi \in \mathcal{P}(\mathcal{C},4)} \Tilde{X}_{\pi_1\pi_2\pi_3\pi_4} \Tilde{U}_{\pi_1\pi_2\pi_3\pi_4} \\
    &= \frac{1}{{N \choose 4}} \sum_{\mathcal{C} \in \mathcal{C}(\mathcal{N},4)} \frac{1}{4!} \sum_{\pi \in \mathcal{P}(\mathcal{C},4)} ((X_{\pi_1\pi_2} - X_{\pi_1\pi_3}) - (X_{\pi_4\pi_2} - X_{\pi_4\pi_3}))  ((U_{\pi_1\pi_2} - U_{\pi_1\pi_3}) - (U_{\pi_4\pi_2} - U_{\pi_4\pi_3})). \nonumber
\end{align}
Even though this term resembles a U-statistic, it is not strictly speaking. However, we can adapt tools used in the literature of U-statistics to obtain the properties of the term $U_N$. Namely, we employ a Hoeffding decomposition \citep{hoeffding1948central} to obtain the variance of the term $U_N$, and we also propose two possibilities of H\'{a}jek projections of this term, which we prove both to be asymptotically equivalent to $U_N$. The reason why obtaining such projections is that the terms on it are i.i.d. such that it is possible to apply laws of large numbers and central limit theorems to obtain the asymptotic properties of $U_N$, and, thus of the proposed estimator.

In this paper we consider asymptotics under one single network growing, i.e., we consider that $N$ (the number of individuals in a network) tends to infinity when obtaining the asymptotic properties of the proposed estimator.

\section{Using U-statistics Tools In Dyadic Settings}

\subsection{The U-statistics}

According to \cite{serfling2009approximation}, the U-statistic is a generalization of the sample mean, i.e., a generalization of the notion of forming an average. The formal definition of the U-statistic is the following:

\begin{definition} \label{def:1}
    Let $W_1, W_2, ... W_n$ be independent observations on a distribution $F$ (which can be vector-valued). Consider a parametric function $\theta = \theta(F)$ for which there is an unbiased estimator:

    $$ \theta(F) = \mathbb{E} [h(W_1,... W_m)] = \int ... \int h(w_1,...,w_m) dF(w_1)... dF(w_m) $$

    \noindent for some function $h = h(x_1,...,x_m)$ called a kernel. It is assumed without loss of generality that $h$ is symmetric. Then, for any kernel $h$, the corresponding U-statistic for estimation of $\theta$ on the basis of a sample $X_1,...X_n$ of size $n \geq m$ is obtained by averaging the kernel symmetrically over the observations:

    $$ U_n = U(W_1,...W_n) = \frac{1}{{n \choose m}} \sum_c h(W_{i_1},...,W_{i_m}) $$

    \noindent where $\sum_c$ denotes summation over the ${n \choose m}$ combinations of $m$ distinct elements $\{i_1,...,i_m\}$ from $\{1,...,n\}$. An important property is that $U_n$ is an unbiased estimate of $\theta$.
\end{definition}

We can see that the term $U_N$, as defined in Equation \eqref{eq:UN}, contains elements of a U-statistic, resembling one at a first glace. However, it is not formally one given the definition above.

The shared properties to a U-statistic are related to having a similar dependence structure, such that it consists of a sum over all combinations of quadruples of individuals, evaluated at some given function, analogous to a fourth-order U-process. We can define the symmetric kernel for a given combination $\mathcal{C} = \{i,j,k,l\}$ in our case to be:
\begin{align} \label{eq:kernel}
    s_{ijkl} &:= \frac{1}{4!} \sum_{\pi \in \mathcal{P}(\mathcal{C},4)} ((X_{\pi_1\pi_2} - X_{\pi_1\pi_3}) - (X_{\pi_4\pi_2} - X_{\pi_4\pi_3})) 
     ((U_{\pi_1\pi_2} - U_{\pi_1\pi_3}) - (U_{\pi_4\pi_2} - U_{\pi_4\pi_3})), 
\end{align}
\noindent which is essentially the score of our estimator (being the reason why we denote by $s$, and not $h$). Note once again that the indices $k_1, k_2, k_3, k_4$ denote the elements of the permutations of a given combination of individuals ${i,j,k,l}$. Then, we can also see that another shared property with the U-statistic is that the kernel is permutation invariant and that the arguments of it, namely, the random variables $X_{ij}$ and $U_{ij}$, are identically distributed from Assumptions \ref{ass:1} and \ref{ass:2}.

Another important property that the term $U_N$ has in common to a U-statistic is that, if we define a parametric function $\theta$ to be:
\begin{align} \label{eq:unbiased}
    U & =\theta(F) \\ & =\mathbb{E}_F\left[\frac{1}{4 !} \sum_{\pi \in \mathcal{P}(\mathcal{C},4)}\left(\left(X_{\pi_1 \pi_2}-X_{\pi_1 \pi_3}\right)-\left(X_{\pi_4 \pi_2}-X_{\pi_4 \pi_3}\right)\right)\left(\left(U_{\pi_1 \pi_2}-U_{\pi_1 \pi_3}\right)-\left(U_{\pi_4 \pi_2}-U_{\pi_4 \pi_3}\right)\right)\right] \nonumber \\ & =0, \nonumber
\end{align}
\noindent then, we also have in our context that $U_N$ is an estimator of $\theta$, and it is also unbiased, since,
\begin{align}
    \mathbbm{E}_F [U_N] &= \mathbbm{E}_F \left[ \frac{1}{{N \choose 4}} \sum_{\mathcal{C} \in \mathcal{C}(\mathcal{N},4)} \frac{1}{4!} \sum_{\pi \in \mathcal{P}(\mathcal{C},4)} \Tilde{X}_{\pi_1\pi_2\pi_3\pi_4} \Tilde{U}_{\pi_1\pi_2\pi_3\pi_4} \right] \\
    &= \frac{1}{{N \choose 4}} \sum_{\mathcal{C} \in \mathcal{C}(\mathcal{N},4)} \mathbbm{E}_F \left[ \frac{1}{4!} \sum_{\pi \in \mathcal{P}(\mathcal{C},4)} \Tilde{X}_{\pi_1\pi_2\pi_3\pi_4} \Tilde{U}_{\pi_1\pi_2\pi_3\pi_4}\right] \nonumber \\
    &= \theta = 0, \nonumber
\end{align}
\noindent where the second equality follows from linearity of expectations.

In spite of these similarities, the statistic $U_N$ is not an U-statistic as conventionally defined, since its kernel includes random variables at both the individual (since $X_{ij} = f(A_i, B_j)$) and dyad level ($U_{ij}$). Therefore, single-index U-statistics as the one defined in the definition above are not well-suited, hence, the tools need to be slightly modified to accomodate the dyadic structure. 

Even more crucial is the fact that the observations $\{X_{ij}\}_{i=1,j\neq i}^N$ are not independent, due to the common individual characteristics $A_i$ or $B_j$. However, the fact that $U_{ij}$ is i.i.d. and that it is independent of $X_{ij}$ allows us to employ tools of U-statistics, such as the Hoeffding decomposition to obtain the variance of $U_N$, and the possibility to define a H\'{a}jek projection to obtain the asymptotic properties of this term (since we will demonstrate that $U_N$ and the projections are asymptotically equivalent). Importantly, to apply LLNs and CLTs to the H\'{a}jek projection, we will exploit the conditional independence structure of the projection. The conditional independence arguments extend straightforwardly to CID models, making them well suited for the use of U-statistic tools.

\subsection{Calculating the variance of $U_N$ using a Hoeffding decomposition}

%\textcolor{red}{[MY NOTES:]}
%Include:
%\begin{itemize}
%    \item if uses the definition of $h_c$ maybe it is easier to show that: $\Delta_c = \text{Var} [h_c()] = \mathbb{E} [\tilde{h}^2_c ()]$ and then we have that $0 = \Delta_0 = \Delta_1 \leq \Delta_2 \leq \Delta_3 \leq \Delta_4 = \text{Var} [h] < \infty$ 
%    \item look at last part of notes on part 2
%\end{itemize}
%\textcolor{red}{[END OF MY NOTES]}

To derive the variance, we first use some arguments provided by \cite{serfling2009approximation}, that are also employed by, for instance, \cite{graham2017econometric}. First, we define:
\begin{definition} \label{def:2}
    Consider two sets of combinations, say $\{i,j,k,l\}$ and $\{m,n,o,p\}$, of four distinct individuals from the set $\mathcal{N} = \{1,\dots,N\}$. Then, let $q \in \{0,1,2,3,4\}$ be the number of common individuals in the two combinations. Then, it follows by symmetry of the kernel function $s$, and by Assumptions \ref{ass:1} and \ref{ass:2}, that:
    $$ \Delta_q := \text{Cov}[\Tilde{s}_{ijkl}, \Tilde{s}_{mnop}] = \mathbbm{E} [\Tilde{s}_{ijkl} \Tilde{s}_{mnop}], $$
    \noindent where $\Tilde{s}_{ijkl} = s_{ijkl} - \theta$.
\end{definition}

Notice that independently of which pairs of combinations of quadruples we look at from the sampled individuals, the covariance between the two kernels evaluated at such combinations will only depend on the number of common individuals that the combinations share, namely, $q$. This follows from the fact that from Assumptions \ref{ass:1} and \ref{ass:2}, $U_{ij}$, $A_i$ and $B_j$ are i.i.d. and that the kernel (score) $s$ is symmetric on its arguments. By working out further the expression for the covariance $\Delta_q$, one can see that the nonzero terms in the expression are mainly driven by the covariance between the idiosyncratic errors. This is due to: (i) $\{U_{ij}\}_{i=1,j \neq i}^N$ being independent of $\{X_{ij}\}_{i=1,j \neq i}^N$, and to (ii) the idiosyncratic errors being independent of each other, while, for instance, $X_{ij}$ is correlated with $X_{ik}$ due to the common individual factor $A_i$. Therefore, if there is no common dyad in the expressions of the kernels for both combinations, the covariance between them will be zero, since $U_{ij}$ is i.i.d. (importantly, for instance, $U_{ij}$ and $U_{ik}$ are independent). This argument will become clearer in the Appendix A.

Due to the dyadic structure and the fact that $U_{ij}$ is i.i.d., $\text{Cov} (s_{ijkl}, s_{mnop}) = 0$ whenever the quadruples share zero or only one individual in common. Therefore, $\Delta_0 = \Delta_1 = 0$ indicates that $U_N$ exhibits degeneracy of order one. As long as the combinations have two or more indices in common, since the kernels sums over all permutations of the combinations, the same idiosyncratic error (with the same indices $i$ and $j$) appears in both terms $s_{ijkl}$ and $s_{mnop}$, leading to a non-zero covariance.

Assuming further that:
\begin{assumption} \label{ass:4}
    The symmetric kernel $s_{ijkl}$ satisfies:
    $$ \mathbb{E} [s_{ijkl}^2] < \infty. $$
\end{assumption}

Since the covariances $\Delta_q$ are constant across pairs of combinations sharing $q$ individuals in common, we can obtain the variance of $U_N$ through the Hoeffding decomposition \citep{hoeffding1948central}, as the following Lemma states:

\begin{lemma} \label{lemma:1}
    The variance of $U_N$ is given by:
    $$\text{Var}(U_N) = {N \choose 4}^{-1} \sum_{q=0}^4 {4 \choose q} {N-4 \choose 4-q} \Delta_q.$$
    And it satisfies, given Assumption \ref{ass:4}:
    $$\text{Var}(U_N) < \infty.$$
\end{lemma}

\noindent \textit{Proof.} Provided in Appendix B.1.

Given the result provided by Lemma \ref{lemma:1}, we can rescale the statistic $U_N$ by the factor $\sqrt{N(N-1)}$, and by taking into account that $\Delta_0 = \Delta_1 = 0$,
and  denoting:
    $$ \bar{s}_{ij} = \mathbb{E}[s_{ijkl} \rvert A_i, B_j, U_{ij}] \quad \text{and} \quad \bar{s}_{ji} = \mathbb{E}[s_{ijkl} \rvert A_j, B_i, U_{ji}], $$ 
    $$ \delta_2 = \mathbb{E} [\bar{s}_{ij}^2] = \mathbb{E} [\bar{s}_{ji}^2], $$
we arrive at the following result:

\begin{theorem} \label{theorem:1}
    Given the result in Lemma \ref{lemma:1}, and under Assumptions \ref{ass:1}-\ref{ass:3} and \ref{ass:4}:
    $$ \text{Var} (\sqrt{N(N-1)}U_N) = \mathcal{O}(1) + \mathcal{O} \left(\frac{1}{N}\right) + \mathcal{O} \left(\frac{1}{N^2}\right). $$
    The term related to $\Delta_2$ asymptotically dominates the expression, such that the variance of the rescaled statistic $U_N$ converges to:
    $$ \text{Var} (\sqrt{N(N-1)} U_N) \xrightarrow[]{N \xrightarrow{} \infty} 72 \Delta_2 = 144 \delta_2. $$
\end{theorem}
\noindent \textit{Proof.} Provided in Appendix B.2. 

Where the terms of order $\mathcal{O}(1)$ in the above expression relates to the term $\Delta_2$, $\mathcal{O}\left(\frac{1}{N}\right)$ relates to the term $\Delta_3$ and $\mathcal{O}\left(\frac{1}{N^2}\right)$ relates to the term $\Delta_4$. Furthermore, given our simplified model, it is possible to further pin down the expression for $\Delta_2$. This result can be found in Appendix C.

\subsection{Deriving the H\'{a}jek projection of $U_N$}

As explained by \cite{serfling2009approximation}, the appealing feature of a U-statistic as given by Definition \ref{def:1}, is its simple structure as a sum of identically distributed random variables. But, even in the simpler context of a single-index U-statistic, if the kernel $h$ has a dimension $m>1$, then the summands in the statistic $U_N$ are not all independent, as the sample sampled observations are taken into account in different combinations. Therefore, it is not possible to directly employ LLNs and CLTs for sums of independent random variables, as it is customary done. However, \cite{serfling2009approximation} and other textbooks on U-statistics show that it is possible to obtain a projection to which the U-statistic can be approximated to. The advantage is that such projection is a sum of i.i.d. random variables, to which classical limit theory can be applied.

In the following we will present the formal definition of this projection, the H\'{a}jek projection, and explain how this concept can be applied in our context. We also highlight that there are considerable differences between our approach and the classical textbook projection. Again, the main difference is that, while the standard definitions account for single-index variables, in our case of a dyadic setting the random variables forming the U-statistics have double-indices. Moreover, the pairwise differences structure in the kernel are formed by random variables reflecting the dyadic interactions originated by four individuals.

Therefore, to obtain the H\'{a}jek projection, instead of conditioning on a single-indexed random variable alone as it is done in textbooks, we condition on both the individual and dyad-level random variables given by the dyad indices. We will show that by doing so, we will still obtain a projection where the summands are conditionally independent, even if the sequence $\{X_{ij}\}_{i=1,j\neq i}^N$ is not formed by independent variables, since the idiosyncratic errors $U_{ij}$ are i.i.d. and independent of the former sequence. This relies on the previously mentioned arguments of conditional independence of CID models.

Besides, in the general textbook case or single-index variables, it is stated that the projection has no purpose in the case where $\Delta_1 = 0$, however, we will see that in our case, due to the dyadic structure, the projection proves to be useful even when $\Delta_1 = 0$ holds.

The most important result in this section is that we can propose two different forms of H\'{a}jek projections, depending whether we condition on all random variables generated by the combination $\{i,j\}$ of a dyad, or if we condition on the random variables generated by the permutation $\{i,j\}$, where the ordering of the indices matter.

\subsubsection{The textbook definition of a H\'{a}jek projection}

According to \cite{serfling2009approximation}, and following the same notation and framework as in Definition \ref{def:1}, we have the following definition for a H\'{a}jek projection:

\begin{definition} \label{def:3}
    Assume $E_{F}|h|<\infty .$ The projection of the U-statistic $U_{n}$ is defined as
    $$
    \hat{U}_{n} := \sum_{i=1}^{n} E_{F}\left\{U_{n} \mid W_{i}\right\}-(n-1) \theta.
    $$
    Notice that, in the context of \cite{serfling2009approximation} it is exactly a sum of i.i.d. random variables. 
\end{definition}

It is important to notice that, in the definition of $\hat{U}_n$ above, when taking the expectation of $U_n$ conditioning on each different $W_i$ for each summand, we are left with a sum of i.i.d. random variables, since $W_i$ are i.i.d. themselves.

\subsubsection{First H\'{a}jek projection, $\hat{U}_{N,1}$}
%\textcolor{red}{[MY NOTES:]} \\
%\noindent - after definition of $\hat{U}_{N,1}$ say that expectation in the summand is by law of iterated expectations and conditional independence zero if i,j does not belong to permutations $\{i,j\} \not\subset \{perm\}$ and equals to $\bar{s}_{ij}$ if so. mean independence holds true even if share only one index in common:
%$$ \mathbb{E} [\bar{s}_{ij} \rvert X_k, X_l, A_k, A_l, U_{kl}] $$
%forget about before and say that holds true even if they share one index in common to explain the final variance result.

%\noindent - also if consider two quadruples $\{i,j,k,l\} and \{i,j,m,p\}$ that shares the dyad $\{i,j\}$ in common, by conditional independence the kernels are independent conditional on stuff (check which is assumption 3 graham - assumption of random sampling Uij iid mainly):
%$$ \mathbb{E} [s_{ijkl} s_{ijmp} \rvert X_i, X_j, A_i, A_j, U_{ij}] = \bar{s}_{ij} \bar{s}_{ij} $$
%Iterated expectations then imply that $\Delta_2 = \bar{s}_{ij} \bar{s}_{ij}$. this is graham in our case like this is equal to $\delta_2$ or conditioning on more it is equal to $\Delta_2$.

%\noindent - second point of part 1

%\noindent - Graham conditions on FE as well, should we do that?

%\noindent - Rewrite part of $U = \theta(F) = 0$ as in page 12 of serfling book

%\textcolor{red}{[END OF MY NOTES]}

We denote the first proposed H\'{a}jek projection by $\hat{U}_{N,1}$. In our context, we already derived before that $\theta = \mathbbm{E}[s_{ijkl}] = 0$ (see Equation \eqref{eq:unbiased}), therefore, we only need to focus now on deriving the first term of the similar projection proposed in Definition \ref{def:3}. In addition, we are working in a context of dyads, therefore, the sum is over the expected value of the statistic $U_N$ conditional on each of the dyad characteristics, namely, for a given dyad $\{i',j'\}$ we condition on $\{A_i', B_j', U_{i'j'}\}$. Therefore, we sum over all the possible dyads $N(N-1)$. Notice that the order of the indices in the dyad matter, since we have a directed network. 

\begin{definition} \label{def:4}
    Given the statistic in Equation \eqref{eq:UN}, we define the first H\'{a}jek projection as:
    \begin{align}
    \hat{U}_{N,1} &= \sum_{i' = 1}^N \sum_{j' \neq i'} \mathbbm{E} [ U_N \rvert A_{i'}, B_{j'}, U_{i'j'}] \\
    &= \sum_{i' = 1}^N \sum_{j' \neq i'} \mathbbm{E} \left[ \frac{1}{{N \choose 4}} \sum_{\mathcal{C} \in \mathcal{C}(\mathcal{N},4)} \frac{1}{4!} \sum_{\pi \in \mathcal{P}(\mathcal{C},4)} \Tilde{X}_{\pi_1\pi_2\pi_3\pi_4} \Tilde{U}_{\pi_1\pi_2\pi_3\pi_4}  \rvert A_{i'}, B_{j'},, U_{i'j'}\right] \nonumber \\
    &= \sum_{i' = 1}^N \sum_{j' \neq i'} \mathbbm{E} \left[ \frac{1}{{N \choose 4}} \sum_{\mathcal{C} \in \mathcal{C}(\mathcal{N},4)} s_{ijkl} \rvert A_{i'}, B_{j'}, U_{i'j'}\right] \nonumber\\
    &= \sum_{i' = 1}^N \sum_{j' \neq i'} \frac{1}{{N \choose 4}} \sum_{\mathcal{C} \in \mathcal{C}(\mathcal{N},4)} \mathbbm{E}[s_{ijkl} \rvert A_{i'}, B_{j'}, U_{i'j'}]. \nonumber
    \end{align}
\end{definition}

The main idea behind this projection is that the double sum $\sum_{i' = 1}^N \sum_{j' \neq i'}$ fixes the two indices of a dyad, and refers to the individual-level $\{A_{i'}, B_{j'}\}$ and the dyad level characteristics $U_{i'j'}$, which we condition the statistic $U_N$ on. In this case, the order of the indices $(i',j')$ matter to determine on which random variables we condition on. 

For each summand of the double sum $\sum_{i' = 1}^N \sum_{j' \neq i'}$ we take the expectation of the statistic $U_N$ conditional on the variables described above. The statistic is essentially an average of the scores $s_{ijkl}$ evaluated at all possible combinations of quadruples $\{i,j,k,l\}$ from the set $\mathcal{N}$. From Assumption \ref{ass:1}, and more precisely, the fact that $U_{ij}$ is independent from the sequence $\{X_{ij}\}_{i=1, j \neq i}^N$, leads to the fact that the only non-zero summands are the ones where the combination $\mathcal{C}$ contains the elements $i'$ and $j'$, and any other two remaining elements. Since the kernel contains all permutations of the combination, inevitably the term $U_{i'j'}$ will appear in the expression for the kernel $s_{ijkl}$ in this case (where $\{i',j'\} \subset \{i,j,k,l\}$, with, for instance, $i=i'$ and $j=j'$), leading to a non-zero conditional expectation.

We can further boil down the expression of the projection $\hat{U}_{N,1}$ by first noting that, as shown in Appendix C, for a given combination $\{i',j',k,l\}$ for any value of $k$ and $l$, the conditional expected value of the kernel evaluated at such combination is of the form:
$$ \mathbbm{E}[s_{i'j'kl} \rvert A_{i'}, B_{j'}, U_{i'j'}] = 8 [(X_{i'j'} - \mathbbm{E}[X_{i'j'} \rvert A_{i'}] - \mathbbm{E}[X_{i'j'} \rvert B_{j'}] + \mathbbm{E}[X_{i'j'}])U_{i'j'}]. $$
Moreover, there will be ${N-2 \choose 2}$ possible combinations of four elements of the set $\mathcal{N}$ containing the individuals $i'$ and $j'$. Then, we can rewrite the projection as:
\begin{align}
    \hat{U}_{N,1} &= \sum_{i' = 1}^N \sum_{j' \neq i'} \frac{1}{{N \choose 4}} \frac{1}{4!} {N-2 \choose 2} 8 [(X_{i'j'} - \mathbbm{E}[X_{i'j'} \rvert A_{i'}] - \mathbbm{E}[X_{i'j'} \rvert B_{j'}] + \mathbbm{E}[X_{i'j'}])U_{i'j'}] \\
    &= \frac{4}{N(N-1)} \sum_{i' = 1}^N \sum_{j' \neq i'} [(X_{i'j'} - \mathbbm{E}[X_{i'j'} \rvert A_{i'}] - \mathbbm{E}[X_{i'j'} \rvert B_{j'}] + \mathbbm{E}[X_{i'j'}])U_{i'j'}]. \nonumber
\end{align}
In order to show in the following sections that the statistic $U_N$ and the projection $\hat{U}_{N,1}$ are asymptotically equivalent, we first need to derive the variance of the projection. 

\begin{lemma} \label{lemma:2}
    Under Assumptions \ref{ass:1} and \ref{ass:2}, the variance of the first H\'{a}jek projection given by Definition \ref{def:4} is:
    $$ \text{Var} ( \hat{U}_{N,1} ) = \frac{144}{N(N-1)} \delta_2.$$
    \noindent Therefore, by rescaling the projection by the factor $\sqrt{N(N-1)}$, we have:
    $$ \text{Var} (\sqrt{N(N-1)} \hat{U}_{N,1}) = 144 \delta_2. $$
\end{lemma}

\noindent \textit{Proof.} Proof provided in Appendix B.3.

\subsubsection{Second H\'{a}jek projection, $\hat{U}_{N,2}$}

Before defining the second possibility for the H\'{a}jek projection, notice first that, conditioning on individual and dyad-level random variables related to a directed dyad $(i',j')$ is different than conditioning on all individual and dyad-level random variables related to a combination $\{i',j'\}$. More specifically, the first comprises of the elements $A_{i'}$, $B_{j'}$ and $U_{i'j'}$, while the second comprises of $A_{i'}$, $A_{j'}$, $B_{i'}$, $B_{j'}$, $U_{i'j'}$ and $U_{j'i'}$.

Therefore, in this second proposed projection, instead of summing over all possible directed dyads, we sum over all possible combinations of indices $i'$ and $j'$, which amounts to $\frac{N(N-1)}{2}$ combinations. We therefore condition on all characteristics of these both indices:

\begin{definition} \label{def:5}
    Given the statistic in Equation \eqref{eq:UN}, we define the second H\'{a}jek projection as:
\begin{align}
    \hat{U}_{N,2} & := \sum_{i'=1}^N \sum_{j' > i'} \mathbbm{E} [ U_N \rvert A_{i'}, B_{j'}, U_{i'j'}, A_{j'}, B_{i'}, U_{j'i'} ] \\
    &= \sum_{i'=1}^N \sum_{j' > i'} \mathbbm{E} \left[ \frac{1}{{N \choose 4}} \sum_{\mathcal{C} \in \mathcal{C}(\mathcal{N},4)} \frac{1}{4!} \sum_{\pi \in \mathcal{P}(\mathcal{C},4)} \Tilde{X}_{\pi_1\pi_2\pi_3\pi_4} \Tilde{U}_{\pi_1\pi_2\pi_3\pi_4} \rvert A_{i'}, B_{j'}, U_{i'j'}, A_{j'}, B_{i'}, U_{j'i'} \right]  \nonumber \\
    &= \sum_{i'=1}^N \sum_{j' > i'}  \mathbbm{E} \left[\frac{1}{{N \choose 4}} \sum_{\mathcal{C} \in \mathcal{C}(\mathcal{N},4)} s_{ijkl} \rvert A_{i'}, B_{j'}, U_{i'j'}, A_{j'}, B_{i'}, U_{j'i'}\right] \nonumber\\
    &= \sum_{i'=1}^N \sum_{j' > i'}  \frac{1}{{N \choose 4}}  \sum_{\mathcal{C} \in \mathcal{C}(\mathcal{N},4)} \mathbbm{E}[s_{ijkl} \rvert A_{i'}, B_{j'}, U_{i'j'}, A_{j'}, B_{i'}, U_{j'i'}]. \nonumber
\end{align}
\end{definition}

Again, the double sum, $\sum_{i'=1}^N \sum_{j > i}$, fixes two indices $i'$ and $j'$ of the possible tetrads, and it runs over the conditioning terms. Then, we take the expectation of the statistic $U_N$ conditional on the terms described above. The structure of the second H\'{a}jek projection is essentially the same as the first, apart from which terms we condition on. Therefore, again, we will have that for all combinations of quadruples for which we take the average of the conditional expectation of the score function (kernel), only ${N-2 \choose 2}$ combinations will lead to non-zero expectated values. Those combinations refer again to the ones containing the elements $i'$ and $j'$.

The difference with respect to the previous projection, that is induced by the extra conditioning terms, boils down to the terms in the score function $s_{ijkl}$ (where $\{i',j'\} \subset \{i,j,k,l\}$, with, for instance, $i=i'$ and $j=j'$) that will be non-zero, since now the permutations that both the terms  $U_{i'j'}$ and $U_{j'i'}$ will have non-zero terms.

Once again, we can further boil down the expression of the projection $\hat{U}_{N,2}$ by first noting that, as shown in Appendix C, for a given combination $\{i',j',k,l\}$ for any value of $k$ and $l$, the conditional expected value of the kernel evaluated at such combination is of the form:
\begin{align*}
    &\mathbbm{E}[s_{i'j'kl} \rvert A_{i'}, B_{j'}, U_{i'j'}, A_{j'}, B_{i'}, U_{j'i'}] \\ &= 8 [(X_{i'j'} - \mathbbm{E}[X_{i'j'} \rvert A_{i'}] - \mathbbm{E}[X_{i'j'} \rvert B_{j'}] + \mathbbm{E}[X_{i'j'}])U_{i'j'}] \nonumber + 8 [(X_{j'i'} - \mathbbm{E}[X_{j'i'} \rvert A_{j'}] - \mathbbm{E}[X_{j'i'} \rvert B_{i'}] + \mathbbm{E}[X_{j'i'}])U_{j'i'}].
\end{align*}
Such that we can further simplify the expression for the projection:
\begin{align}
    \hat{U}_N &= \sum_{i=1'}^N \sum_{j' > i'}  \frac{1}{{N \choose 4}} \frac{1}{4!} {N-2 \choose 2} \Big[ 8 [(X_{i'j'} - \mathbbm{E}[X_{i'j'} \rvert A_{i'}] - \mathbbm{E}[X_{i'j'} \rvert B_{j'}] + \mathbbm{E}[X_{i'j'}])U_{i'j'}] \nonumber \\
    &+ 8 [(X_{j'i'} - \mathbbm{E}[X_{j'i'} \rvert A_{j'}] - \mathbbm{E}[X_{j'i'} \rvert B_{i'}] + \mathbbm{E}[X_{j'i'}])U_{j'i'}]\Big]\\
    &= \frac{4}{N(N-1)} \sum_{i'=1}^N \sum_{j' > i'}  \Big[(X_{i'j'} - \mathbbm{E}[X_{i'j'} \rvert A_{i'}] - \mathbbm{E}[X_{i'j'} \rvert B_{j'}] + \mathbbm{E}[X_{i'j'}])U_{i'j'} \nonumber \\ &+ (X_{j'i'} - \mathbbm{E}[X_{j'i'} \rvert A_{j'}] - \mathbbm{E}[X_{j'i'} \rvert B_{i'}] + \mathbbm{E}[X_{j'i'}])U_{j'i'}\Big]. \nonumber
\end{align}
Again, we proceed by deriving the variance of this second projection, which should be equivalent to the variance of the first proposed projection.

\begin{lemma} \label{lemma:3}
    Under Assumptions \ref{ass:1} and \ref{ass:2}, the variance of the second H\'{a}jek projection given by Definition \ref{def:5} is:
    $$ \text{Var} ( \hat{U}_{N,2} ) = \frac{72}{N(N-1)} \Delta_2. $$
    \noindent Therefore, by rescaling the projection by the factor $\sqrt{N(N-1)}$, we have:
    $$ \text{Var} (\sqrt{N(N-1)} \hat{U}_{N,2}) = 72 \Delta_2.  $$
\end{lemma}

\noindent \textit{Proof.} Provided in Appendix B.4.
\subsection{Showing the asymptotic equivalence of $U_N$ and $\hat{U}_{N,1}$ or $\hat{U}_{N,2}$}

The main idea of defining a H\'{a}jek projection is to obtain a statistic that is approximatelly and asymptotically close enough to the U-statistic to which central limit theorems and laws of large numbers can be applied.

\cite{serfling2009approximation} provides readily applicable results for the asymptotic equivalence of the U-statistic given by Definition \ref{def:1} and the H\'{a}jek projection given by Definition \ref{def:2}, and, consequently, to the asymptotic properties of the U-statistics, since in this case the projection is an average of i.i.d. random variables. However, in our case, as the statistic $U_N$ is not formally an U-statistic, such results cannot be immediately used.

To derive the asymptotic equivalence between $U_N$ and the two proposed projections, $\hat{U}_{N,1}$ and $\hat{U}_{N,2}$, we follow closely the arguments in \cite{graham2017econometric}.

\begin{remark} \label{lemma:4}
    According to \cite{graham2017econometric}, the asymptotic equivalence\footnote{This result is also provided in \cite{serfling2009approximation}.} of $\sqrt{N(N-1)} U_N$ and of $\sqrt{N(N-1)} \hat{U}_N$ follows if:
$$ N(N-1) \mathbbm{E}[(\hat{U}_N - U_N)^2] \quad \text{is} \quad {o}(1)$$.
\end{remark}

Following up this Remark, we have the following result:

\begin{theorem} \label{theorem:2}
    Given the definitions of the statistic $U_N$ given by Equation \eqref{eq:UN} and the proposed H\'{a}jek projections $\hat{U}_{N,1}$, in Definition \ref{def:4}, and $\hat{U}_{N,2}$, in Definition \ref{def:5}, $U_N$ is asymptotically equivalent to $\hat{U}_{N,1}$ and $\hat{U}_{N,2}$ under Assumptions \ref{ass:1}-\ref{ass:3} and \ref{ass:4}.
\end{theorem}

\noindent \textit{Proof.} Provided in Appendix B.5.

Hence, even though the statistic $U_N$ is not properly defined as an U-statistic, we still have the result that, under the assumptions needed for the results above, its limit distribution coincides with that of the proposed H\'{a}jek projections. This property is key to define the asymptotic properties of the pairwise differences estimator in the next section.

\section{Asymptotic properties of the Pairwise Differences estimator and Estimation}

\subsection{Asymptotic properties of the Pairwise Differences estimator}

Considering the rewritten estimator, defined before:
\begin{align} 
  \hat{\beta}_{1, PD} = \beta_1 + &\left[ \frac{1}{{N \choose 4}} \sum_{\mathcal{C} \in \mathcal{C}(\mathcal{N},4)} \frac{1}{4!} \sum_{\pi \in \mathcal{P}(\mathcal{C},4)} \Tilde{X}_{\pi_1\pi_2\pi_3\pi_4} \Tilde{X}_{\pi_1\pi_2\pi_3\pi_4}' \right]^{-1} \left[ \frac{1}{{N \choose 4}} \sum_{\mathcal{C} \in \mathcal{C}(\mathcal{N},4)} \frac{1}{4!} \sum_{\pi \in \mathcal{P}(\mathcal{C},4)} \Tilde{X}_{\pi_1\pi_2\pi_3\pi_4} \Tilde{U}_{\pi_1\pi_2\pi_3\pi_4} \right].
\end{align}
We note that to obtain the asymptotic properties, it is key to obtain the convergence of the Hessian:
\begin{align} \label{eq:hessian}
  \left[ \frac{1}{{N \choose 4}} \sum_{\mathcal{C} \in \mathcal{C}(\mathcal{N},4)} \frac{1}{4!} \sum_{\pi \in \mathcal{P}(\mathcal{C},4)} \Tilde{X}_{\pi_1\pi_2\pi_3\pi_4} \Tilde{X}_{\pi_1\pi_2\pi_3\pi_4}' \right].
\end{align}
While in most of the studies on dyadic regressions and U-statistics tools associated with it the convergence of this Hessian is assumed (for instance, in \cite{graham2019network}), we instead proof such convergence result. Observe that, even though this term also resembles a U-statistic, or, at least a term to which U-statitics tools can be applied to, it is not the case. This follows from not having a term such as $U_{ij}$ which is i.i.d. in the dyad level in such statistic, which would guarantee the conditional independence of summands. Therefore, the same tools applied to the statistic $U_N$ cannot be carried over here. Instead, our approach relies on deriving the variance of such term, and proving the convergence in probability through the Chebyshev's inequality. 

\begin{proposition} \label{lemma:5}
  Under the assumption that :
  $$ \mathbb{E} [ \rvert X_{ij} X_{i'j'} \rvert] < \infty \quad \forall \quad i,i',j,j' $$
  It follows that:
  $$\left[ \frac{1}{{N \choose 4}} \sum_{\mathcal{C} \in \mathcal{C}(\mathcal{N},4)} \frac{1}{4!} \sum_{\pi \in \mathcal{P}(\mathcal{C},4)} \Tilde{X}_{\pi_1\pi_2\pi_3\pi_4} \Tilde{X}_{\pi_1\pi_2\pi_3\pi_4}' \right] \xrightarrow[]{p} \Gamma := \mathbb{E} [\Tilde{X}_{\pi_1\pi_2\pi_3\pi_4} \Tilde{X}_{\pi_1\pi_2\pi_3\pi_4}'],$$
  \noindent where $\Gamma$ is finite and invertible.
\end{proposition}

\noindent \textit{Proof.} Provided in Appendix B.6. 

Given the result of the Proposition above, we can rewrite, by rescaling the expression of the rewritten estimator by $\sqrt{N(N-1)}$:
\begin{align} \label{eq:maineq}
\sqrt{N(N-1)} (\hat{\beta}_{1,PD} - \beta_1) = \Gamma^{-1} \sqrt{N(N-1)} U_N + o_p(1),
\end{align}
\noindent which follows by the continuous mapping theorem. Therefore, the asymptotic sampling properties of $\sqrt{N(N-1)} (\hat{\beta}_{1,PD} - \beta_1)$ will be driven by the behaviour of $\sqrt{N(N-1)} U_N$.

From Theorem \ref{theorem:2}, we have that the statistic $U_N$ is asymptotically equivalent to the projections $\hat{U}_{N,1}$ and $\hat{U}_{N,2}$. Therefore, the asymptotic properties of those carry over to the asymptotic properties of $U_N$. Notice that, from Equation \eqref{eq:uncorrelated1} and its analogous for the second proposed projection, we have that the summands of the projections are uncorrelated, but not necessarily independently distributed. The dependence structure remains since the same individual characteristics $A_i$ and $B_j$ for a given $i$ and $j$ are still present in different summands, as for instance, we can have terms such as $\mathbb{E} [X_{ij'} \rvert B_{j'}]$ in one summand and $\mathbb{E} [X_{ik'} \rvert B_{k'}]$ in another summand.

However, as pointed out by \cite{graham2017econometric} and \cite{jochmans2018semiparametric} in their contexts, by law of iterated expectations, we can rewrite:

\begin{align}
  \hat{U}_{N,1} &= \sum_{i' = 1}^N \sum_{j' \neq i'} \frac{1}{{N \choose 4}} \sum_{\mathcal{C} \in \mathcal{C}(\mathcal{N},4)} \mathbbm{E}[s_{ijkl} \rvert A_{i'}, B_{j'}, U_{i'j'}] \nonumber \\
  &= \sum_{i' = 1}^N \sum_{j' \neq i'} \frac{1}{{N \choose 4}} \sum_{\mathcal{C} \in \mathcal{C}(\mathcal{N},4)} \mathbbm{E}[\mathbbm{E}[s_{ijkl} \rvert A_{i'}, B_{j'}, U_{i'j'}] \mathbf{A},\mathbf{B}] .  
  \end{align}
Such that the summands of the projections are conditionally independent when conditional on all indices $\{A_i\}_{i=1}^N$ and $\{B_j\}_{j=1}^N$, which is a characteristic of CID models, and that carries over to this context. Given this conditional independence of the random variables, we can assert:

\begin{lemma} \label{lemma:6}
  From a conditional version of the strong law of large numbers and a conditional version of the Lyapunov's central limit theorem, given by \cite{rao2009conditional}, it follows that: \\
\\
  \noindent (i) $\hat{U}_{N,1} \xrightarrow[]{p} 0$ and $\hat{U}_{N,2} \xrightarrow[]{p} 0$ \\
  (ii) $\sqrt{N(N-1)} \hat{U}_{N,1} \xrightarrow[]{d} N(0,144\delta_2)$ and $\sqrt{N(N-1)} \hat{U}_{N,2} \xrightarrow[]{d} N(0,72\Delta_2).$ \\
\\
  Since the expectation of the H\'{a}jek projections are zero, and their variances are defined by Lemma \ref{lemma:2} and Lemma \ref{lemma:3}.

  Moreover, since $U_N$ and the projections are asymptotically equivalent, that is, $\rvert \rvert \hat{U}_{N,1} - U_N \rvert \rvert \xrightarrow[]{p} 0$, and $\rvert \rvert \hat{U}_{N,2} - U_N \rvert \rvert \xrightarrow[]{p} 0$, we also have that: \\
\\
  \noindent (i) $U_N \xrightarrow[]{p} 0$ \\
  (ii) $\sqrt{N(N-1)} U_N \xrightarrow[]{d} N(0,\sigma_U^2)$, where $\sigma^2_U = 144\delta_2 = 72\Delta_2$.
\end{lemma}

\textit{Proof.} Available in the next versions of this paper.

Following Proposition \ref{lemma:5} and Lemma \ref{lemma:6}, we can establish first the consistency of the estimator, which is provided in the following theorem.

\begin{theorem} \label{theorem:3}
  Given the results of Proposition \ref{lemma:5} and Lemma \ref{lemma:6}, and its associated assumptions, we have that $\hat{\beta}_{1,PD}$ is a \textbf{consistent estimator} of $\beta_1$:
  $$ \hat{\beta}_{1,PD} \xrightarrow[]{p} \beta_1. $$
\end{theorem}

\noindent \textit{Proof.} Provided in Appendix B.7.

From the same Proposition and Lemma, the asymptotic normality and the associated asymptotic variance of the estimator can be established. Also note that the estimator is asymptotically unbiased according to the following theorem.

\begin{theorem} \label{theorem:4}
  Using the representation in Equation \eqref{eq:maineq}:
  $$ \sqrt{N(N-1)} (\hat{\beta}_{1,PD} - \beta_1) = \Gamma^{-1} \sqrt{N(N-1)} U_N + o_p(1).
$$
And under the results of Lemma \ref{lemma:6}, it follows by the Slutsky theorem:
$$ \sqrt{N(N-1)} (\hat{\beta}_{1,PD} - \beta_1) \xrightarrow[]{d} N(0, \Gamma^{-1} 144 \Gamma^{-1}) = N(0, \Gamma^{-1} 72 \Delta_2 \Gamma^{-1}).$$
Therefore, the estimator $\hat{\beta}_{1,PD}$ is \textbf{normally distributed} and \textbf{asymptotically unbiased}.
\end{theorem}

\subsection{An estimator for the asymptotic variance of $\hat{\beta}_{1,PD}$}

From Theorem \ref{theorem:4}, it trivially follows that:
$$ \hat{\beta}_{1,PD} \overset{a}{\sim} N \left( \beta_1, \frac{1}{N(N-1)} \Gamma^{-1} 144 \delta_2 \Gamma^{-1}\right) $$
$$ \hat{\beta}_{1,PD} \overset{a}{\sim} N \left( \beta_1, \frac{1}{N(N-1)} \Gamma^{-1} 72 \Delta_2 \Gamma^{-1}\right) $$
Therefore, the asymptotic variance of the estimator $\hat{\beta}_{1,PD}$ can be estimated as:
$$ \widehat{\text{AVar}}(\hat{\beta}_{1,PD}) =  \frac{1}{N(N-1)} \hat{\Gamma}^{-1} 144 \hat{\delta}_2 \hat{\Gamma}^{-1} $$
$$ \widehat{\text{AVar}}(\hat{\beta}_{1,PD}) =  \frac{1}{N(N-1)} \hat{\Gamma}^{-1} 72 \hat{\Delta}_2 \hat{\Gamma}^{-1}, $$
\noindent where: 
$$ \hat{\Gamma} = \frac{1}{{N \choose 4}} \sum_{\mathcal{C} \in \mathcal{C}(\mathcal{N},4)} \frac{1}{4!} \sum_{\pi \in \mathcal{P}(\mathcal{C},4)} \Tilde{X}_{\pi_1\pi_2\pi_3\pi_4} \Tilde{X}_{\pi_1\pi_2\pi_3\pi_4}',$$
\noindent where we then have that the asymptotic variance can be estimated using either a consistent estimator for $\delta_2$ or a consistent estimator for $\Delta_2$. In the following subsections we will propose consistent estimators for both.

\subsubsection{A consistent estimator of $\delta_2$}

As mentioned before, the definition of $\delta_2$ is:
\begin{align}
    \delta_2 = \mathbbm{E} [\bar{s}_{ij} \bar{s}_{ij}'], \nonumber
\end{align}
\noindent where:
$$\bar{s}_{ij} = \mathbbm{E}[s_{ijkl} \rvert A_{i}, B_{j}, U_{ij}]$$
Importantly, the elements which we condition on, namely, $A_{i}, B_{j}, U_{ij}$ have indices $i$ and $j$ that necessarily are in the combinations $\{i,j,k,l\}$ for any other elements $k$ and $l$. This reflects the fact that the term $\delta_2$ originates from the expression of the variance of the statistic $U_N$, considering the components in such variance that has two elements in common. To obtain a consistent estimator $\hat{\delta}_2$, we also need a consistent estimator $\hat{\bar{s}}_{ij}$. 

\cite{graham2017econometric} suggests that, for an undirected network the consistent estimators are:
$$ \hat{\Delta}_{2,G} = \frac{1}{n} \sum_{i<j} \hat{\bar{s}}_{ij} \hat{\bar{s}}_{ij}', $$
$$ \hat{\bar{s}}_{ij,G} = \frac{1}{n - 2(N-1) + 1} \sum_{k<l, \{i,j\} \cap \{k,l\} = \emptyset} s_{ijkl}, $$
\noindent where $n = \frac{N(N-1)}{2}$ is the number of undirected dyads, therefore the expression for $\hat{\Delta}_{2,G}$ considers the average over all undirected dyads. The sum $\sum_{k<l, \{i,j\} \cap \{k,l\}}$ explicity means that, given two fixed indices $i$ and $j$ for the first dyad, we take the sum over all possible remaining different indiced $k$ and $l$, such that $k<l$, since in the context of \cite{graham2017econometric} we have an undirected network, and therefore only the different combinations $\{k,l\}$ matters, but not its different permutations. Moreover, notice that $n - 2(N-1) + 1$ coincides with the ${N-2 \choose 2}$ tetrads that contain a fixed $i$ and $j$. Therefore, the expression for $\hat{\bar{s}}_{ij,G}$ averages over all the kernels of the combinations that contain $i$ and $j$. 

As in our case we are looking at a directed network, some adjustments seem to be necessary. Especially, notice that, for a directed network we have that not necessarily $\bar{s}_{ij} = \bar{s}_{ji}$, since:
$$\mathbbm{E}[s_{ijkl} \rvert A_i, B_j, U_{ij}] \neq \mathbbm{E}[s_{ijkl} \rvert A_j, B_i, U_{ji}]$$
That means that in the expression for the consistent estimator of $\delta_2$ we should average over all possible directed dyads:
\begin{align} \label{eq:delta2}
  \hat{\delta}_{2} = \frac{1}{N(N-1)} \sum_{i=1}^N \sum_{j \neq i} \hat{\bar{s}}_{ij} \hat{\bar{s}}_{ij}' .
\end{align}
One possibility is to work out further the expression for $\bar{s}_{ji}$, such that it does not simply boil down to, when estimated, the average over the kernels.

To be more precise, we can see that, when taking the expectation over the kernel $s_{ijkl}$ conditioning on the characteristics of a single dyad $\{i,j\}$, only some of its permutations (that are inside the kernel, and namely the ones that contain the idiosyncratic error term $U_{ij}$) will have a conditional expectation different than zero:
    \begin{align}
        \bar{s}_{ij} &= \mathbbm{E}[s_{ijkl} \rvert A_i, B_j, U_{ij}] \\
        &= \mathbbm{E}[\frac{1}{4!} \sum_{\pi \in \mathcal{P}(\mathcal{C},4)} ((X_{\pi_1\pi_2} - X_{\pi_1\pi_3})-(X_{\pi_4\pi_2}-X_{\pi_4\pi_3})) ((U_{\pi_1\pi_2} - U_{\pi_1\pi_3})-(U_{\pi_4\pi_2}-U_{\pi_4\pi_3})) \rvert A_i, B_j, U_{ij}] \nonumber\\
        &= \frac{1}{4!} \Big( \mathbbm{E}[((X_{ij} - X_{ik}) - (X_{lj} - X_{lk}))((U_{ij} - U_{ik}) - (U_{lj} - U_{lk}))\rvert A_i, B_j, U_{ij}] \nonumber \\
        &+ \mathbbm{E} [((X_{ik} - X_{ij}) - (X_{lk} - X_{lj}))((U_{ik} - U_{ij}) - (U_{lk} - U_{lj})) \rvert A_i, B_j, U_{ij}] \nonumber \\
        &+ \mathbbm{E}[((X_{kj} - X_{kl}) - (X_{ij} - X_{il}))((U_{kj} - U_{kl}) - (U_{ij} - U_{il}))\rvert A_i, B_j, U_{ij}] \nonumber \\
        &+ \mathbbm{E}[((X_{lk} - X_{lj}) - (X_{ik} - X_{ij}))((U_{lk} - U_{lj}) - (U_{ik} - U_{ij}))\rvert A_i, B_j, U_{ij}] \nonumber \\
        &+ \mathbbm{E}[((X_{kl} - X_{kj}) - (X_{il} - X_{ij}))((U_{kl} - U_{kj}) - (U_{il} - U_{ij}))\rvert A_i, B_j, U_{ij}] \nonumber \\
        &+ \mathbbm{E}[((X_{lj} - X_{lk}) - (X_{ij} - X_{ik}))((U_{lj} - U_{lk}) - (U_{ij} - U_{ik}))\rvert A_i, B_j, U_{ij}] \nonumber \\
        &+ \mathbbm{E}[((X_{ij} - X_{il}) - (X_{kj} - X_{kl}))((U_{ij} - U_{il}) - (U_{kj} - U_{kl}))\rvert A_i, B_j, U_{ij}] \nonumber \\
        &+ \mathbbm{E}[((X_{il} - X_{ij}) - (X_{kl} - X_{kj}))((U_{il} - U_{ij}) - (U_{kl} - U_{kj}))\rvert A_i, B_j, U_{ij}] \Big),  \nonumber
    \end{align}
\noindent which is different than:
    \begin{align}
        \bar{s}_{ji} &= \mathbbm{E}[s_{ijkl} \rvert A_j, B_i, U_{ji}] \\
        &= \mathbbm{E}[\frac{1}{4!} \sum_{\pi \in \mathcal{P}(\mathcal{C},4)} ((X_{\pi_1\pi_2} - X_{\pi_1\pi_3})-(X_{\pi_4\pi_2}-X_{\pi_4\pi_3})) ((U_{\pi_1\pi_2} - U_{\pi_1\pi_3})-(U_{\pi_4\pi_2}-U_{\pi_4\pi_3})) \rvert A_j, B_i, U_{ji}] \nonumber\\
        &= \frac{1}{4!} \Big( \mathbbm{E}[((X_{ji} - X_{jk}) - (X_{li} - X_{lk}))((U_{ji} - U_{jk}) - (U_{li} - U_{lk}))\rvert A_j, B_i, U_{ji}] \nonumber \\
        &+ \mathbbm{E} [((X_{jk} - X_{ji}) - (X_{lk} - X_{li}))((U_{jk} - U_{ji}) - (U_{lk} - U_{li})) \rvert A_j, B_i, U_{ji}] \nonumber \\
        &+ \mathbbm{E}[((X_{ki} - X_{kl}) - (X_{ji} - X_{jl}))((U_{ki} - U_{kl}) - (U_{ji} - U_{jl}))\rvert A_j, B_i, U_{ji}] \nonumber \\
        &+ \mathbbm{E}[((X_{lk} - X_{li}) - (X_{jk} - X_{ji}))((U_{lk} - U_{li}) - (U_{jk} - U_{ji}))\rvert A_j, B_i, U_{ji}] \nonumber \\
        &+ \mathbbm{E}[((X_{kl} - X_{ki}) - (X_{jl} - X_{ji}))((U_{kl} - U_{ki}) - (U_{jl} - U_{ji}))\rvert A_j, B_i, U_{ji}] \nonumber \\
        &+ \mathbbm{E}[((X_{li} - X_{lk}) - (X_{ji} - X_{ik}))((U_{li} - U_{lk}) - (U_{ji} - U_{ik}))\rvert A_j, B_i, U_{ji}] \nonumber \\
        &+ \mathbbm{E}[((X_{ji} - X_{jl}) - (X_{ki} - X_{kl}))((U_{ji} - U_{jl}) - (U_{ki} - U_{kl}))\rvert A_j, B_i, U_{ji}] \nonumber \\
        &+ \mathbbm{E}[((X_{jl} - X_{ji}) - (X_{kl} - X_{ki}))((U_{jl} - U_{ji}) - (U_{kl} - U_{ki}))\rvert A_j, B_i, U_{ji}] \Big) . \nonumber
    \end{align}
Therefore, if we take the sample analogue of those expressions applied to a given combination $\{i,j,k,l\}$ that contain the fixed elements $i, j$ and any elements $k, l$:
    \begin{align}
        \hat{\bar{s}}_{ij} &= \frac{1}{n - 2(N-1) + 1} \sum_{k<l, \{i,j\} \cap \{k,l\} = \emptyset} \frac{1}{4!} \Big(((X_{ij} - X_{ik}) - (X_{lj} - X_{lk})) \hat{\Tilde{U}}_{ijkl} \nonumber \\
        &+ ((X_{ik} - X_{ij}) - (X_{lk} - X_{lj})) \hat{\Tilde{U}}_{ikjl} + ((X_{kj} - X_{kl}) - (X_{ij} - X_{il})) \hat{\Tilde{U}}_{kjli} \nonumber \\
        &+ ((X_{lk} - X_{lj}) - (X_{ik} - X_{ij})) \hat{\Tilde{U}}_{lkji} + ((X_{kl} - X_{kj}) - (X_{il} - X_{ij})) \hat{\Tilde{U}}_{klji} \nonumber \\
        &+ ((X_{lj} - X_{lk}) - (X_{ij} - X_{ik})) \hat{\Tilde{U}}_{ljki} + ((X_{ij} - X_{il}) - (X_{kj} - X_{kl})) \hat{\Tilde{U}}_{ijlk} \nonumber \\
        &+ ((X_{il} - X_{ij}) - (X_{kl} - X_{kj})) \hat{\Tilde{U}}_{iljk} \Big),  \nonumber
    \end{align}
    \begin{align}
        \hat{\bar{s}}_{ji} &=  \frac{1}{n - 2(N-1) + 1} \sum_{k<l, \{i,j\} \cap \{k,l\} = \emptyset} \frac{1}{4!} \Big(((X_{ji} - X_{jk}) - (X_{li} - X_{lk})) \hat{\Tilde{U}}_{jikl} \nonumber \\
        & +((X_{jk} - X_{ji}) - (X_{lk} - X_{li}))\hat{\Tilde{U}}_{jkil} + ((X_{ki} - X_{kl}) - (X_{ji} - X_{jl})) \hat{\Tilde{U}}_{kilj} \nonumber \\
        &+ ((X_{lk} - X_{li}) - (X_{jk} - X_{ji})) \hat{\Tilde{U}}_{lkij} + ((X_{kl} - X_{ki}) - (X_{jl} - X_{ji})) \hat{\Tilde{U}}_{klij} \nonumber \\
        &+ ((X_{li} - X_{lk}) - (X_{ji} - X_{jk})) \hat{\Tilde{U}}_{likj} + ((X_{ji} - X_{jl}) - (X_{ki} - X_{kl})) \hat{\Tilde{U}}_{jilk} \nonumber \\
        &+ ((X_{jl} - X_{ji}) - (X_{kl} - X_{ki})) \hat{\Tilde{U}}_{jlik} \Big).  \nonumber
    \end{align}
In the expressions above we plugged in the estimates of the idiosyncratic error terms, obtained from the estimated coefficient $\hat{\beta}_{1,PD}$, such that:
$$ \hat{\Tilde{U}}_{ijkl} = \Tilde{Y}_{ijkl}  - \hat{\beta}_{1,PD} \Tilde{X}_{ijkl}, $$
\noindent for any indices $i,j,k,l$.

Then, for these proposed consistent estimators we would have that $\hat{\bar{s}}_{ij} \neq \hat{\bar{s}}_{ji}$.

\subsubsection{A consistent estimator of $\Delta_2$}

In this case, we have that the previous definition of $\Delta_2$ is:
\begin{align}
  \Delta_2 = \text{Cov} (s_{ijkl}, s_{ijmp}) &= \mathbbm{E}[s_{ijkl}s_{ijmp}'] - \mathbbm{E}[s_{ijkl}] \mathbbm{E}[s_{ijmp}]' \\
  &= \mathbbm{E}[ \mathbbm{E}[s_{ijkl}s_{ijmp}' \rvert A_i, A_j, B_i, B_j, U_{ij}, U_{ji}]] \nonumber \\
  &= \mathbbm{E} [\bar{s}_{ij,2} \bar{s}_{ij,2}'] \nonumber \\
  &= \mathbbm{E} [\bar{s}_{ij,2}^2 ]. \nonumber
\end{align}
As the H\'{a}jek projection in this case was obtained by summing all combinations (and not permutations) of indices $i$ and $j$, we have that the consistent estimator of $\Delta_2$ should average over all these possible combinations:
\begin{align} \label{eq:Delta2}
  \hat{\Delta}_2 = \frac{2}{N(N-1)} \sum_{i=1}^N \sum_{j>i} \hat{\bar{s}}_{ij,2}^2 .
\end{align}
Moreover, remembering that $\bar{s}_{ij,2}$ is the kernel conditioning on all characteristics of $i$ and $j$, we have that its estimator, $\hat{\bar{s}}_{ij,2}$ is given by:
  \begin{align}
    \hat{\bar{s}}_{ij,2} &= \frac{1}{n - 2(N-1) + 1} \sum_{k<l, \{i,j\} \cap \{k,l\} = \emptyset} \frac{1}{4!} \Big(((X_{ij} - X_{ik}) - (X_{lj} - X_{lk})) \hat{\Tilde{U}}_{ijkl} \nonumber \\
    &+ ((X_{ik} - X_{ij}) - (X_{lk} - X_{lj})) \hat{\Tilde{U}}_{ikjl} + ((X_{kj} - X_{kl}) - (X_{ij} - X_{il})) \hat{\Tilde{U}}_{kjli} \nonumber \\
    &+ ((X_{lk} - X_{lj}) - (X_{ik} - X_{ij})) \hat{\Tilde{U}}_{lkji} + ((X_{kl} - X_{kj}) - (X_{il} - X_{ij})) \hat{\Tilde{U}}_{klji}  \nonumber \\
    &+ ((X_{lj} - X_{lk}) - (X_{ij} - X_{ik})) \hat{\Tilde{U}}_{ljki} + ((X_{ij} - X_{il}) - (X_{kj} - X_{kl})) \hat{\Tilde{U}}_{ijlk} \nonumber \\
    &+ ((X_{il} - X_{ij}) - (X_{kl} - X_{kj})) \hat{\Tilde{U}}_{iljk} + ((X_{ji} - X_{jk}) - (X_{li} - X_{lk})) \hat{\Tilde{U}}_{jikl} \nonumber \\
    &+ ((X_{jk} - X_{ji}) - (X_{lk} - X_{li})) \hat{\Tilde{U}}_{jkil}  + ((X_{ki} - X_{kl}) - (X_{ji} - X_{jl})) \hat{\Tilde{U}}_{kilj} \nonumber \\
    &+ ((X_{lk} - X_{li}) - (X_{jk} - X_{ji})) \hat{\Tilde{U}}_{lkij} + ((X_{kl} - X_{ki}) - (X_{jl} - X_{ji})) \hat{\Tilde{U}}_{klij} \nonumber \\
    &+ ((X_{li} - X_{lk}) - (X_{ji} - X_{i'k})) \hat{\Tilde{U}}_{likj} + ((X_{ji} - X_{jl}) - (X_{ki} - X_{kl})) \hat{\Tilde{U}}_{jilk} \nonumber \\
    &+ ((X_{jl} - X_{ji}) - (X_{kl} - X_{ki})) \hat{\Tilde{U}}_{jlik} \Big),  \nonumber
  \end{align}
\noindent where, again, in the expression above we plugged in the estimates of the idiosyncratic error terms, obtained from the estimated coefficient $\hat{\beta}_{1,PD}$, such that.

With both consistent estimates of the covariances, it is then possible to conduct valid inference. Moreover, in the next Section we investigate the finite sample performance of both analytical estimates.

\section{Simulations}

In this section, we explore the finite sample properties of the estimator $\hat{\beta}_{1,PD}$ through a Monte Carlo simulation exercise. We also aim to evaluate the finite sample properties of the estimator of the asymptotic variance of $\hat{\beta}_{1,PD}$, and the associated t-tests using both the consistent estimator $\hat{\delta}_2$, based on the first obtained Hájek projection, and the consistent estimator $\hat{\Delta}_2$, based on the second. In a nutshell, we find that: (i) the estimated slope parameters are unbiased in general, even when the fixed effects are correlated with the covariates; (ii) the estimated asymptotic variances using either estimators are very close to each other, which was expected; and (iii) the size of the t-tests are correct, indicating a valid inference procedure. 

\subsection{Data Generating Processes}

For simplifying purposes, for now, we consider the case of a single regressor $X_{ij}$ in the different proposed designs. In general, I follow closely the DGP specifications proposed by \cite{jochmans2018semiparametric} and \cite{charbonneau2017multiple}, who also consider a directed network model. Note, however, that in their cases, they consider a binary outcome variable, while we consider a continuous dependent variable.

The DGP in general follows:

$$ Y_{ij} = \beta_1 X_{ij} + \theta_i + \xi_j + U_{ij} $$

In all different designs, we take $\beta_1 =0$. The idiosyncratic error terms $U_{ij}$ are independently drawn from a standard normal distribution, $U_{ij} \sim N(0,1)$. In our case of a directed network, we specifically have that $U_{ij} \neq U_{ji}$, therefore for a simulation considering $N$ nodes we draw from the standard normal $N(N-1)$ idiosyncratic errors.
The fixed effects $\theta_i$ and $\xi_j$ are also drawn from standard normal distributions.

The difference among the designs relies on how the regressor $X_{ij}$ is drawn.

\subsubsection{Design 1}

Here we follow essentially the same DGP as proposed by \cite{jochmans2018semiparametric}. We generate the single regressor as:

$$X_{ij} = - \rvert A_i - B_j \rvert $$

\noindent where $ A_i = V_i - \frac{1}{2}$, for $V_i \sim \text{Beta}(2,2)$, and the same for $B_j$. The covariate thus is generated in such a way that is dependent across both senders and receivers in the dyadic relation. The difference to the DGP proposed by \cite{jochmans2018semiparametric} relies on the fact that we consider the individual effect of the \textit{alter}, $A_i$, to be different and drawn independently from that of the \textit{ego}, $B_j$, while \cite{jochmans2018semiparametric} considers $B_j = A_j$. 

\subsubsection{Design 2}

We introduce a correlation between the regressor $X_{ij}$ and the fixed effects $\theta_i$ and $\xi_j$, such that:

$$X_{ij} = - \rvert A_i - B_j \rvert + \theta_i + \xi_j $$

\noindent where $ A_i = V_i - \frac{1}{2}$, for $V_i \sim \text{Beta}(2,2)$. Also, $B_j$ is drawn independently from $A_i$, such that $ B_j = V_j - \frac{1}{2}$, for $V_j \sim \text{Beta}(2,2)$. Note that the manner in which we introduce a correlation between the regressor and the fixed effects is similar to that of \cite{charbonneau2017multiple}.

\subsubsection{Design 3}

We now consider a binary regressor that is uncorrelated with the fixed effects. We generate the regressor according to:

$$X_{ij} = \mathbbm{1} \{ A_i - B_j > 0 \}$$

\noindent where $A_i$ and $B_j$ are drawn according to Designs 1 and 2.

\subsubsection{Design 4}

We again consider a binary regressor, however, now it is correlated with the fixed effects, such that:

$$X_{ij} = \mathbbm{1} \{ A_i - B_j + \theta_i + \xi_j > 0 \}$$

\noindent where, again, $A_i$ and $B_j$ are drawn according to Designs 1 and 2.

\subsection{Results of Monte Carlo simulations}

We propose several settings of Monte Carlo simulations. For each design, we run simulations for $S \in \{1000, 5000, 10000\}$, where $S$ refers to the number of simulations, and for $N \in \{10, 20, 30, 50\}$.

\subsubsection{Results for the estimator $\hat{\beta}_{1,PD}$ and its estimated asymptotic variance}

In the tables below we show the results for the estimator $\hat{\beta}_{1,PD}$ in terms of biasedness, as well as  its variance across the simulations. We also present the results for the average of the estimated asymptotic variance considering both the estimations taking into account $\hat{\delta}_2$, according to Equation \ref{eq:delta2}, and taking into account $\hat{\Delta}_2$, according to Equation \ref{eq:Delta2}.

\begin{table}[H]
    \centering
    \begin{tabular}{rrrrrrrrrrrrrrrrrrrrrrrr}
      \hline
       Simulations & N & bias($\hat{\beta}_1$) & var($\hat{\beta}_1$) & $\text{mean}(\hat{\text
       {var}}(\hat{\beta}_1))_{\hat{\delta}_{2}}$ & $\text{mean}(\hat{\text
       {var}}(\hat{\beta}_1))_{\hat{\Delta}_{2}}$ \\
      \hline
      1000 & 10 & -0.007 & 0.165 & 0.225 & 0.197 \\
      1000 & 20 &  0.001 & 0.039 & 0.042 & 0.040 \\
      1000 & 30 & -0.001 & 0.013 & 0.017 & 0.016 \\
      1000 & 50 & -0.000 & 0.005 & 0.006 & 0.005 \\ 
      5000 & 10 & -0.002 & 0.180 & 0.224 & 0.197 \\
      5000 & 20 & -0.001 & 0.035 & 0.042 & 0.040 \\
      5000 & 30 & -0.001 & 0.015 & 0.017 & 0.016 \\
      5000 & 50 & -0.001 & 0.005 & 0.005 & 0.005 \\
     10000 & 10 &  0.007 & 0.176 & 0.224 & 0.197 \\
     10000 & 20 &  0.000 & 0.035 & 0.042 & 0.040 \\
     10000 & 30 & -0.000 & 0.015 & 0.017 & 0.016 \\
     10000 & 50 & -0.001 & 0.005 & 0.005 & 0.005 \\ 
      \hline
  \end{tabular}
    \caption{Results of the Monte Carlo Simulation of the Pairwise Differences estimators obtained for the first data generating process}
    \label{tab:results1}
\end{table} 

\begin{table}[H]
  \centering
  \begin{tabular}{rrrrrrrrrrrrrrrrrrrrrr}
    \hline
    Simulations & N & bias($\hat{\beta}_1$) & var($\hat{\beta}_1$) & $\text{mean}(\hat{\text
    {var}}(\hat{\beta}_1))_{\hat{\delta}_{2}}$ & $\text{mean}(\hat{\text
    {var}}(\hat{\beta}_1))_{\hat{\Delta}_{2}}$ \\
   \hline
   1000 & 10 &  0.021 & 0.181 & 0.222 & 0.194 \\
   1000 & 20 & -0.004 & 0.036 & 0.042 & 0.041 \\
   1000 & 30 &  0.000 & 0.015 & 0.017 & 0.016 \\
   1000 & 50 &  0.002 & 0.006 & 0.006 & 0.005 \\
   5000 & 10 &  0.001 & 0.179 & 0.224 & 0.196 \\
   5000 & 20 & -0.004 & 0.036 & 0.041 & 0.040 \\
   5000 & 30 & -0.001 & 0.016 & 0.017 & 0.016 \\
   5000 & 50 & -0.000 & 0.005 & 0.005 & 0.005 \\ 
  10000 & 10 &  0.002 & 0.174 & 0.225 & 0.198 \\
  10000 & 20 & -0.001 & 0.035 & 0.042 & 0.040 \\
  10000 & 30 &  0.001 & 0.015 & 0.017 & 0.016 \\
  10000 & 50 & -0.000 & 0.005 & 0.006 & 0.005 \\
      \hline
  \end{tabular}
  \caption{Results of the Monte Carlo Simulation of the Pairwise Differences estimators obtained for the second data generating process}
  \label{tab:results2}
\end{table} 

  \begin{table}[H]
  \centering
  \begin{tabular}{rrrrrrrrrrrrrrrrrrrrrr}
      \hline
      Simulations & N & bias($\hat{\beta}_1$) & var($\hat{\beta}_1$) & $\text{mean}(\hat{\text
      {var}}(\hat{\beta}_1))_{\hat{\delta}_{2}}$ & $\text{mean}(\hat{\text
      {var}}(\hat{\beta}_1))_{\hat{\Delta}_{2}}$ \\
     \hline
     1000 & 10 & -0.004 & 0.180 & 0.221 & 0.192 \\
     1000 & 20 &  0.004 & 0.033 & 0.042 & 0.040 \\
     1000 & 30 &  0.004 & 0.015 & 0.017 & 0.016 \\
     1000 & 50 &  0.002 & 0.005 & 0.005 & 0.005 \\
     5000 & 10 &  0.004 & 0.179 & 0.224 & 0.197 \\
     5000 & 20 & -0.001 & 0.034 & 0.042 & 0.040 \\
     5000 & 30 &  0.001 & 0.015 & 0.017 & 0.016 \\
     5000 & 50 & -0.001 & 0.005 & 0.005 & 0.005 \\
    10000 & 10 & -0.004 & 0.175 & 0.224 & 0.197 \\
    10000 & 20 & -0.003 & 0.036 & 0.042 & 0.040 \\
    10000 & 30 & -0.001 & 0.015 & 0.017 & 0.016 \\
    10000 & 50 &  0.000 & 0.005 & 0.005 & 0.005 \\
        \hline
    \end{tabular}
  \caption{Results of the Monte Carlo Simulation of the Pairwise Differences estimators obtained for the third data generating process}
  \label{tab:results3}
\end{table} 

  \begin{table}[H]
  \centering
  \begin{tabular}{rrrrrrrrrrrrrrrrrrrrrrr}
    \hline
    Simulations & N & bias($\hat{\beta}_1$) & var($\hat{\beta}_1$) & $\text{mean}(\hat{\text
    {var}}(\hat{\beta}_1))_{\hat{\delta}_{2}}$ & $\text{mean}(\hat{\text
    {var}}(\hat{\beta}_1))_{\hat{\Delta}_{2}}$ \\
    \hline
    1000 & 10 & -0.001 & 0.173 & 0.229 & 0.203 \\
    1000 & 20 &  0.003 & 0.035 & 0.041 & 0.040 \\
    1000 & 30 & -0.003 & 0.015 & 0.017 & 0.016 \\
    1000 & 50 & -0.001 & 0.005 & 0.005 & 0.005 \\ 
    5000 & 10 & -0.001 & 0.170 & 0.223 & 0.197 \\
    5000 & 20 &  0.002 & 0.035 & 0.042 & 0.040 \\
    5000 & 30 & -0.001 & 0.015 & 0.017 & 0.016 \\
    5000 & 50 &  0.000 & 0.005 & 0.005 & 0.005 \\
   10000 & 10 & -0.004 & 0.177 & 0.223 & 0.196 \\
   10000 & 20 & -0.002 & 0.036 & 0.042 & 0.040 \\
   10000 & 30 &  0.000 & 0.015 & 0.017 & 0.016 \\
   10000 & 50 & -0.000 & 0.005 & 0.005 & 0.005 \\
    \hline
\end{tabular}
  \caption{Results of the Monte Carlo Simulation of the Pairwise Differences estimators obtained for the fourth data generating process}
  \label{tab:results4}
\end{table}

From the tables above, we point out two results: (i) the estimator $\hat{\beta}_1$ seems to be unbiased, and (ii) the mean of the estimated variaces is basically on spot when compared to the variance of $\hat{\beta}_1$ across the simulations and across the different designs. More specifically, while for all designs (except for Design 3) there is still some bias in the simulations for $N=10$ and $S=1000$, the bias essentially vanishes as we consider larger numbers of nodes $N$, or a larger number of simulations $S$. 

Another feature that was already expected is that the average of the estimated asymptotic variances are very close when comparing to whether the variance was estimated using $\hat{\delta}_2$ or $\hat{\Delta}_2$. Moreover, we notice that in general those averages are almost spot on with the variances of the estimated $\hat{\beta}_1$ across simulations. The only exception are the simulations with $N=10$ for designs 1 and 2, however, as soon as $N$ is increased the results are again essentially the same. This indicates that the variance estimator captures well the small-sample variability in the point estimator, and that inference using such estimators is valid.

We next explore if normality might be a good approximation to the finite sample distribution of the proposed estimator $\hat{\beta}_1$. We present below the histograms and the QQ-plots of the estimated values for Designs 2 and 4, which are considered to be the most relevant, since it allows for correlations between the covariates and the fixed-effects. However, the plots for the other designs can be found in Appendix D.

\begin{figure}[H]
  \centering
  \begin{subfigure}[b]{0.49\textwidth}
  \label{ fig7} 
  \begin{minipage}[b]{0.5\linewidth}
    \centering
    \includegraphics[width=1\linewidth]{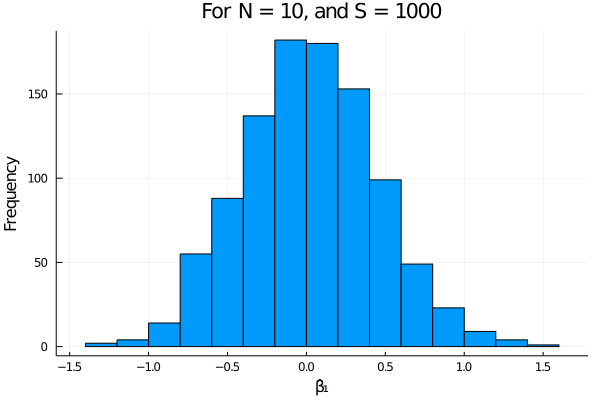} 
  \end{minipage}%%
  \begin{minipage}[b]{0.5\linewidth}
    \centering
    \includegraphics[width=1\linewidth]{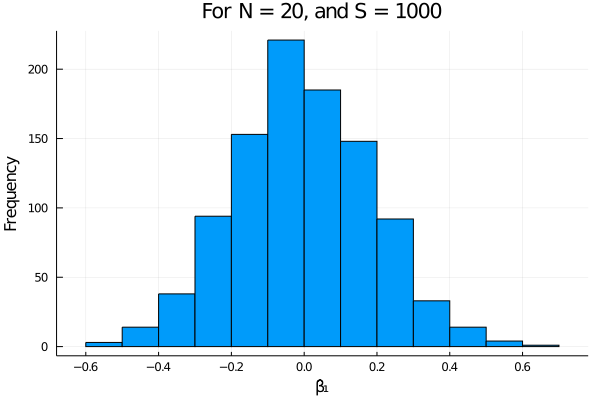} 
  \end{minipage} 
  \begin{minipage}[b]{0.5\linewidth}
    \centering
    \includegraphics[width=1\linewidth]{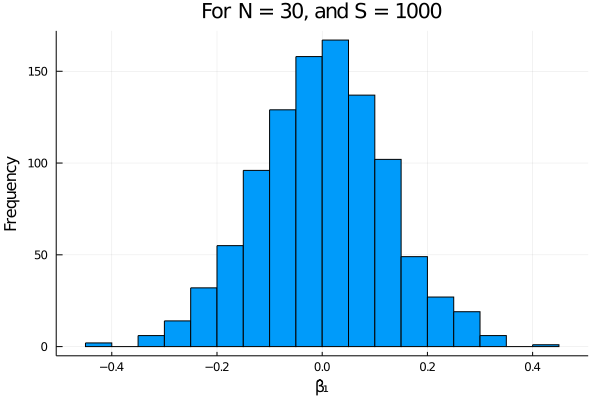} 
  \end{minipage}%% 
  \begin{minipage}[b]{0.5\linewidth}
    \centering
    \includegraphics[width=1\linewidth]{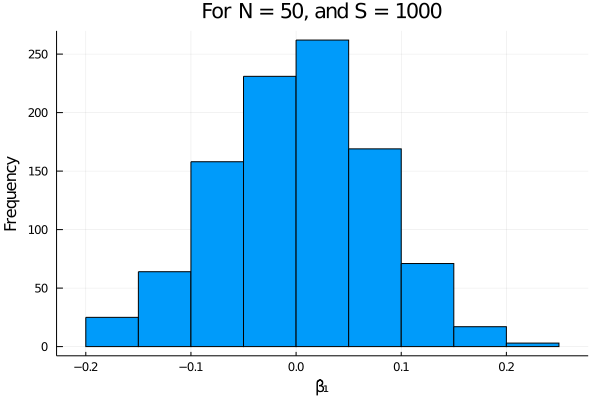} 
  \end{minipage} 
  \caption*{Histograms of $\hat{\beta}_1$ for design 2 and $S=1000$} 
\end{subfigure}
\begin{subfigure}[b]{0.49\textwidth} 
  \label{ fig7} 
  \begin{minipage}[b]{0.5\linewidth}
    \centering
    \includegraphics[width=1\linewidth]{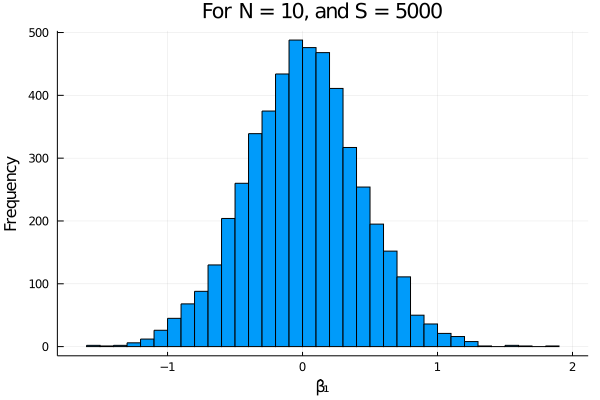} 
  \end{minipage}%%
  \begin{minipage}[b]{0.5\linewidth}
    \centering
    \includegraphics[width=1\linewidth]{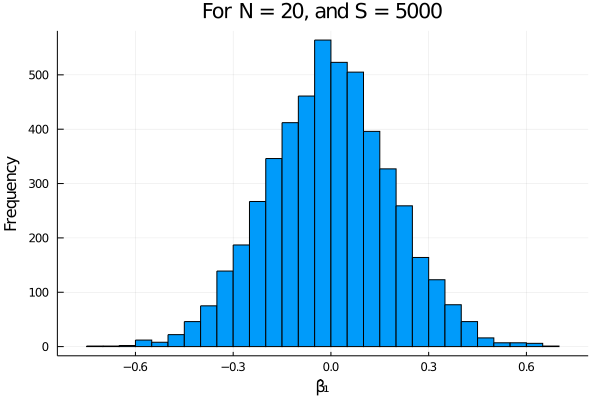} 
  \end{minipage} 
  \begin{minipage}[b]{0.5\linewidth}
    \centering
    \includegraphics[width=1\linewidth]{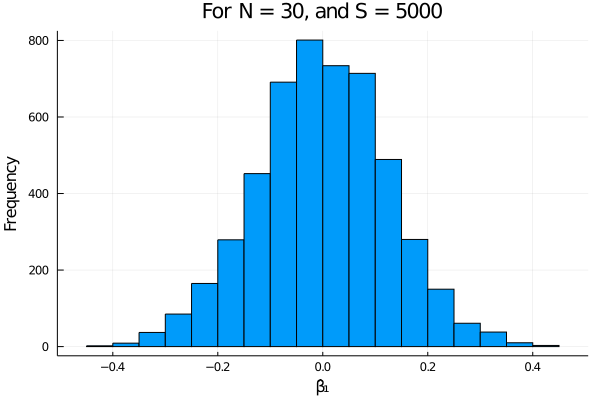} 
  \end{minipage}%% 
  \begin{minipage}[b]{0.5\linewidth}
    \centering
    \includegraphics[width=1\linewidth]{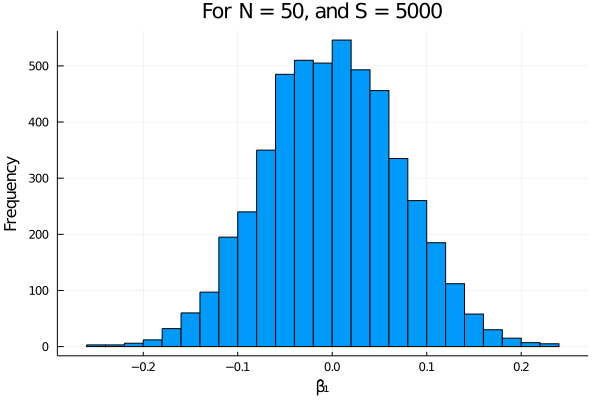} 
  \end{minipage} 
  \caption*{Histograms of $\hat{\beta}_1$ for design 2 and $S=5000$} 
\end{subfigure}
\end{figure}

\begin{figure}[H]
  \centering
  \begin{subfigure}[b]{0.49\textwidth}
   \label{ fig7} 
  \begin{minipage}[b]{0.5\linewidth}
    \centering
    \includegraphics[width=1\linewidth]{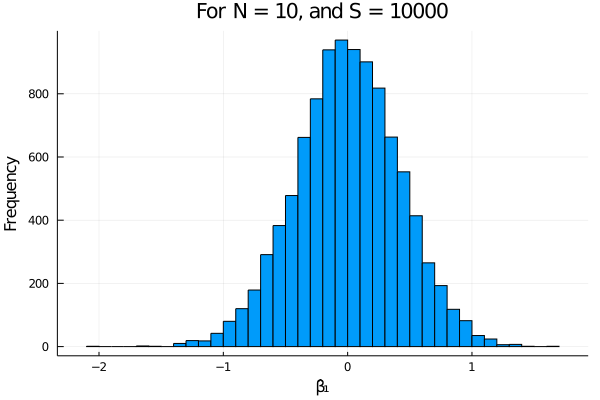} 
  \end{minipage}%%
  \begin{minipage}[b]{0.5\linewidth}
    \centering
    \includegraphics[width=1\linewidth]{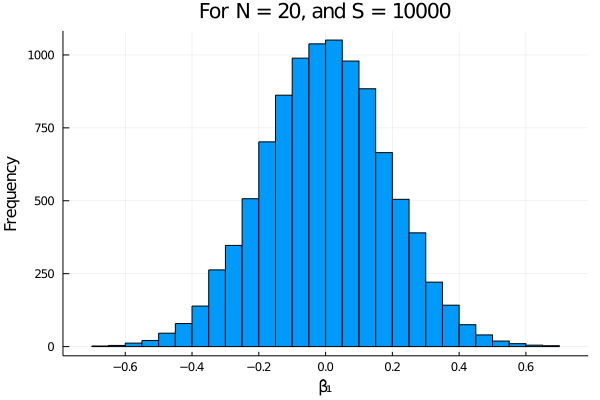} 
  \end{minipage} 
  \begin{minipage}[b]{0.5\linewidth}
    \centering
    \includegraphics[width=1\linewidth]{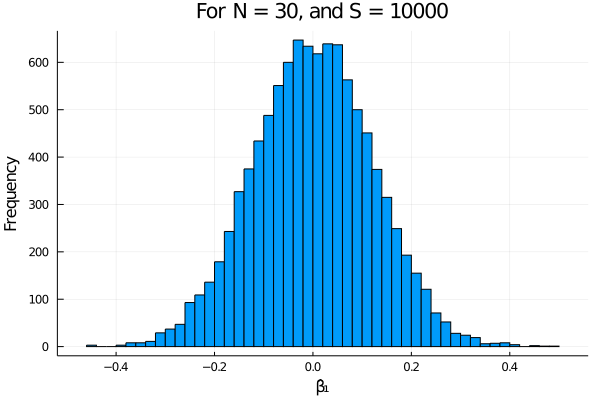} 
  \end{minipage}%% 
  \begin{minipage}[b]{0.5\linewidth}
    \centering
    \includegraphics[width=1\linewidth]{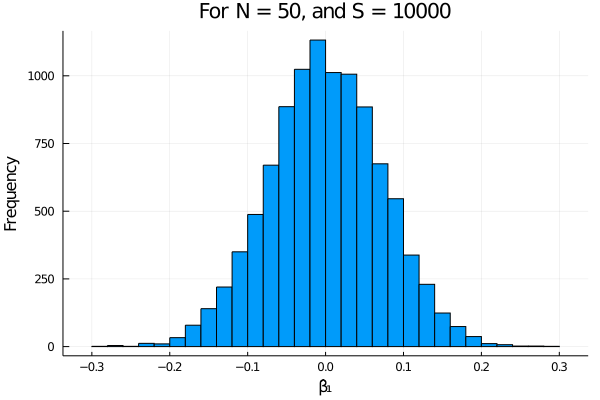} 
  \end{minipage} 
  \caption*{Histograms of $\hat{\beta}_1$ for design 2 and $S=10000$} 
\end{subfigure}
\begin{subfigure}[b]{0.49\textwidth}  
  \label{ fig7} 
  \begin{minipage}[b]{0.5\linewidth}
    \centering
    \includegraphics[width=1\linewidth]{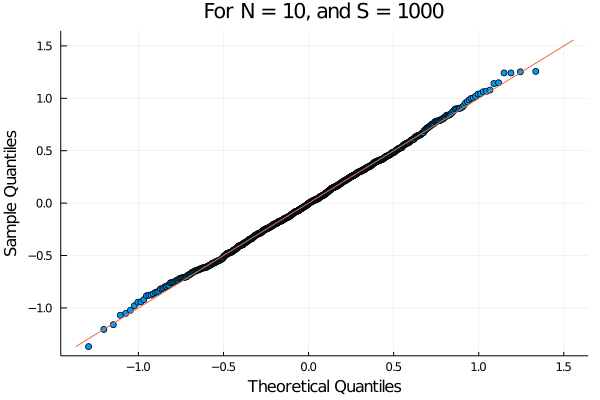} 
  \end{minipage}%%
  \begin{minipage}[b]{0.5\linewidth}
    \centering
    \includegraphics[width=1\linewidth]{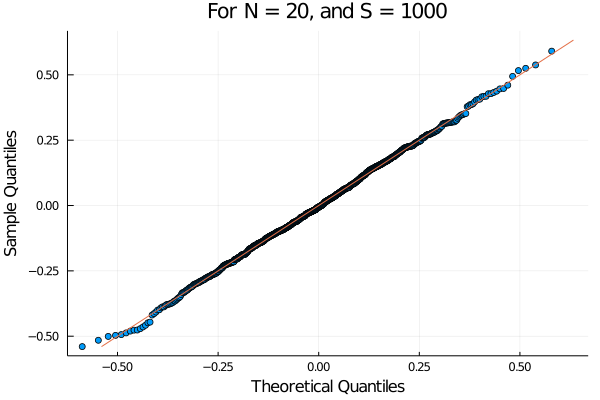} 
  \end{minipage} 
  \begin{minipage}[b]{0.5\linewidth}
    \centering
    \includegraphics[width=1\linewidth]{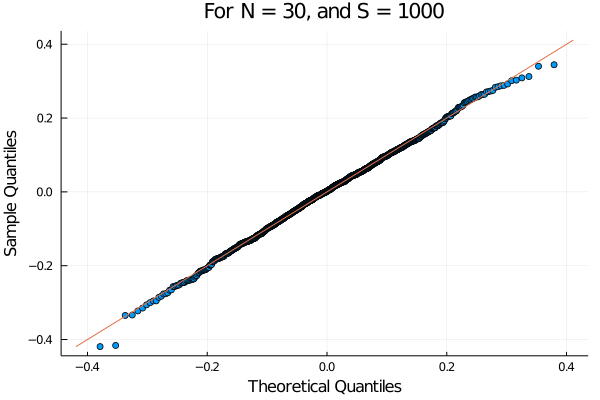} 
  \end{minipage}%% 
  \begin{minipage}[b]{0.5\linewidth}
    \centering
    \includegraphics[width=1\linewidth]{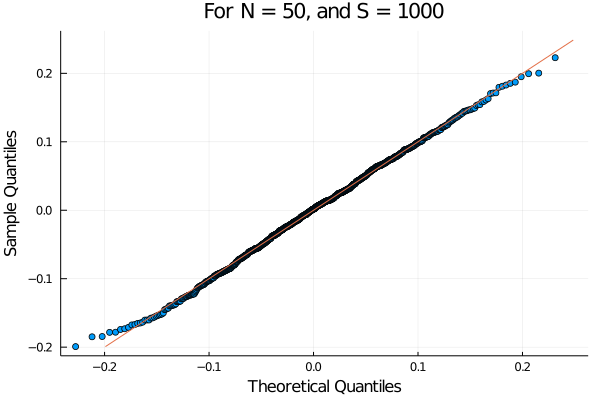} 
  \end{minipage} 
  \caption*{QQ-plot of $\hat{\beta}_1$ for design 2 and $S=1000$} 
\end{subfigure}
\end{figure}

\begin{figure}[H]
  \centering
  \begin{subfigure}[b]{0.49\textwidth}
    \label{ fig7} 
  \begin{minipage}[b]{0.5\linewidth}
    \centering
    \includegraphics[width=1\linewidth]{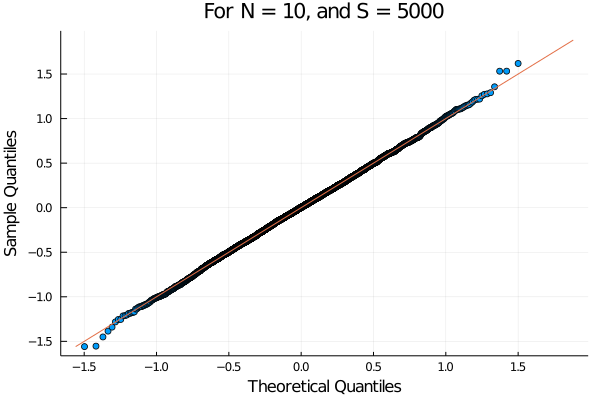} 
  \end{minipage}%%
  \begin{minipage}[b]{0.5\linewidth}
    \centering
    \includegraphics[width=1\linewidth]{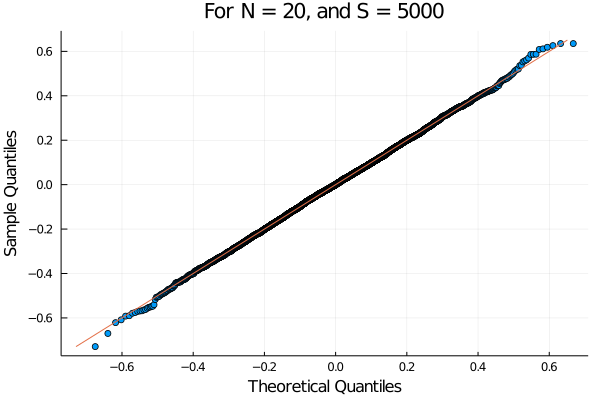} 
  \end{minipage} 
  \begin{minipage}[b]{0.5\linewidth}
    \centering
    \includegraphics[width=1\linewidth]{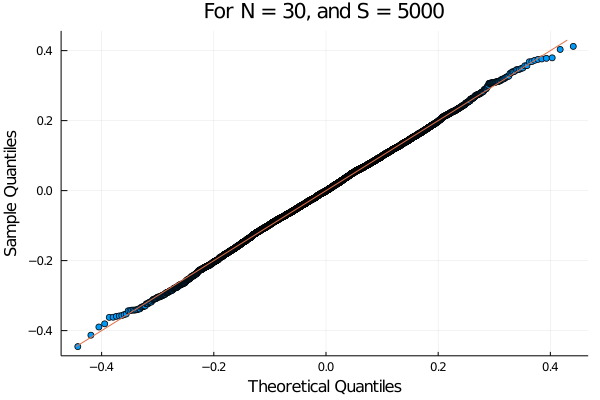} 
  \end{minipage}%% 
  \begin{minipage}[b]{0.5\linewidth}
    \centering
    \includegraphics[width=1\linewidth]{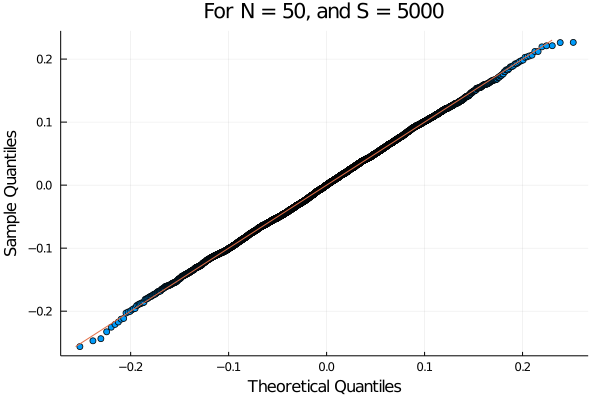} 
  \end{minipage} 
  \caption*{QQ-plot of $\hat{\beta}_1$ for design 2 and $S=5000$} 
\end{subfigure}
\begin{subfigure}[b]{0.49\textwidth} 
  \label{ fig7} 
  \begin{minipage}[b]{0.5\linewidth}
    \centering
    \includegraphics[width=1\linewidth]{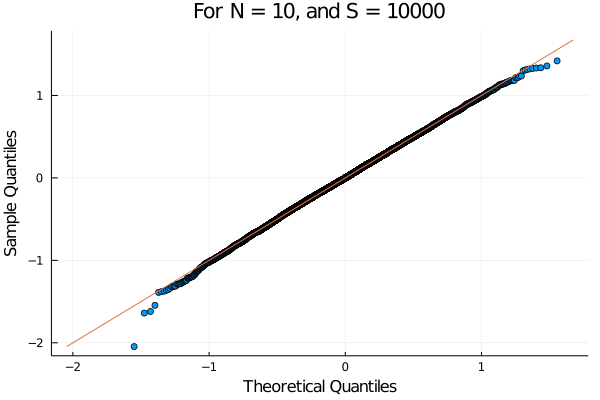} 
  \end{minipage}%%
  \begin{minipage}[b]{0.5\linewidth}
    \centering
    \includegraphics[width=1\linewidth]{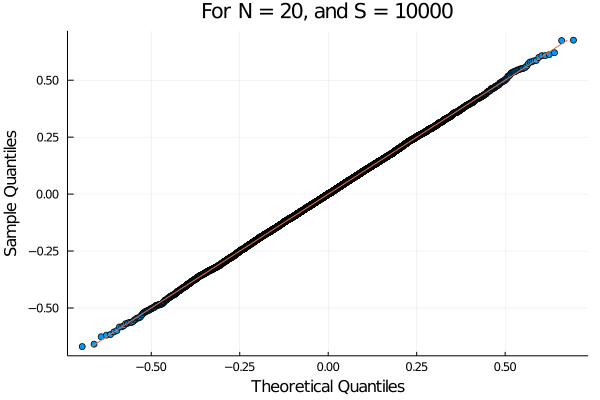} 
  \end{minipage} 
  \begin{minipage}[b]{0.5\linewidth}
    \centering
    \includegraphics[width=1\linewidth]{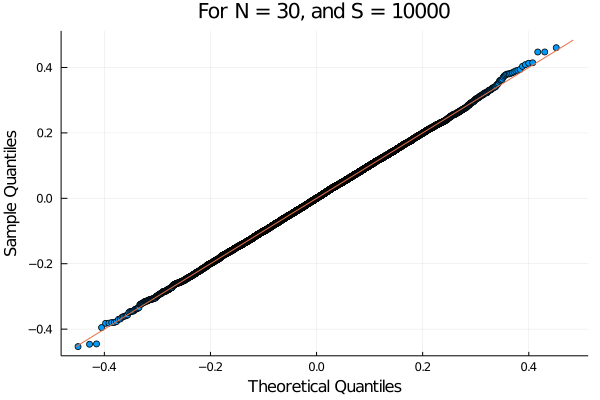} 
  \end{minipage}%% 
  \begin{minipage}[b]{0.5\linewidth}
    \centering
    \includegraphics[width=1\linewidth]{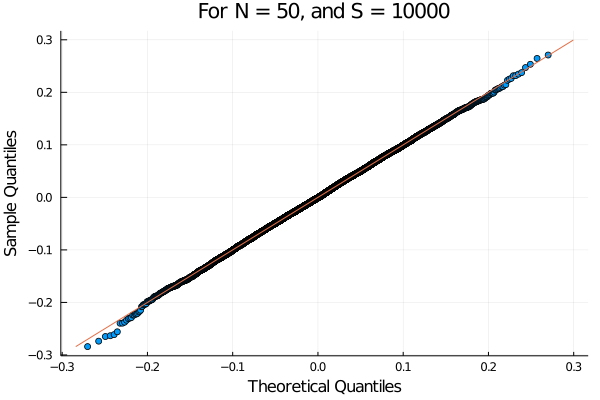} 
  \end{minipage} 
  \caption*{QQ-plot of $\hat{\beta}_1$ for design 2 and $S=10000$} 
\end{subfigure}
\end{figure}

\begin{figure}[H]
  \centering
  \begin{subfigure}[b]{0.49\textwidth}
      \label{ fig7} 
  \begin{minipage}[b]{0.5\linewidth}
    \centering
    \includegraphics[width=1\linewidth]{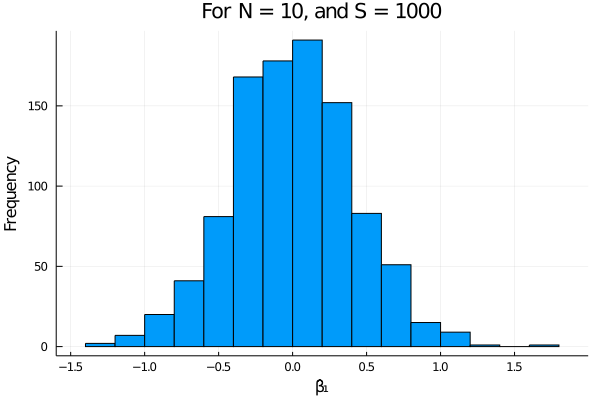} 
  \end{minipage}%%
  \begin{minipage}[b]{0.5\linewidth}
    \centering
    \includegraphics[width=1\linewidth]{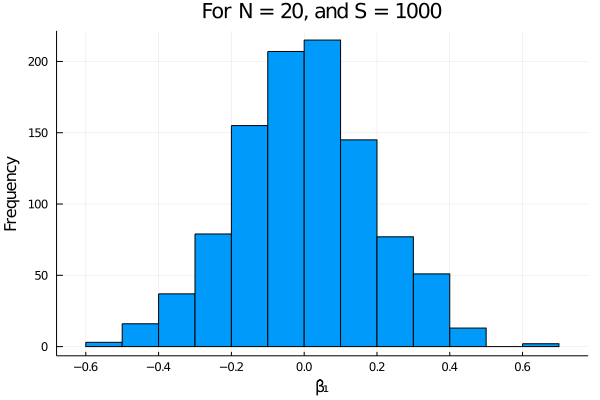} 
  \end{minipage} 
  \begin{minipage}[b]{0.5\linewidth}
    \centering
    \includegraphics[width=1\linewidth]{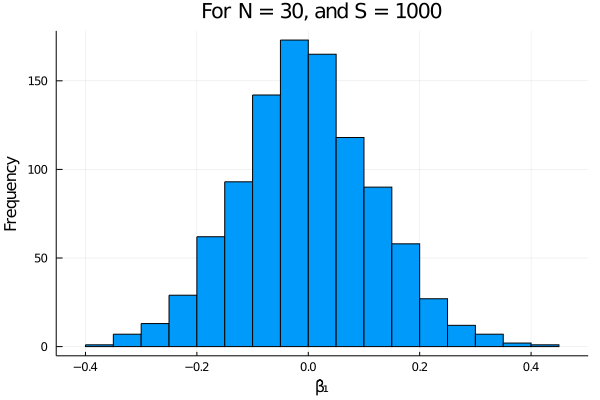} 
  \end{minipage}%% 
  \begin{minipage}[b]{0.5\linewidth}
    \centering
    \includegraphics[width=1\linewidth]{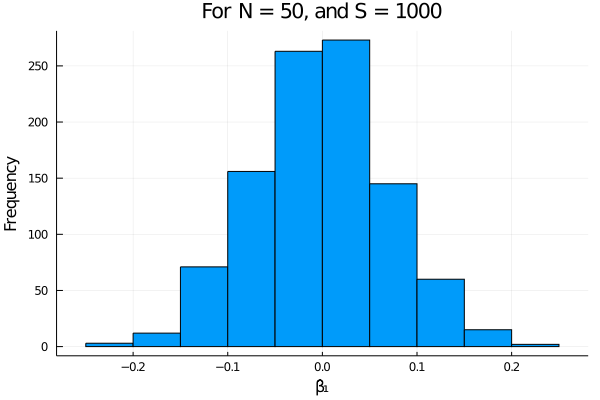} 
  \end{minipage} 
  \caption*{Histograms of $\hat{\beta}_1$ for design 4 and $S=1000$} 
\end{subfigure}
\begin{subfigure}[b]{0.49\textwidth} 
  \label{ fig7} 
  \begin{minipage}[b]{0.5\linewidth}
    \centering
    \includegraphics[width=1\linewidth]{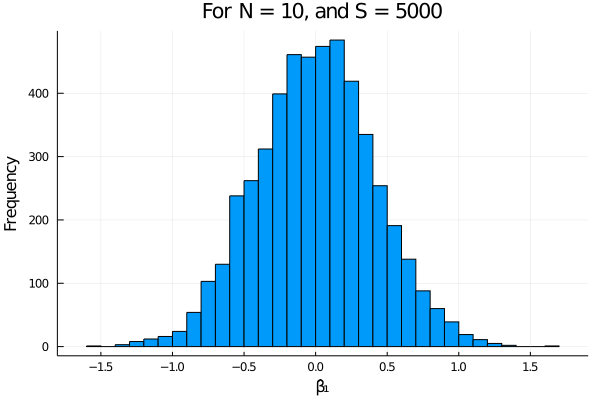} 
  \end{minipage}%%
  \begin{minipage}[b]{0.5\linewidth}
    \centering
    \includegraphics[width=1\linewidth]{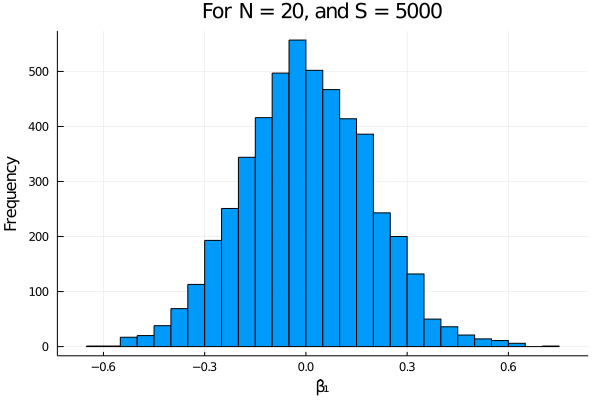} 
  \end{minipage} 
  \begin{minipage}[b]{0.5\linewidth}
    \centering
    \includegraphics[width=1\linewidth]{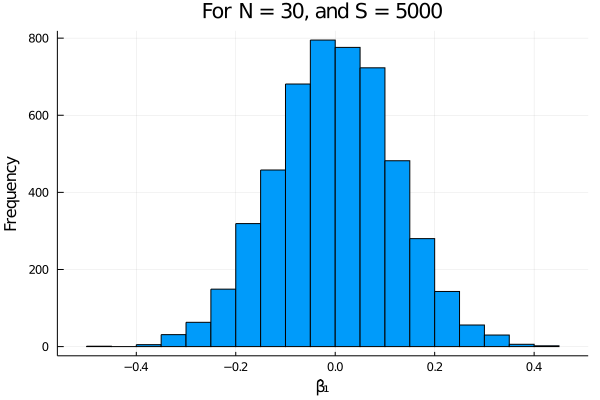} 
  \end{minipage}%% 
  \begin{minipage}[b]{0.5\linewidth}
    \centering
    \includegraphics[width=1\linewidth]{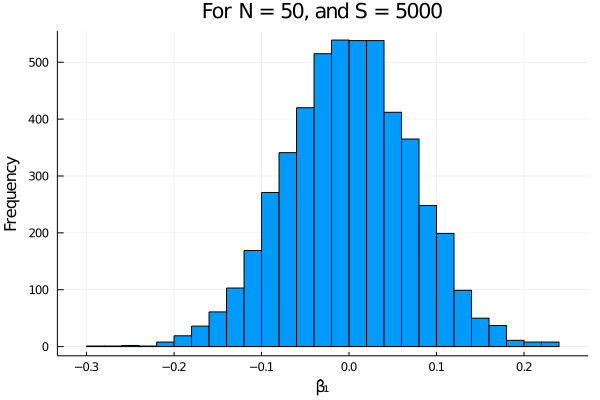} 
  \end{minipage} 
  \caption*{Histograms of $\hat{\beta}_1$ for design 4 and $S=5000$} 
\end{subfigure}
\end{figure}

\begin{figure}[H]
  \centering
  \begin{subfigure}[b]{0.49\textwidth}
      \label{ fig7} 
  \begin{minipage}[b]{0.5\linewidth}
    \centering
    \includegraphics[width=1\linewidth]{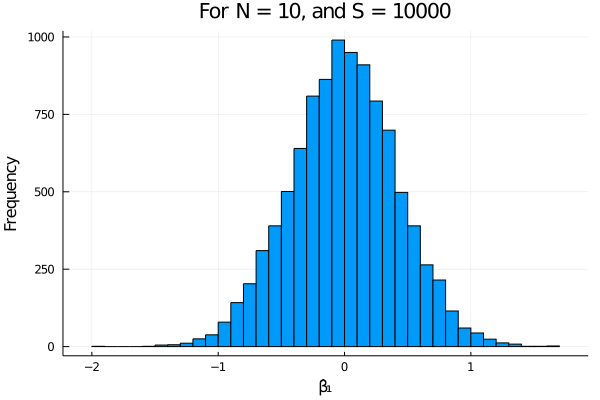} 
  \end{minipage}%%
  \begin{minipage}[b]{0.5\linewidth}
    \centering
    \includegraphics[width=1\linewidth]{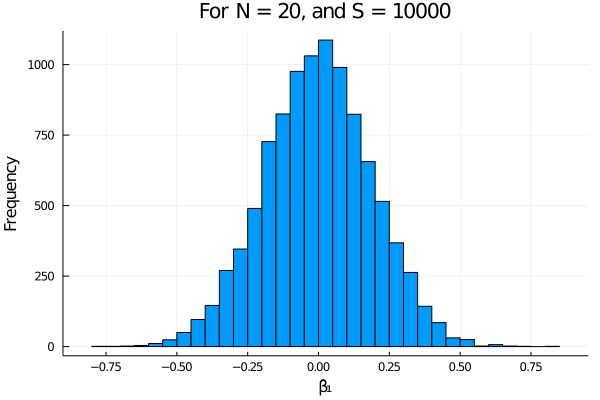} 
  \end{minipage} 
  \begin{minipage}[b]{0.5\linewidth}
    \centering
    \includegraphics[width=1\linewidth]{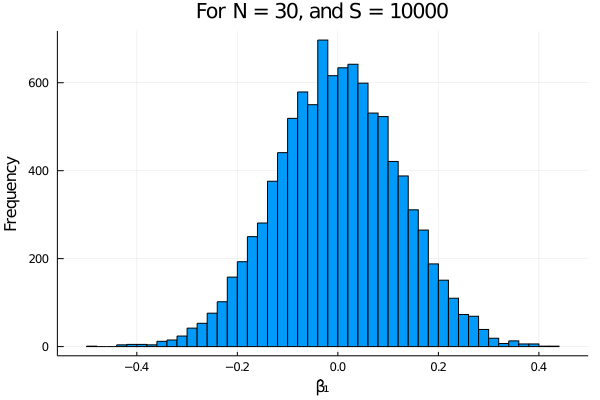} 
  \end{minipage}%% 
  \begin{minipage}[b]{0.5\linewidth}
    \centering
    \includegraphics[width=1\linewidth]{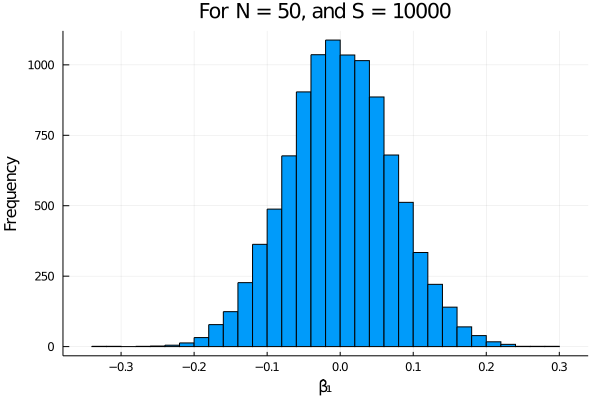} 
  \end{minipage} 
  \caption*{Histograms of $\hat{\beta}_1$ for design 4 and $S=10000$} 
\end{subfigure}
\begin{subfigure}[b]{0.49\textwidth} 
  \label{ fig7} 
  \begin{minipage}[b]{0.5\linewidth}
    \centering
    \includegraphics[width=1\linewidth]{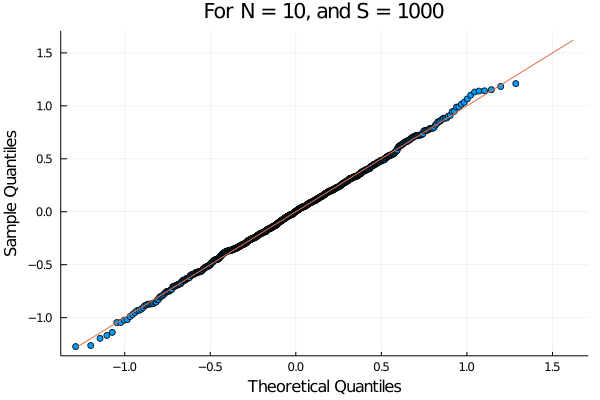} 
  \end{minipage}%%
  \begin{minipage}[b]{0.5\linewidth}
    \centering
    \includegraphics[width=1\linewidth]{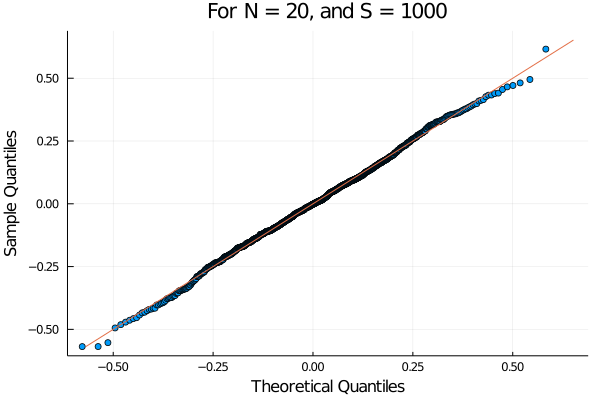} 
  \end{minipage} 
  \begin{minipage}[b]{0.5\linewidth}
    \centering
    \includegraphics[width=1\linewidth]{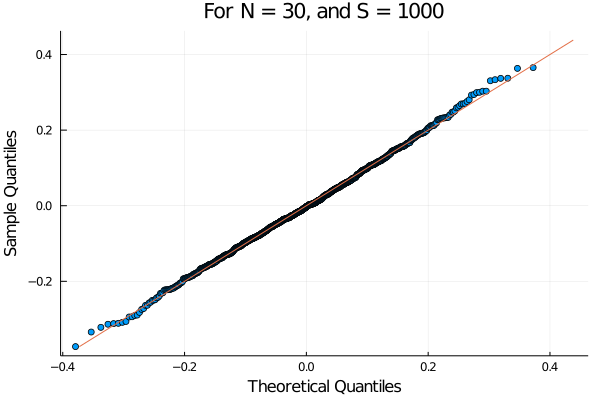} 
  \end{minipage}%% 
  \begin{minipage}[b]{0.5\linewidth}
    \centering
    \includegraphics[width=1\linewidth]{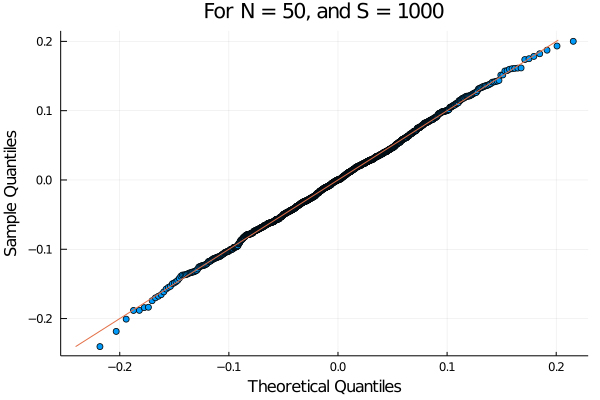} 
  \end{minipage} 
  \caption*{QQ-plot of $\hat{\beta}_1$ for design 4 and $S=1000$}
\end{subfigure} 
\end{figure}

\begin{figure}[H]
  \centering
  \begin{subfigure}[b]{0.49\textwidth}
      \label{ fig7} 
  \begin{minipage}[b]{0.5\linewidth}
    \centering
    \includegraphics[width=1\linewidth]{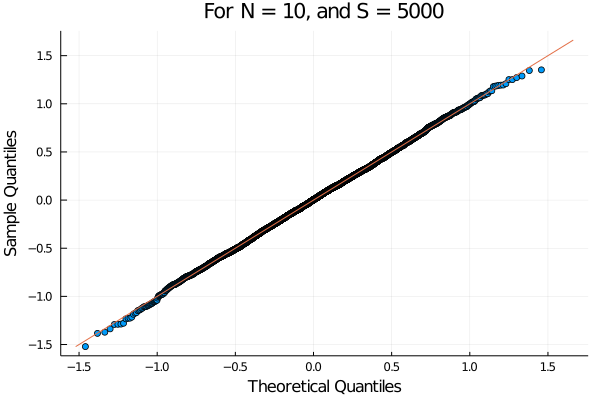} 
  \end{minipage}%%
  \begin{minipage}[b]{0.5\linewidth}
    \centering
    \includegraphics[width=1\linewidth]{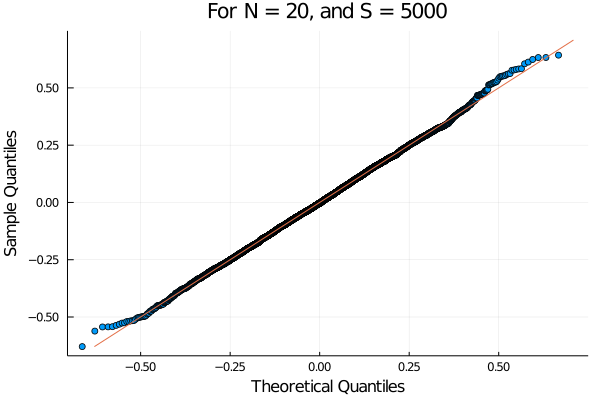} 
  \end{minipage} 
  \begin{minipage}[b]{0.5\linewidth}
    \centering
    \includegraphics[width=1\linewidth]{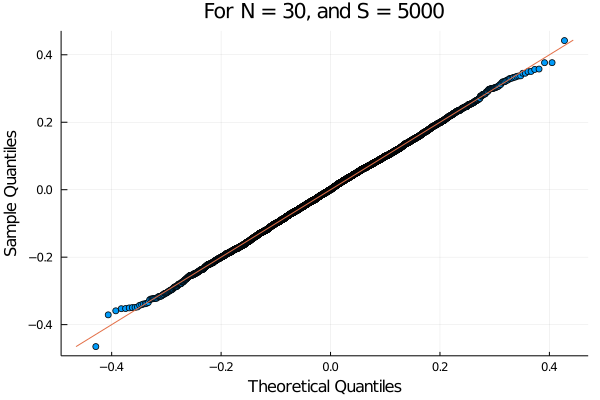} 
  \end{minipage}%% 
  \begin{minipage}[b]{0.5\linewidth}
    \centering
    \includegraphics[width=1\linewidth]{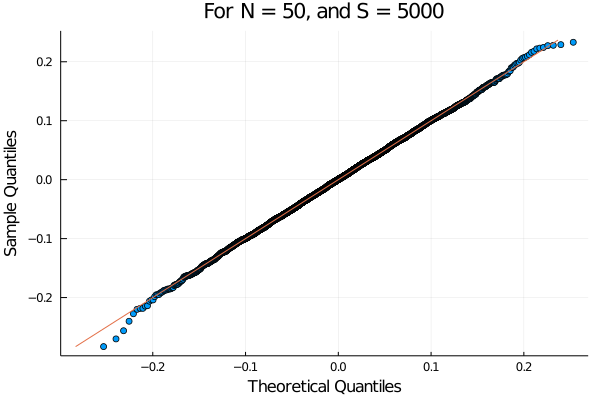} 
  \end{minipage} 
  \caption*{QQ-plot of $\hat{\beta}_1$ for design 4 and $S=5000$} 
\end{subfigure}
\begin{subfigure}[b]{0.49\textwidth} 
  \label{ fig7} 
  \begin{minipage}[b]{0.5\linewidth}
    \centering
    \includegraphics[width=1\linewidth]{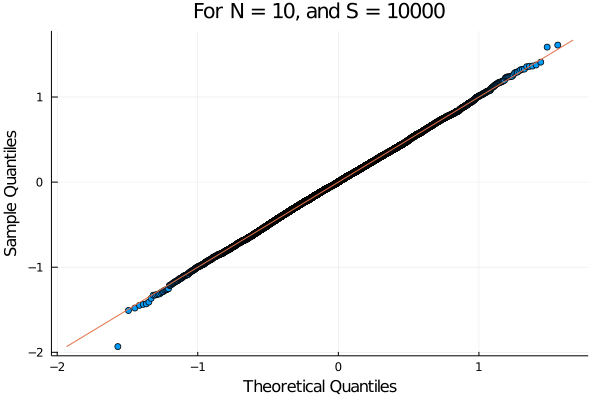} 
  \end{minipage}%%
  \begin{minipage}[b]{0.5\linewidth}
    \centering
    \includegraphics[width=1\linewidth]{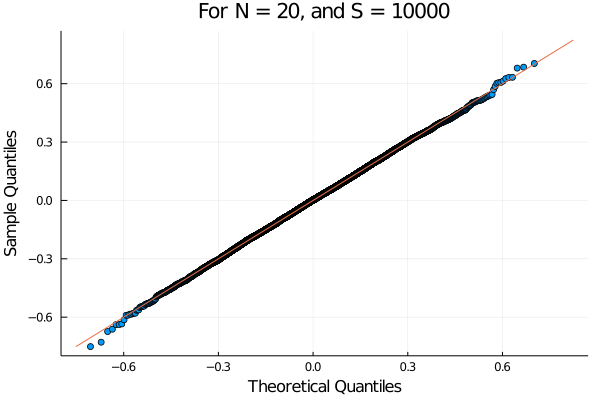} 
  \end{minipage} 
  \begin{minipage}[b]{0.5\linewidth}
    \centering
    \includegraphics[width=1\linewidth]{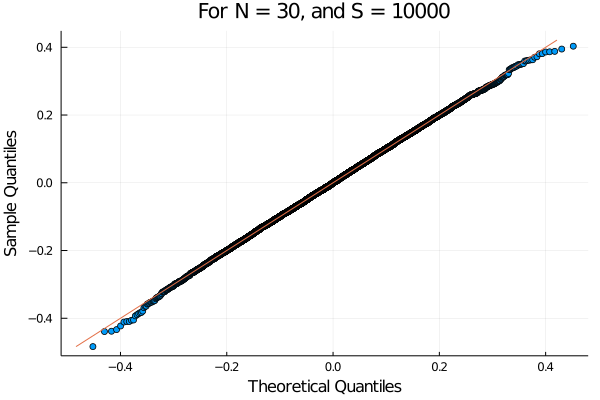} 
  \end{minipage}%% 
  \begin{minipage}[b]{0.5\linewidth}
    \centering
    \includegraphics[width=1\linewidth]{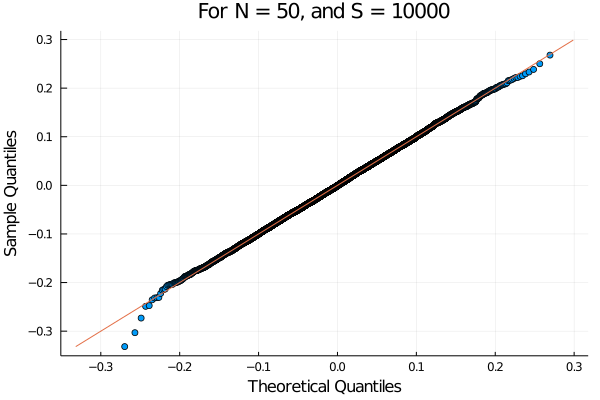} 
  \end{minipage} 
  \caption*{QQ-plot of $\hat{\beta}_1$ for design 4 and $S=10000$} 
\end{subfigure}

\end{figure}

From the histograms above there is evidence that the estimator of $\beta_1$ is normally distributed as the size of $N$ increases for all the number of simulations $S$. More specifically, for smaller values of $N$, we can see that the range of the histogram is wider than the one from a normal distribution. This is corroborated by the QQ-plots, that show that for any number of simulations $S$, for lower values of $N$, the distribution seems to have fatter tails than a normal distribution, but as $N$ increases it seems to be distributed as a normal. 

We next examine the size of the \textit{t-tests} where the test statistic use the asymptotic variance estimator proposed before. We test the null hypothesis that the coefficient $\beta_1$ is equal to its true value, $\beta_1 = 0$. The tables below shows the fractions of samples for which the null hypothesis is rejected at the $5\%$ statistical significance level.\\
\\

\begin{table}[H]
  \begin{minipage}[t]{0.45\linewidth}
    \centering
  \begin{tabular}{rrrrrrrrrrrrrrrrrrrrrrrrrr}
    \hline
      Simulations & N & $\hat{\text
      {var}}(\hat{\beta}_1)_{\hat{\delta}_{2}}$ & $\hat{\text
      {var}}(\hat{\beta}_1)_{\hat{\Delta}_{2}}$\\
      \hline
      1000 & 10 & 0.033 & 0.037 \\
      1000 & 20 & 0.043 & 0.050 \\
      1000 & 30 & 0.030 & 0.032 \\
      1000 & 50 & 0.040 & 0.040 \\
      5000 & 10 & 0.040 & 0.054 \\
      5000 & 20 & 0.032 & 0.038 \\
      5000 & 30 & 0.039 & 0.042 \\
      5000 & 50 & 0.046 & 0.047 \\
     10000 & 10 & 0.037 & 0.050 \\
     10000 & 20 & 0.034 & 0.038 \\
     10000 & 30 & 0.039 & 0.041 \\
     10000 & 50 & 0.042 & 0.043 \\
      \hline
  \end{tabular}
    \caption{Results of the Monte Carlo Simulation of the size of the t-test of the Pairwise Differences estimators obtained for the first data generating process}
    \label{tab:results5}
\end{minipage}\hfill
\begin{minipage}[t]{0.45\linewidth}
  \centering

    \begin{tabular}{rrrrrrrrrrrrrrrrrrrrrrr}
      \hline
      Simulations & N & $\hat{\text
      {var}}(\hat{\beta}_1)_{\hat{\delta}_{2}}$ & $\hat{\text
      {var}}(\hat{\beta}_1)_{\hat{\Delta}_{2}}$\\
      \hline
      1000 & 10 & 0.036 & 0.049 \\
      1000 & 20 & 0.036 & 0.038 \\
      1000 & 30 & 0.048 & 0.049 \\ 
      1000 & 50 & 0.055 & 0.057 \\
      5000 & 10 & 0.041 & 0.058 \\
      5000 & 20 & 0.035 & 0.038 \\
      5000 & 30 & 0.047 & 0.048 \\
      5000 & 50 & 0.040 & 0.040 \\
      10000 & 10 & 0.038 & 0.055 \\
      10000 & 20 & 0.034 & 0.038 \\
      10000 & 30 & 0.035 & 0.038 \\
      10000 & 50 & 0.046 & 0.047 \\
      \hline
  \end{tabular}
    \caption{Results of the Monte Carlo Simulation of the size of the t-test of the Pairwise Differences estimators obtained for the second data generating process}
    \label{tab:results6}
\end{minipage}
\end{table} 

% DUPLICATE
\begin{table}[H]
  \begin{minipage}[t]{0.45\linewidth}
    \centering
  \begin{tabular}{rrrrrrrrrrrrrrrrrrrrrrrrrr}
    \hline
      Simulations & N & $\hat{\text
      {var}}(\hat{\beta}_1)_{\hat{\delta}_{2}}$ & $\hat{\text
      {var}}(\hat{\beta}_1)_{\hat{\Delta}_{2}}$\\
      \hline
      1000 & 10 & 0.035 & 0.051 \\
      1000 & 20 & 0.025 & 0.031 \\
      1000 & 30 & 0.044 & 0.046 \\
      1000 & 50 & 0.047 & 0.049 \\
      5000 & 10 & 0.038 & 0.051 \\
      5000 & 20 & 0.034 & 0.038 \\
      5000 & 30 & 0.040 & 0.042 \\
      5000 & 50 & 0.041 & 0.042 \\
      10000 & 10 & 0.038 & 0.052 \\
      10000 & 20 & 0.037 & 0.043 \\
      10000 & 30 & 0.040 & 0.041 \\
      10000 & 50 & 0.043 & 0.042 \\
      \hline
  \end{tabular}
  \caption{Results of the Monte Carlo Simulation of the size of the t-test of the Pairwise Differences estimators obtained for the third data generating process}
  \label{tab:results7}
\end{minipage}\hfill
\begin{minipage}[t]{0.45\linewidth}
  \centering

    \begin{tabular}{rrrrrrrrrrrrrrrrrrrrrrr}
      \hline
      Simulations & N & $\hat{\text
      {var}}(\hat{\beta}_1)_{\hat{\delta}_{2}}$ & $\hat{\text
      {var}}(\hat{\beta}_1)_{\hat{\Delta}_{2}}$\\
      \hline
      1000 & 10 & 0.040 & 0.053 \\
      1000 & 20 & 0.039 & 0.045 \\
      1000 & 30 & 0.041 & 0.043 \\
      1000 & 50 & 0.039 & 0.037 \\
      5000 & 10 & 0.034 & 0.049 \\
      5000 & 20 & 0.038 & 0.042 \\
      5000 & 30 & 0.039 & 0.042 \\
      5000 & 50 & 0.041 & 0.043 \\
      10000 & 10 & 0.038 & 0.057 \\
      10000 & 20 & 0.036 & 0.040 \\
      10000 & 30 & 0.039 & 0.042 \\
      10000 & 50 & 0.046 & 0.047 \\
      \hline
  \end{tabular}
  \caption{Results of the Monte Carlo Simulation of the size of the t-test of the Pairwise Differences estimators obtained for the fourth data generating process}
  \label{tab:results8}
\end{minipage}
\end{table} 

When we look at the size of the t-test for the different variance estimators, we see that, as expected from the previous findings, the sizes for the estimators are  close to 0.05, however the estimates using $\hat{\Delta}_2$ are somewhat closer than those using  $\hat{\delta}_2$.
\section{Conclusion and Further Research}

In this paper we showed how one can adapt U-statistics tools to show the asymptotic properties of linear dyadic models for network data. More especifically, we proposed a linear model with two-way fixed effects that enter additively in the specification. While the usual two-way fixed effects estimator is consistent and asymptotically unbiased for this particular model, we propose an estimator that relies on pairwise differences, that completely eliminate the fixed effects from the objective (and influence) function(s). This choice of estimator was done with the purpose of demonstrating step-by-step in a simple model how one can adapt tools from U-statistics to this particular dyadic setting (with a pairwise differences estimator) to obtain an analytical form and an estimator for the asymptotic variance. 

These specific tools are needed because the pairwise differencing approach introduces a dependence structure in the summands of the influence function of the estimator. A similar set of tools are also used in non-linear models that employ a similar estimation method, in particular in \cite{charbonneau2017multiple} and in \cite{graham2017econometric}. For non-linear models, differencing out the fixed effects is desirable to eliminate the incidental parameter problem, which would lead to asymptotically biased estimates of the coefficients of the covariates. 

With a Monte Carlo exercise, we showed that the obtained estimates for the slope coefficients are unbiased in finite samples, and the estimated asymptotic variance delivers the correct size for the t-test. However, the model assumed in this paper can still be relaxed to allow for a richer dependence structure in the network in future research. For instance, we also could allow for dependencies across outcomes $Y_{ij}$ and $Y_{ji}$ by relaxing Assumption \ref{ass:1} such that the idiosyncratic terms $U_{ij}$ and $U_{ji}$ are allowed to covary. In practice, this would have implications for how the tools of U-statistics are employed in this dyadic framework. Namely, the Hoeffding decomposition for the variance of the U-statistic and the H\'{a}jek projection would have to be modified to allow for these dependencies. 

Another avenue of interest is that we could allow for, in Assumption \ref{ass:2}, the individual-level observed characteristics $A_i$ and $B_i$ to covary. That would allow, for instance, that exports from Japan to Korea might covary with those from Korea to Thailand. In our derivations, this would have implications for the probability limit of the Hessian of the proposed estimator.

Finally, the main computational challenge is the estimate of $\Delta_2$. It could also be of pratical use, in future research, to explore possible bootstrap procedures to obtain inference, such as in \cite{graham2019network} and in \cite{menzel2018bootstrap}.

\bibliography{library.bib}

\begin{thebibliography}{22}
\newcommand{\enquote}[1]{``#1''}
\expandafter\ifx\csname natexlab\endcsname\relax\def\natexlab#1{#1}\fi

\bibitem[\protect\citeauthoryear{Anderson and Van~Wincoop}{Anderson and
  Van~Wincoop}{2003}]{anderson2003gravity}
\textsc{Anderson, J.~E. and E.~Van~Wincoop} (2003): \enquote{Gravity with
  gravitas: A solution to the border puzzle,} \emph{American economic review},
  93, 170--192.

\bibitem[\protect\citeauthoryear{Charbonneau}{Charbonneau}{2017}]{charbonneau2017multiple}
\textsc{Charbonneau, K.~B.} (2017): \enquote{Multiple fixed effects in binary
  response panel data models,} \emph{The Econometrics Journal}, 20, S1--S13.

\bibitem[\protect\citeauthoryear{Chen, Fern{\'a}ndez-Val, and Weidner}{Chen
  et~al.}{2021}]{chen2021nonlinear}
\textsc{Chen, M., I.~Fern{\'a}ndez-Val, and M.~Weidner} (2021):
  \enquote{Nonlinear factor models for network and panel data,} \emph{Journal
  of Econometrics}, 220, 296--324.

\bibitem[\protect\citeauthoryear{Dzemski}{Dzemski}{2019}]{dzemski2019empirical}
\textsc{Dzemski, A.} (2019): \enquote{An empirical model of dyadic link
  formation in a network with unobserved heterogeneity,} \emph{Review of
  Economics and Statistics}, 101, 763--776.

\bibitem[\protect\citeauthoryear{Ferguson, Shapley, and MacQueen}{Ferguson
  et~al.}{2005}]{ferguson2005u}
\textsc{Ferguson, T.~S., L.~Shapley, and J.~MacQueen} (2005):
  \enquote{U-statistics,} \emph{Probability and Game Theory, Papers in Honor of
  David Blackwell, Institute of Mathematical Statistics Lecture Notes-Monograph
  Series}, 30.

\bibitem[\protect\citeauthoryear{Fern{\'a}ndez-Val and
  Weidner}{Fern{\'a}ndez-Val and Weidner}{2016}]{fernandez2016individual}
\textsc{Fern{\'a}ndez-Val, I. and M.~Weidner} (2016): \enquote{Individual and
  time effects in nonlinear panel models with large N, T,} \emph{Journal of
  Econometrics}, 192, 291--312.

\bibitem[\protect\citeauthoryear{Graham}{Graham}{2017}]{graham2017econometric}
\textsc{Graham, B.~S.} (2017): \enquote{An econometric model of network
  formation with degree heterogeneity,} \emph{Econometrica}, 85, 1033--1063.

\bibitem[\protect\citeauthoryear{Graham}{Graham}{2019}]{graham2019network}
---\hspace{-.1pt}---\hspace{-.1pt}--- (2019): \enquote{Network data,} Tech.
  rep., National Bureau of Economic Research.

\bibitem[\protect\citeauthoryear{Graham}{Graham}{2020}]{graham2020dyadic}
---\hspace{-.1pt}---\hspace{-.1pt}--- (2020): \enquote{Dyadic regression,} in
  \emph{The Econometric Analysis of Network Data}, Elsevier, 23--40.

\bibitem[\protect\citeauthoryear{Hoeffding, Robbins et~al.}{Hoeffding
  et~al.}{1948}]{hoeffding1948central}
\textsc{Hoeffding, W., H.~Robbins, et~al.} (1948): \enquote{The central limit
  theorem for dependent random variables,} \emph{Duke Mathematical Journal},
  15, 773--780.

\bibitem[\protect\citeauthoryear{Jackson and L{\'o}pez-Pintado}{Jackson and
  L{\'o}pez-Pintado}{2013}]{jackson2013diffusion}
\textsc{Jackson, M.~O. and D.~L{\'o}pez-Pintado} (2013): \enquote{Diffusion and
  contagion in networks with heterogeneous agents and homophily,} \emph{Network
  Science}, 1, 49--67.

\bibitem[\protect\citeauthoryear{Jochmans}{Jochmans}{2017}]{jochmans2017two}
\textsc{Jochmans, K.} (2017): \enquote{Two-way models for gravity,}
  \emph{Review of Economics and Statistics}, 99, 478--485.

\bibitem[\protect\citeauthoryear{Jochmans}{Jochmans}{2018}]{jochmans2018semiparametric}
---\hspace{-.1pt}---\hspace{-.1pt}--- (2018): \enquote{Semiparametric analysis
  of network formation,} \emph{Journal of Business \& Economic Statistics}, 36,
  705--713.

\bibitem[\protect\citeauthoryear{Juodis}{Juodis}{2021}]{juodis2020shock}
\textsc{Juodis, A.} (2021): \enquote{This shock is different: Estimation and
  inference in misspecified two-way fixed effects panel regressions,} Tech.
  rep., Working Paper.

\bibitem[\protect\citeauthoryear{Menzel}{Menzel}{2018}]{menzel2018bootstrap}
\textsc{Menzel, K.} (2018): \enquote{Bootstrap with cluster-dependence in two
  or more dimensions,} \emph{ArXiv eprints, New York University}.

\bibitem[\protect\citeauthoryear{Neyman and Scott}{Neyman and
  Scott}{1948}]{neyman1948consistent}
\textsc{Neyman, J. and E.~L. Scott} (1948): \enquote{Consistent estimates based
  on partially consistent observations,} \emph{Econometrica: Journal of the
  Econometric Society}, 1--32.

\bibitem[\protect\citeauthoryear{Rao}{Rao}{2009}]{rao2009conditional}
\textsc{Rao, B.~P.} (2009): \enquote{Conditional independence, conditional
  mixing and conditional association,} \emph{Annals of the Institute of
  Statistical Mathematics}, 61, 441--460.

\bibitem[\protect\citeauthoryear{Serfling}{Serfling}{2009}]{serfling2009approximation}
\textsc{Serfling, R.~J.} (2009): \emph{Approximation theorems of mathematical
  statistics}, vol. 162, John Wiley \& Sons.

\bibitem[\protect\citeauthoryear{Shalizi}{Shalizi}{2016}]{shalizi2016cid}
\textsc{Shalizi, C.} (2016): \enquote{Lecture 1: Conditionally-independent dyad
  models,} Tech. rep., Lecture note, Carnegie Mellon University.

\bibitem[\protect\citeauthoryear{Silva and Tenreyro}{Silva and
  Tenreyro}{2006}]{silva2006log}
\textsc{Silva, J.~S. and S.~Tenreyro} (2006): \enquote{The log of gravity,}
  \emph{The Review of Economics and statistics}, 88, 641--658.

\bibitem[\protect\citeauthoryear{Tinbergen}{Tinbergen}{1962}]{tinbergen1962shaping}
\textsc{Tinbergen, J.} (1962): \emph{Shaping the World Economy; Suggestions for
  an International Economic Policy}, Twentieth Century Fund, New York.

\bibitem[\protect\citeauthoryear{Van~der Vaart}{Van~der
  Vaart}{2000}]{van2000asymptotic}
\textsc{Van~der Vaart, A.~W.} (2000): \emph{Asymptotic statistics}, vol.~3,
  Cambridge university press.

\end{thebibliography}

\appendix
\section{Application to $N=5$}
To demonstrate how the \cite{hoeffding1948central} decomposition is obtained, we now look at at an application for the Pairwise Differences estimator in for the case $N=5$, meaning that we have $N (N-1)=5 \times 4 = 20$ dyads in the dataset.

\subsection{Showing some correspondences between ways to write the U-statistic}
The first element we can look at is that indeed the equivalence below holds:
\begin{align}
\left[ \frac{1}{N(N-1)(N-2)(N-3)}\sum_{i=1}^N \sum_{j \neq i} \sum_{k \neq i,j} \sum_{l \neq i,j,k} \Tilde{X}_{ijkl} \Tilde{U}_{ijkl} \right] = \left[ \frac{1}{{N \choose 4}}  \sum_{\mathcal{C} \in \mathcal{C}(\mathcal{N},4)} \frac{1}{4!} \sum_{\pi \in \mathcal{P}(\mathcal{C},4)} \Tilde{X}_{\pi_1\pi_2\pi_3\pi_4} \Tilde{U}_{\pi_1\pi_2\pi_3\pi_4}  \right] \nonumber
\end{align}
Note that the first quadruple of sums boils down to summing over $5 \times 4 \times 3 \times 2 = 120$ elements. The second double sum boils down to summing over ${5 \choose 4} 4! = \frac{5 \times 4 \times 3 \times 2}{4 \times 3 \times 2 \times 1} 4! = 120$ elements. Now, we only need to check that those elements are the same.

To do so, we can construct the elements in the first sum by taking first $i$ fixed to for instance 1, then considering the subsequent element $j$ where $j \neq i$, and so on. Then, we repeat for all possible values of $i \in \{ 1,2,3,4,5\}$:

Now we can compare to the elements taken into account in the second term of the previous equation. The first sum is over the combinations, and we have that for $N=5$, there are ${N \choose 4} = {5 \choose 4} = 5$ combinations, more precisely, the following: $\{ 2, 3, 4, 5 \}$,  $\{ 1, 3, 4, 5\}$, $\{ 1, 2, 4, 5\}$, $\{ 1, 2, 3, 5\}$ and $\{ 1, 2, 3, 4\}$. For each of these combinations, there are $4! = 24$ permutations. Therefore, we have $5 \times 24 = 120$ elements. In the following table, for instance, each column contains the permutations for a given combination. If we look at both ways of defining the sums in the equation above, we will sum over all these rows and columns for N=5.

\begin{table}[h!]
  \centering
  \begin{tabular}{|cccc|cccc|cccc|cccc|}
    \hline
      $\mathcal{C}_1$ & & & & $\mathcal{C}_2$ & & & & $\mathcal{C}_3$ & & & &  $\mathcal{C}_4$ & & &  \\
     \hline
     2 & 3 & 4 & 5 & 1 & 3 & 4 & 5 & 1 & 2 & 4 & 5 & 1 & 2 & 3 & 5 \\
     3 & 2 & 4 & 5 & 3 & 1 & 4 & 5 & 2 & 1 & 4 & 5 & 2 & 1 & 3 & 5 \\
     4 & 2 & 3 & 5 & 4 & 1 & 3 & 5 & 4 & 1 & 2 & 5 & 3 & 1 & 2 & 5 \\
     2 & 4 & 3 & 5 & 1 & 4 & 3 & 5 & 1 & 4 & 2 & 5 & 1 & 3 & 2 & 5 \\
     3 & 4 & 2 & 5 & 3 & 4 & 1 & 5 & 2 & 4 & 1 & 5 & 2 & 3 & 1 & 5 \\
     4 & 3 & 2 & 5 & 4 & 3 & 1 & 5 & 4 & 2 & 1 & 5 & 3 & 2 & 1 & 5 \\
     4 & 3 & 5 & 2 & 4 & 3 & 5 & 1 & 4 & 2 & 5 & 1 & 3 & 2 & 5 & 1 \\
     3 & 4 & 5 & 2 & 3 & 4 & 5 & 1 & 2 & 4 & 5 & 1 & 2 & 3 & 5 & 1 \\
     5 & 4 & 3 & 2 & 5 & 4 & 3 & 1 & 5 & 4 & 2 & 1 & 5 & 3 & 2 & 1 \\
     4 & 5 & 3 & 2 & 4 & 5 & 3 & 1 & 4 & 5 & 2 & 1 & 3 & 5 & 2 & 1 \\
     3 & 5 & 4 & 2 & 3 & 5 & 4 & 1 & 2 & 5 & 4 & 1 & 2 & 5 & 3 & 1 \\
     5 & 3 & 4 & 2 & 5 & 3 & 4 & 1 & 5 & 2 & 4 & 1 & 5 & 2 & 3 & 1 \\
     5 & 2 & 4 & 3 & 5 & 1 & 4 & 3 & 5 & 1 & 4 & 2 & 5 & 1 & 3 & 2 \\
     2 & 5 & 4 & 3 & 1 & 5 & 4 & 3 & 1 & 5 & 4 & 2 & 1 & 5 & 3 & 2 \\
     4 & 5 & 2 & 3 & 4 & 5 & 1 & 3 & 4 & 5 & 1 & 2 & 3 & 5 & 1 & 2 \\
     5 & 4 & 2 & 3 & 5 & 4 & 1 & 3 & 5 & 4 & 1 & 2 & 5 & 3 & 1 & 2 \\
     2 & 4 & 5 & 3 & 1 & 4 & 5 & 3 & 1 & 4 & 5 & 2 & 1 & 3 & 5 & 2 \\
     4 & 2 & 5 & 3 & 4 & 1 & 5 & 3 & 4 & 1 & 5 & 2 & 3 & 1 & 5 & 2 \\
     3 & 2 & 5 & 4 & 3 & 1 & 5 & 4 & 2 & 1 & 5 & 4 & 2 & 1 & 3 & 5 \\
     2 & 3 & 5 & 4 & 1 & 3 & 5 & 4 & 1 & 2 & 5 & 4 & 1 & 2 & 3 & 5 \\
     5 & 3 & 2 & 4 & 5 & 3 & 1 & 4 & 5 & 2 & 1 & 4 & 5 & 2 & 3 & 1 \\
     3 & 5 & 2 & 4 & 3 & 5 & 1 & 4 & 2 & 5 & 1 & 4 & 2 & 5 & 3 & 1 \\
     2 & 5 & 3 & 4 & 1 & 5 & 3 & 4 & 1 & 5 & 2 & 4 & 1 & 5 & 2 & 3 \\
     5 & 2 & 3 & 4 & 5 & 1 & 3 & 4 & 5 & 1 & 2 & 4 & 5 & 1 & 2 & 3 \\
    \hline
  \end{tabular}
  \caption{Combinations and permutations for $N=5$}
  \label{tab:comb}
\end{table}
\newpage
\subsection{Calculating the variance of the U-statistics for $N=5$}

To calculate the variance of $U_N$, we first note that, by definition of the variance:
\begin{align*}
    \text{Var}(U_N) &= \text{Var} \left( \frac{1}{{N \choose 4}} \sum_{\mathcal{C} \in \mathcal{C}(\mathcal{N},4)} \frac{1}{4!} \sum_{\pi \in \mathcal{P}(\mathcal{C},4)} \Tilde{X}_{\pi_1\pi_2\pi_3\pi_4} \Tilde{U}_{\pi_1\pi_2\pi_3\pi_4} \right) \\
    &= \left( \frac{1}{{N \choose 4}} \right)^2 \text{Var} \left(\sum_{\mathcal{C} \in \mathcal{C}(\mathcal{N},4)} \frac{1}{4!} \sum_{\pi \in \mathcal{P}(\mathcal{C},4)} \Tilde{X}_{\pi_1\pi_2\pi_3\pi_4} \Tilde{U}_{\pi_1\pi_2\pi_3\pi_4} \right) \nonumber \\
    &= \left( \frac{1}{{N \choose 4}} \right)^2 \text{Cov} \left( \sum_{\mathcal{C} \in \mathcal{C}(\mathcal{N},4)} \frac{1}{4!} \sum_{\pi \in \mathcal{P}(\mathcal{C},4)} \Tilde{X}_{\pi_1\pi_2\pi_3\pi_4} \Tilde{U}_{\pi_1\pi_2\pi_3\pi_4}, \sum_{\mathcal{C} \in \mathcal{C}(\mathcal{N},4)} \frac{1}{4!} \sum_{\pi \in \mathcal{P}(\mathcal{C},4)} \Tilde{X}_{\pi_1\pi_2\pi_3\pi_4} \Tilde{U}_{\pi_1\pi_2\pi_3\pi_4} \right) \nonumber
\end{align*}
We can denote further by $\pi(\mathcal{C})$ the set of tetrads corresponding to the permutations of a given combination $c$ (given in the previous table). Then, we can rewrite:
\begin{align}
\left[ \frac{1}{{N \choose 4}} \sum_{\mathcal{C} \in \mathcal{C}(\mathcal{N},4)} \frac{1}{4!} \sum_{\pi \in \mathcal{P}(\mathcal{C},4)} \Tilde{X}_{\pi_1\pi_2\pi_3\pi_4} \Tilde{U}_{\pi_1\pi_2\pi_3\pi_4} \right] = \left[ \frac{1}{{N \choose 4}} \sum_{\mathcal{C} \in \mathcal{C}(\mathcal{N},4)} \frac{1}{4!} \sum_{\pi(\mathcal{C})} \Tilde{X}_{\pi(\mathcal{C})} \Tilde{U}_{\pi(\mathcal{C})}\right]
\end{align}
In the case of $N=5$, we have ${5 \choose 4} = \frac{5!}{4! (5-4)!} = 5$ combinations, with $\mathcal{C} \in \{\mathcal{C}_1,\mathcal{C}_2,\mathcal{C}_3,\mathcal{C}_4,\mathcal{C}_5\}$. Therefore:
\begin{gather}
\begin{align*}
&\text{Var} (U_N) = \\ &\left( \frac{1}{5} \right)^2 \text{Cov} \Big( \frac{1}{4!} \sum_{\pi(\mathcal{C}_1)} \Tilde{X}_{\pi(\mathcal{C}_1)} \Tilde{U}_{\pi(\mathcal{C}_1)} + \frac{1}{4!} \sum_{\pi(\mathcal{C}_2)} \Tilde{X}_{\pi(\mathcal{C}_2)} \Tilde{U}_{\pi(\mathcal{C}_2)} + \frac{1}{4!} \sum_{\pi(\mathcal{C}_3)} \Tilde{X}_{\pi(\mathcal{C}_3)} \Tilde{U}_{\pi(\mathcal{C}_3)} + \frac{1}{4!} \sum_{\pi(\mathcal{C}_4)} \Tilde{X}_{\pi(\mathcal{C}_4)} \Tilde{U}_{\pi(\mathcal{C}_4)} + \frac{1}{4!} \sum_{\pi(\mathcal{C}_5)} \Tilde{X}_{\pi(\mathcal{C}_5)} \Tilde{U}_{\pi(\mathcal{C}_5)}, \nonumber \\
& \frac{1}{4!} \sum_{\pi(1)} \Tilde{X}_{\pi(\mathcal{C}_1)} \Tilde{U}_{\pi(\mathcal{C}_1)} + \frac{1}{4!} \sum_{\pi(\mathcal{C}_2)} \Tilde{X}_{\pi(\mathcal{C}_2)} \Tilde{U}_{\pi(\mathcal{C}_2)} + \frac{1}{4!} \sum_{\pi(\mathcal{C}_3)} \Tilde{X}_{\pi(\mathcal{C}_3)} \Tilde{U}_{\pi(\mathcal{C}_3)} + \frac{1}{4!} \sum_{\pi(\mathcal{C}_4)} \Tilde{X}_{\pi(\mathcal{C}_4)} \Tilde{U}_{\pi(\mathcal{C}_4)} + \frac{1}{4!} \sum_{\pi(\mathcal{C}_5)} \Tilde{X}_{\pi(\mathcal{C}_5)} \Tilde{U}_{\pi(\mathcal{C}_5)}\Big) \nonumber
\end{align*}
\end{gather}

Note that each term corresponds to the average of the permutations, which also corresponds to the previously defined kernel $s_{ijkl}$. We can now focus on the covariance of each separate summand in the first term with the entire second term, that is:
\begin{gather}
\begin{align}
\text{Cov} \Big( &\frac{1}{4!} \sum_{\pi(\mathcal{C})} \Tilde{X}_{\pi(\mathcal{C})} \Tilde{U}_{\pi(\mathcal{C})}, \\ & \frac{1}{4!} \sum_{\pi(\mathcal{C}_1)} \Tilde{X}_{\pi(\mathcal{C}_1)} \Tilde{U}_{\pi(\mathcal{C}_1)} + \frac{1}{4!} \sum_{\pi(\mathcal{C}_2)} \Tilde{X}_{\pi(\mathcal{C}_2)} \Tilde{U}_{\pi(\mathcal{C}_2)} + \frac{1}{4!} \sum_{\pi(\mathcal{C}_3)} \Tilde{X}_{\pi(\mathcal{C}_3)} \Tilde{U}_{\pi(\mathcal{C}_3)} + \frac{1}{4!} \sum_{\pi(\mathcal{C}_4)} \Tilde{X}_{\pi(\mathcal{C}_4)} \Tilde{U}_{\pi(\mathcal{C}_4)} + \frac{1}{4!} \sum_{\pi(\mathcal{C}_5)} \Tilde{X}_{\pi(\mathcal{C}_5)} \Tilde{U}_{\pi(\mathcal{C}_5)} \Big) \nonumber
\end{align}
\end{gather}
For each $\mathcal{C} \in \{\mathcal{C}_1,\mathcal{C}_2,\mathcal{C}_3,\mathcal{C}_4,\mathcal{C}_5\}$. Looking first at the first summand, taking $\mathcal{C}=\mathcal{C}_1$, we note two facts:
\begin{itemize}
    \item Under the assumptions that (i) $U_{ij}$ is i.i.d., that (ii) $X_{ij} = f(A_i, B_j)$ and that (iii) $A_i$ and $B_j$ are i.i.d., we have that when considering each separate summand in the second term of the covariance, i.e., looking at:
    $$\text{Cov} \left(\frac{1}{4!} \sum_{\pi(\mathcal{C}_1)} \Tilde{X}_{\pi(\mathcal{C}_1)} \Tilde{U}_{\pi(\mathcal{C}_1)}, \frac{1}{4!} \sum_{\pi(\mathcal{C}')} \Tilde{X}_{\pi(\mathcal{C}')} \Tilde{U}_{\pi(\mathcal{C}')}\right)$$ 
    for each $\mathcal{C}' \in \{\mathcal{C}_1,\mathcal{C}_2,\mathcal{C}_3,\mathcal{C}_4,\mathcal{C}_5\}$, the magnitude of such covariance - that takes into account the covariances between the permutations of the first combination, that are in the set $\pi(1)$ and the permutations of the combinations $\mathcal{C}'$, in the sets $\pi(\mathcal{C}')$ - depends solely on the number of common indices that the two combinations have. Importantly, the magnitude of this covariance does not depend on the order $N$, since we are always looking at the covariances between the permutations of two sets of combinations of 4 elements, regardless of the value of N. It also does not depend on which indices we are specifically looking at given the assumptions above.\\
    Therefore, the only the only determinant of such covariance is the number of common indices between the two combinations, denoted by $q$, and we can then define:
    $$ \text{Cov} \left(\frac{1}{4!} \sum_{\pi(\mathcal{C}_1)} \Tilde{X}_{\pi(\mathcal{C}_1)} \Tilde{U}_{\pi(\mathcal{C}_1)}, \frac{1}{4!} \sum_{\pi(\mathcal{C}')} \Tilde{X}_{\pi(\mathcal{C}')} \Tilde{U}_{\pi(\mathcal{C}')}\right) = \Delta_q $$
    Note that this definition corresponds to the definition given in the Hoeffding decomposition. 
    \item Given the previous fact, the entire covariance:
    \begin{gather}
    \begin{align}
        \text{Cov} \Big( &\frac{1}{4!} \sum_{\pi(\mathcal{C}_1)} \Tilde{X}_{\pi(\mathcal{C}_1)} \Tilde{U}_{\pi(\mathcal{C}_1)}, \\ & \frac{1}{4!} \sum_{\pi(\mathcal{C}_1)} \Tilde{X}_{\pi(\mathcal{C}_1)} \Tilde{U}_{\pi(\mathcal{C}_1)} + \frac{1}{4!} \sum_{\pi(\mathcal{C}_2)} \Tilde{X}_{\pi(\mathcal{C}_2)} \Tilde{U}_{\pi(\mathcal{C}_2)} + \frac{1}{4!} \sum_{\pi(\mathcal{C}_3)} \Tilde{X}_{\pi(\mathcal{C}_3)} \Tilde{U}_{\pi(\mathcal{C}_3)} + \frac{1}{4!} \sum_{\pi(\mathcal{C}_4)} \Tilde{X}_{\pi(\mathcal{C}_4)} \Tilde{U}_{\pi(\mathcal{C}_4)} + \frac{1}{4!} \sum_{\pi(\mathcal{C}_5)} \Tilde{X}_{\pi(\mathcal{C}_5)} \Tilde{U}_{\pi(\mathcal{C}_5)} \Big) \nonumber
    \end{align}
\end{gather}
    will depend on how many combinations in the second term have $q$ elements in common with the first combination $\mathcal{C}_1$. Remembering that $q$ can take values in $\{0,1,2,3,4\}$, we can look at the different combinations for the case $N=5$ and find how many elements in common each have with $\mathcal{C}_1$. 
\end{itemize}

For the case of the first combination for N=5, we have that the set $\mathcal{C}_1 = \{2,3,4,5\}$, then comparing to the other combinations $\mathcal{C}'$ we have:

\begin{itemize}
    \item $\mathcal{C}' = \mathcal{C}_1 = \{ 2,3,4,5 \}$: there are 4 elements in common, therefore: \\
    $$ \text{Cov} \left(\frac{1}{4!} \sum_{\pi(\mathcal{C}_1)} \Tilde{X}_{\pi(\mathcal{C}_1)} \Tilde{U}_{\pi(\mathcal{C}_1)}, \frac{1}{4!} \sum_{\pi(\mathcal{C}_1)} \Tilde{X}_{\pi(\mathcal{C}_1)} \Tilde{U}_{\pi(\mathcal{C}_1)}  \right) =  \Delta_4$$
    \item $\mathcal{C}' = \mathcal{C}_2 = \{ 1,3,4,5 \}$ : there are 3 elements in common, therefore: \\
    $$ \text{Cov} \left(\frac{1}{4!} \sum_{\pi(\mathcal{C}_1)} \Tilde{X}_{\pi(\mathcal{C}_1)} \Tilde{U}_{\pi(\mathcal{C}_1)}, \frac{1}{4!} \sum_{\pi(\mathcal{C}_2)} \Tilde{X}_{\pi(\mathcal{C}_2)} \Tilde{U}_{\pi(\mathcal{C}_2)}  \right) = \Delta_3$$
    \item $\mathcal{C}'= \mathcal{C}_3 = \{ 1,2,4,5 \}$: there are 3 elements in common, therefore:
    $$ \text{Cov} \left(\frac{1}{4!} \sum_{\pi(\mathcal{C}_1)} \Tilde{X}_{\pi(\mathcal{C}_1)} \Tilde{U}_{\pi(\mathcal{C}_1)}, \frac{1}{4!} \sum_{\pi(\mathcal{C}_3)} \Tilde{X}_{\pi(\mathcal{C}_3)} \Tilde{U}_{\pi(\mathcal{C}_3)}  \right)= \Delta_3$$
    \item $\mathcal{C}' = \mathcal{C}_4 = \{ 1,2,3,5 \}$: there are 3 elements in common, therefore:
    $$ \text{Cov} \left(\frac{1}{4!} \sum_{\pi(\mathcal{C}_1)} \Tilde{X}_{\pi(\mathcal{C}_1)} \Tilde{U}_{\pi(\mathcal{C}_1)}, \frac{1}{4!} \sum_{\pi(\mathcal{C}_4)} \Tilde{X}_{\pi(\mathcal{C}_4)} \Tilde{U}_{\pi(\mathcal{C}_4)}  \right)  = \Delta_3$$
    \item $\mathcal{C}' = \mathcal{C}_5 = \{ 1,2,3,4 \}$: there are 3 elements in common, therefore:
    $$ \text{Cov} \left(\frac{1}{4!} \sum_{\pi(\mathcal{C}_1)} \Tilde{X}_{\pi(\mathcal{C}_1)} \Tilde{U}_{\pi(\mathcal{C}_1)}, \frac{1}{4!} \sum_{\pi(\mathcal{C}_5)} \Tilde{X}_{\pi(\mathcal{C}_5)} \Tilde{U}_{\pi(\mathcal{C}_5)}  \right)  = \Delta_3$$
\end{itemize}
Note that these results are in line with what we find taking into account the combinatorial formula to find how many other combinations have $q$ elements in common when comparing to the fixed $\mathcal{C}_1$. The formula is such that the number of combinations with $q$ elements in common is given by:
\begin{align} \label{numbercomb}
    {4 \choose q} {N-4 \choose 4-q} = \frac{4!}{q! (4-q)!} \frac{(N-4)!}{(4-q)!(N-4-4+q)!}
\end{align} 
By plugging in $N=5$, we find that for $q=0,1,2$ there are 0 combinations, for $q=3$, there are 4 combinations and for $q=4$, there is 1 combination.

Therefore, we have that:
\begin{gather}
\begin{align*}
    \text{Cov} \Big( &\frac{1}{4!} \sum_{\pi(\mathcal{C}_1)} \Tilde{X}_{\pi(\mathcal{C}_1)} \Tilde{U}_{\pi(\mathcal{C}_1)}, \\ & \frac{1}{4!} \sum_{\pi(\mathcal{C}_1)} \Tilde{X}_{\pi(\mathcal{C}_1)} \Tilde{U}_{\pi(\mathcal{C}_1)} + \frac{1}{4!} \sum_{\pi(\mathcal{C}_2)} \Tilde{X}_{\pi(\mathcal{C}_2)} \Tilde{U}_{\pi(\mathcal{C}_2)} + \frac{1}{4!} \sum_{\pi(\mathcal{C}_3)} \Tilde{X}_{\pi(\mathcal{C}_3)} \Tilde{U}_{\pi(\mathcal{C}_3)} + \frac{1}{4!} \sum_{\pi(\mathcal{C}_4)} \Tilde{X}_{\pi(\mathcal{C}_4)} \Tilde{U}_{\pi(\mathcal{C}_4)} + \frac{1}{4!} \sum_{\pi(\mathcal{C}_5)} \Tilde{X}_{\pi(\mathcal{C}_5)} \Tilde{U}_{\pi(\mathcal{C}_5)} \Big) \nonumber \\
&= \Delta_4 + 4 \Delta_3 \nonumber
\end{align*}
\end{gather}
Once this is done for $\mathcal{C}_1$, we need to look at the covariance for the remaining summands in the first term of the entire covariance term, meaning that now we would take the same calculations for $\mathcal{C}_2,\mathcal{C}_3,\mathcal{C}_4,\mathcal{C}_5$.

Based on the fact that the terms $\Delta_q$ are only determined by the number of elements in common between the combinations, and given the fact that the same number of combinations with $q$ elements in common is found for any $\mathcal{C}$, the variance of $U_N$ is given by a function of the number of combinations times the previous found covariance for only one combination $\Delta_4 + 4 \Delta_3$:
\begin{align*}
    &\text{Var} (U_N) = \left( \frac{1}{5} \right)^2 5 \times (\Delta_4 + 4 \Delta_3) = \frac{1}{5} (\Delta_4 + 4 \Delta_3) \nonumber
\end{align*}

\section{Proofs}

\subsection{Proof of Lemma \ref{lemma:1}}
We provide an heuristic proof for the Hoeffding decomposition in this case, but, more details can be seen in \cite{hoeffding1948central}. First, we can expand the expression for the variance of $U_N$ as follows:
\begin{align} \label{eq:hoeffvar}
    \text{Var}(U_N) &= \text{Var} \left( \frac{1}{{N \choose 4}} \sum_{\mathcal{C} \in \mathcal{C}(\mathcal{N},4)} \frac{1}{4!} \sum_{\pi \in \mathcal{P}(\mathcal{C},4)} \Tilde{X}_{\pi_1\pi_2\pi_3\pi_4} \Tilde{U}_{\pi_1\pi_2\pi_3\pi_4} \right) \\
    &= \left( \frac{1}{{N \choose 4}} \right)^2 \text{Var} \left( \sum_{\mathcal{C} \in \mathcal{C}(\mathcal{N},4)} \frac{1}{4!} \sum_{\pi \in \mathcal{P}(\mathcal{C},4)}\Tilde{X}_{\pi_1\pi_2\pi_3\pi_4} \Tilde{U}_{\pi_1\pi_2\pi_3\pi_4} \right) \nonumber \\
    &= \left( \frac{1}{{N \choose 4}} \right)^2 \text{Cov} \left( \sum_{\mathcal{C} \in \mathcal{C}(\mathcal{N},4)} \frac{1}{4!} \sum_{\pi \in \mathcal{P}(\mathcal{C},4)} \Tilde{X}_{\pi_1\pi_2\pi_3\pi_4} \Tilde{U}_{\pi_1\pi_2\pi_3\pi_4}, \sum_{\mathcal{C} \in \mathcal{C}(\mathcal{N},4)} \frac{1}{4!} \sum_{\pi \in \mathcal{P}(\mathcal{C},4)} \Tilde{X}_{\pi_1\pi_2\pi_3\pi_4} \Tilde{U}_{\pi_1\pi_2\pi_3\pi_4}\right) \nonumber \\
    &= \left( \frac{1}{{N \choose 4}} \right)^2 \text{Cov} \left( \sum_{\mathcal{C} \in \mathcal{C}(\mathcal{N},4)} \frac{1}{4!} \sum_{\pi \in \mathcal{P}(\mathcal{C},4)} \Tilde{X}_{\pi_1\pi_2\pi_3\pi_4} \Tilde{U}_{\pi_1\pi_2\pi_3\pi_4}, \sum_{\mathcal{C} \in \mathcal{C}(\mathcal{N},4)} \frac{1}{4!} \sum_{\pi \in \mathcal{P}(\mathcal{C},4)} \Tilde{X}_{\pi_1\pi_2\pi_3\pi_4} \Tilde{U}_{\pi_1\pi_2\pi_3\pi_4} \right). \nonumber
\end{align}

The main idea is that, if we look at the expression in the last equality, we can first focus on the first combination in the sum over combinations for the first term of the expression, and its covariance with the entire second term of the expression, namely, given a first combination $\mathcal{C}_1$:
$$ \text{Cov} \left( \frac{1}{4!}  \sum_{\pi \in \mathcal{P}(\mathcal{C}_1,4)} \Tilde{X}_{\pi_1\pi_2\pi_3\pi_4} \Tilde{U}_{\pi_1\pi_2\pi_3\pi_4}, \sum_{\mathcal{C} \in \mathcal{C}(\mathcal{N},4)} \frac{1}{4!}  \sum_{\pi \in \mathcal{P}(\mathcal{C},4)} \Tilde{X}_{\pi_1\pi_2\pi_3\pi_4} \Tilde{U}_{\pi_1\pi_2\pi_3\pi_4} \right). $$
This expression boils down to the sum of the covariances between the kernel evaluated at this first combination of the first term and the kernels evaluated at all combinations that comes from the second term. As mentioned before, these covariances $\Delta_q$ depend only on the number of common indices $q$ between the combinations (it is noteworthy that $\Delta_q$ does not depend on $N$). For the remaining combinations in the first term, the same pattern would appear.

Thus, the expression in the last equality in Equation \eqref{eq:hoeffvar} is equal to the sum of all covariances between the cross terms in the first term with the second term (by opening up both summations over combinations), similar to \cite{ferguson2005u}:
\begin{align} \label{eq:hoeffvarprelim}
    \text{Var} (U_N) &= {N \choose 4}^{-2} \sum_{\mathcal{C}_1 \in \mathcal{C}(\mathcal{N},4)} \sum_{\mathcal{C}_2 \in \mathcal{C}(\mathcal{N},4)} \text{Cov} \Big( \frac{1}{4!} \sum_{\pi \in \mathcal{P}(\mathcal{C}_1,4)} \Tilde{X}_{\pi_1\pi_2\pi_3\pi_4} \Tilde{U}_{\pi_1\pi_2\pi_3\pi_4},\frac{1}{4!} \sum_{\pi \in \mathcal{P}(\mathcal{C}_2,4)} \Tilde{X}_{\pi_1\pi_2\pi_3\pi_4} \Tilde{U}_{\pi_1\pi_2\pi_3\pi_4}\Big).
\end{align}
Therefore, the main driver of the expression for the variance of $U_N$ boils down to the sum of the number of combinations having $q$ elements in common times $\Delta_q$.

As explained by \cite{ferguson2005u}, the number of pairs of quadruples, say $\{i,j,k,l\}$ and $\{m,n,o,p\}$, where the former appears in the first term of Equation \eqref{eq:hoeffvarprelim} and the latter in the second term, that have exactly $q$ elements in common is given by:
$$ {N \choose 4}{4 \choose q}{N-4 \choose 4 - q}, $$
\noindent which follows from the fact that first, there are ${N \choose 4}$ ways to choose the first combination $\{i,j,k,l\}$ from the set $\mathcal{N}$. Then, for each of those combinations, there are ${4 \choose q}$ ways to choose a subset of size $q$ from it, which will be the individuals in common with the second combination. Then, having fixed these individuals in common, there are then ${N-4 \choose 4-q}$ ways of choosing the $4-q$ individuals in the second combination that are non-common to the first from the $N-4$ individuals left that are not already present in the first combination.

We therefore have that, substituting in Equation \eqref{eq:hoeffvarprelim}:
\begin{align}
    \text{Var} (U_N) &= {N \choose 4}^{-2} \sum_{q=0}^4 {N \choose 4}{4 \choose q}{N-4 \choose 4 - q} \Delta_q \\
    &= {N \choose 4}^{-1} \sum_{q=0}^4 {4 \choose q}{N-4 \choose 4 - q} \Delta_q .\nonumber
\end{align}
Where each of the covariances above have $q=0,\dots,4$ elements in common, and the coefficients on $\Delta_q$ gives the number of covariances with $q$ elements in common. \qed 
\subsection{Proof of Theorem 1}
Supposing that the rate of convergence is given by $\sqrt{N(N-1)}$, we can further write:
$$ \text{Var} (\sqrt{N(N-1)} U_N) = (N(N-1)) \frac{1}{{N \choose 4}} \sum_{q=2}^4 {4 \choose q} {N-4 \choose 4-q} \Delta_q $$
We can then open the combinatorics terms to find that:
$$ \text{Var} (\sqrt{N(N-1)} U_N) = \mathcal{O}(1) + \mathcal{O}\left( \frac{1}{N} \right) + \mathcal{O}\left( \frac{1}{N^2} \right)  $$
Where the term of order $O(1)$ relates to the term $\Delta_2$, $\mathcal{O}\left( \frac{1}{N} \right)$ relates to the term $\Delta_3$ and $\mathcal{O}\left( \frac{1}{N^2} \right) $ relates to the term $\Delta_4$:
\begin{itemize}
    \item The first term of the expression (related to $\Delta_2$) boils down to: \\
    $$ (N(N-1)) \frac{1}{{N \choose 4}} {4 \choose 2} {N-4 \choose 4-2} \Delta_2 = (N(N-1)) \frac{4!}{N(N-1)(N-2)(N-3)} \frac{4 \times 3}{2!} \frac{(N-4)(N-5)}{2!} \Delta_2$$
    which, as $N \xrightarrow[]{} \infty$, is $\mathcal{O}(1)$, converging to $72 \Delta_2$.
    \item The second term of the expression (related to $\Delta_3$) boils down to: \\
    $$ (N(N-1)) \frac{1}{{N \choose 4}} {4 \choose 3} {N-4 \choose 4-3} \Delta_3 = (N(N-1)) \frac{4!}{N(N-1)(N-2)(N-3)} \frac{4 \times 3 \times 2}{3!} \frac{(N-4)}{1!} \Delta_3$$
    which, as $N \xrightarrow[]{} \infty$, is $\mathcal{O} \left( \frac{1}{N} \right)$.
    \item The third term of the expression (related to $\Delta_4$) boils down to: \\
    $$ (N(N-1)) \frac{1}{{N \choose 4}} {4 \choose 4} {N-4 \choose 4-4} \Delta_4 = (N(N-1)) \frac{4!}{N(N-1)(N-2)(N-3)} \frac{4 \times 3 \times 2 \times 1}{4!} \times 1 \Delta_4$$
    which, as $N \xrightarrow[]{} \infty$, is $\mathcal{O} \left( \frac{1}{N^2} \right)$.
\end{itemize}
Therefore, the term related to $\Delta_2$ asymptotically dominates the expression of the variance, and $\text{Var} (\sqrt{N(N-1)} U_N)$ converges to $72 \Delta_2$, or, equivalently to $144 \delta_2$:
$$ \text{Var} (\sqrt{N(N-1)} U_N) \xrightarrow[]{N \xrightarrow{} \infty} 72 \Delta_2 = 144 \delta_2 $$
We can also further pin down the expression for $\Delta_q$, and more specifically for $\Delta_2$. In our specific case we have that:
\begin{align}
    \Delta_q &= \text{Cov} [\tilde{s}_{ijkl} \tilde{s}_{mnop}] \\
    &= \text{Cov} [s_{ijkl} s_{mnop}] \nonumber \\
    &= \mathbbm{E} [s_{ijkl} s_{mnop}] \nonumber \\
    &= \mathbbm{E} [s_{ijkl,q}^2] \nonumber
\end{align}
Where we have that $i,j,k,l$ and $m,n,o,p$ share $q$ indices in common.
The first equality follows by definition, the second and third equalities follows from the fact that, in our case, $\theta = \mathbbm{E} [s_{ijkl}] = \mathbbm{E} [s_{mnop}] = 0$.  Finally, the fourth equation follows from first defining $s_{ijkl,q}$ to be the expected value of $s_{ijkl}$ when conditioning on the characteritics of the $q$ individuals. Since the observations are i.i.d, we have necessarily that $s_{ijkl,q} = s_{mnop,q}$. Further, one can apply the law of iterated expectations on the expression $\mathbbm{E} [s_{ijkl} s_{mnop}]$ by considering the conditioning on the characteritics of the $q$ individuals, arriving to the final term.

To be more clear we can consider the case where $q=2$ and denoting for instance that the two quadruples share the same indices $i$ and $j$, so in the previous expression, for example, $m=i$ and $n=j$:
\begin{align}
    \Delta_2 &= \text{Cov} [\tilde{s}_{ijkl} \tilde{s}_{ijop}] \\
    &= \text{Cov} [s_{ijkl} s_{ijop}] \nonumber \\
    &= \mathbbm{E} [s_{ijkl} s_{ijop}] \nonumber \\
    &= \mathbbm{E} [\mathbbm{E}[s_{ijkl} s_{ijop} \rvert A_i, A_j, B_i, B_j, U_{ij}, U_{ji}]] \nonumber\\
    &= \mathbbm{E} [\mathbbm{E}[s_{ijkl} \rvert A_i, A_j, B_i, B_j, U_{ij}, U_{ji}] \mathbbm{E}[ s_{ijop} \rvert A_i, A_j, B_i, B_j, U_{ij}, U_{ji}]] \nonumber\\
    &= \mathbbm{E} [s_{ijkl,2}^2] \nonumber \\
    &= \mathbbm{E}[\bar{s}_{ij,2}^2]
\end{align}
Where the fifth equality follows from the fact that conditioning on $A_i, A_j, B_i, B_j, U_{ij}, U_{ji}$, $s_{ijkl}$ and $s_{ijop}$ are independent. In the main text we also redefine $s_{ijkl,2} = \bar{s}_{ij,2}$, for simplicity of notation. 

We can further work out the expression for $\Delta_2$ by opening up the expression for $s_{ijkl,2}$:
\begin{gather}
\begin{align}
    s_{ijkl,2} &= \mathbbm{E} [s_{ijkl} \rvert A_i, A_j, B_i, B_j, U_{ij}, U_{ji}] \\
    &= \mathbbm{E} \Big[\frac{1}{4!} \sum_{\pi \in \mathcal{P}(\mathcal{C},4)} ((X_{\pi_1\pi_2} - X_{\pi_1\pi_3})-(X_{\pi_4\pi_2}-X_{\pi_4\pi_3})) ((U_{\pi_1\pi_2} - U_{\pi_1\pi_3})-(U_{\pi_4\pi_2}-U_{\pi_4\pi_3})) \rvert A_i, A_j, B_i, B_j, U_{ij}, U_{ji}\Big]\nonumber \\
    &= \frac{1}{4!} \mathbbm{E} \Big[ \tilde{X}_{ijkl} ((U_{ij} - U_{ik}) - (U_{lj} - U_{lk})) + \tilde{X}_{ikjl} ((U_{ik} - U_{ij}) - (U_{lk} - U_{lj})) \nonumber \\
    &+ \tilde{X}_{kjli} ((U_{kj} - U_{kl}) - (U_{ij} - U_{il})) + \tilde{X}_{lkji} ((U_{lk} - U_{lj}) - (U_{ik} - U_{ij})) \nonumber \\
    &+ \tilde{X}_{klji} ((U_{kl} - U_{kj}) - (U_{il} - U_{ij})) + \tilde{X}_{ljki} ((U_{lj} - U_{lk}) - (U_{ij} - U_{ik})) \nonumber \\
    &+ \tilde{X}_{ijlk} ((U_{ij} - U_{il}) - (U_{kj} - U_{kl})) + \tilde{X}_{iljk} ((U_{il} - U_{ij}) - (U_{kl} - U_{kj})) \nonumber \\
    &+ \tilde{X}_{jikl} ((U_{ji} - U_{jk}) - (U_{li} - U_{lk})) + \tilde{X}_{jkil} ((U_{jk} - U_{ji}) - (U_{lk} - U_{li})) \nonumber \\
    &+ \tilde{X}_{lijk} ((U_{li} - U_{lk}) - (U_{ji} - U_{jk})) + \tilde{X}_{klij} ((U_{kl} - U_{ki}) - (U_{jl} - U_{ji})) \nonumber \\
    &+ \tilde{X}_{lkij} ((U_{lk} - U_{li}) - (U_{jk} - U_{ji})) + \tilde{X}_{kilj} ((U_{ki} - U_{kl}) - (U_{ji} - U_{jl})) \nonumber \\
    &+ \tilde{X}_{jilk} ((U_{ji} - U_{jl}) - (U_{ki} - U_{kl})) + \tilde{X}_{jlik} ((U_{jl} - U_{ji}) - (U_{kl} - U_{ki})) \rvert A_i, A_j, B_i, B_j, U_{ij}, U_{ji} \Big] \nonumber
\end{align}
\end{gather}

Note that we obtain the expression above by observing that terms that do not contain $U_{ij}$ or $U_{ji}$ will have expected value equal to zero. We can further separate the terms where the conditioning on $A_i, B_j$ and $U_{ij}$ matters from the terms where the conditioning on $A_j, B_i$ and $U_{ji}$ matters:
\begin{align}
    s_{ijkl,2} &= \frac{1}{4!} \mathbbm{E} \Big[ \tilde{X}_{ijkl} ((U_{ij} - U_{ik}) - (U_{lj} - U_{lk})) + \tilde{X}_{ikjl} ((U_{ik} - U_{ij}) - (U_{lk} - U_{lj})) \nonumber \\
    &+ \tilde{X}_{kjli} ((U_{kj} - U_{kl}) - (U_{ij} - U_{il})) + \tilde{X}_{lkji} ((U_{lk} - U_{lj}) - (U_{ik} - U_{ij})) \nonumber \\
    &+ \tilde{X}_{klji} ((U_{kl} - U_{kj}) - (U_{il} - U_{ij})) + \tilde{X}_{ljki} ((U_{lj} - U_{lk}) - (U_{ij} - U_{ik})) \nonumber \\
    &+ \tilde{X}_{ijlk} ((U_{ij} - U_{il}) - (U_{kj} - U_{kl})) + \tilde{X}_{iljk} ((U_{il} - U_{ij}) - (U_{kl} - U_{kj})) \rvert A_i, B_j, U_{ij} \Big]  \\
    &+ \frac{1}{4!} \mathbbm{E} \Big[ \tilde{X}_{jikl} ((U_{ji} - U_{jk}) - (U_{li} - U_{lk})) + \tilde{X}_{jkil} ((U_{jk} - U_{ji}) - (U_{lk} - U_{li})) \nonumber \\
    &+ \tilde{X}_{lijk} ((U_{li} - U_{lk}) - (U_{ji} - U_{jk})) + \tilde{X}_{klij} ((U_{kl} - U_{ki}) - (U_{jl} - U_{ji})) \nonumber \\
    &+ \tilde{X}_{lkij} ((U_{lk} - U_{li}) - (U_{jk} - U_{ji})) + \tilde{X}_{kilj} ((U_{ki} - U_{kl}) - (U_{ji} - U_{jl})) \nonumber \\
    &+ \tilde{X}_{jilk} ((U_{ji} - U_{jl}) - (U_{ki} - U_{kl})) + \tilde{X}_{jlik} ((U_{jl} - U_{ji}) - (U_{kl} - U_{ki})) \rvert A_j, B_i, U_{ji} \Big] 
\end{align}

We can denote the first conditional expectation as $\bar{s}_{ij}$, and the second conditional expectation as $\bar{s}_{ji}$. Therefore, $s_{ijkl,2} = \bar{s}_{ij} + \bar{s}_{ji}$, where:
$$ \bar{s}_{ij} = \mathbbm{E} [s_{ijkl} \rvert A_i, B_j, U_{ij}] $$
$$ \bar{s}_{ji} = \mathbbm{E} [s_{ijkl} \rvert A_j, B_i, U_{ji}] $$
Then, we have that:
\begin{align}
    \Delta_2 &= \mathbbm{E} [s_{ijkl,2}^2] = \mathbbm{E} [(\bar{s}_{ij} + \bar{s}_{ji})^2] = \mathbbm{E} [\bar{s}_{ij}^2] + 2 \mathbbm{E} [\bar{s}_{ij} \bar{s}_{ji}] + \mathbbm{E} [\bar{s}_{ji}^2] \\
    &= \mathbbm{E} [\bar{s}_{ij}^2] + \mathbbm{E} [\bar{s}_{ji}^2] \nonumber \\
    &= 2 \delta_2 \nonumber
\end{align}
Where the first line follows from linearity of expectations, the second line follows from independency between $\bar{s}_{ij}$ and $\bar{s}_{ji}$, and the fact that both have expectation equal to zero. And finally, the third line follows from the fact that as the observations and error terms are i.i.d., the expected value of $\bar{s}_{ij}^2$ and $\bar{s}_{ji}^2$ are the same, and we can denote: $\mathbbm{E} [\bar{s}_{ij}^2] = \mathbbm{E} [\bar{s}_{ji}^2] = \delta_2$.

To prove this previous expression, first note that we can rewrite, since the observations and the error terms are independent:
$$ \bar{s}_{ij} = \mathbbm{E} [\tilde{X}_{ijkl} - \tilde{X}_{ikjl} - \tilde{X}_{kjli} + \tilde{X}_{lkji}+ \tilde{X}_{klji}  -  \tilde{X}_{ljki}  + \tilde{X}_{ijlk} -  \tilde{X}_{iljk} \rvert A_i, B_j] U_{ij} = \bar{\tilde{X}}_{ij} U_{ij} $$
$$ \bar{s}_{ji} = \mathbbm{E} [\tilde{X}_{jikl} - \tilde{X}_{jkil} - \tilde{X}_{likj} + \tilde{X}_{klij}+ \tilde{X}_{lkij}  -  \tilde{X}_{kilj}  + \tilde{X}_{jilk} -  \tilde{X}_{jlik} \rvert A_j, B_i] U_{ji} = \bar{\tilde{X}}_{ji} U_{ji} $$
Then, it follows that:
$$ \mathbbm{E} [\bar{s}_{ij} \bar{s}_{ji}] = \mathbbm{E} [\bar{\tilde{X}}_{ij} U_{ij} \bar{\tilde{X}}_{ji} U_{ji}] = \mathbbm{E} [\bar{\tilde{X}}_{ij} \bar{\tilde{X}}_{ji}] \mathbbm{E} [U_{ij}] \mathbbm{E} [U_{ji}] = 0 $$ \qed

\subsection{Proof of Lemma \ref{lemma:2}}
To derive the variance of the H\'{a}jek projection, we will follow closely the steps proposed by \cite{graham2017econometric}.

Denote $\bar{s}_{i'j'} = \mathbbm{E}[s_{ijkl} \rvert A_{i'}, B_{j'}, U_{i'j'}]$, when $\{i',j'\} \subset \{i,j,k,l\}$. From Definition \ref{def:2}, note that we have:
\begin{align}
    \hat{U}_{N,1} &= \sum_{i' = 1}^N \sum_{j' \neq i'} \frac{1}{{N \choose 4}} \sum_{\mathcal{C} \in \mathcal{C}(\mathcal{N},4)} \mathbbm{E}[s_{ijkl} \rvert A_{i'}, B_{j'}, U_{i'j'}] \\
    &= \sum_{i' = 1}^N \sum_{j' \neq i'} \frac{1}{{N \choose 4}} {N-2 \choose 2} \bar{s}_{i'j'} \nonumber \\
    &= \sum_{i' = 1}^N \sum_{j' \neq i'} \frac{12}{N(N-1)} \bar{s}_{i'j'}. \nonumber
\end{align}
As observed by \cite{graham2017econometric}, the random variables $\{\bar{s}_{i'j'}\}_{i' = 1, j' \neq i'}^N$ in the double sum in the last expression are not i.i.d., but they are uncorrelated. This is an implication of conditional independence given $\textbf{A}, \textbf{B}$. For instance, we can show that $\bar{s}_{i'j'}$ and $\bar{s}_{k'l'}$ for at least one index of the dyads being different, i.e., either $i' \neq k'$, or $j' \neq l'$, or both, are uncorrelated. From the law of iterated expectations, we can evaluate:
\begin{align} \label{eq:uncorrelated1}
    \mathbbm{E}[\bar{s}_{i'j'} \bar{s}_{k'l'}'] &= \mathbbm{E}[\mathbbm{E}[s_{i'j'mp} \rvert A_i, B_j, U_{ij}] \mathbbm{E}[s_{klm'p'} \rvert A_k, B_l, U_{kl}]'] \\
    &= \mathbbm{E}[\mathbbm{E}[\mathbbm{E}[s_{ijmp} \rvert A_i, B_j, U_{ij}] \mathbbm{E}[s_{klm'p'} \rvert A_k, B_l, U_{kl}]' \rvert \textbf{A}, \textbf{B} ]] \nonumber \\
    &=  \mathbbm{E}[\mathbbm{E}[\mathbbm{E}[s_{ijmp} \rvert A_i, B_j, U_{ij}] \rvert \textbf{A}, \textbf{B}] \mathbbm{E} [\mathbbm{E}[s_{klm'p'} \rvert A_k, B_l, U_{kl}]' \rvert \textbf{A}, \textbf{B} ]'] \nonumber \\
    &= \mathbbm{E}[\mathbbm{E}[s_{ijmp} \rvert \textbf{A}, \textbf{B}] \mathbbm{E}[s_{klmp}  \rvert \textbf{A}, \textbf{B}]'] \nonumber \\
    &= 0. \nonumber
\end{align}
Given this result and the fact that $\mathbbm{E}[\bar{s}_{i'j'}] = 0$, we have that:
\begin{align}
     \text{Var} ( \hat{U}_{N,1} ) &= \left(\frac{12}{N(N-1)}\right)^2 \sum_{i' = 1}^N \sum_{j' \neq i'} \text{Var} (\bar{s}_{i'j'} ) \\
     &= \left(\frac{12}{N(N-1)}\right)^2 \sum_{i' = 1}^N \sum_{j' \neq i'} \mathbbm{E}[\bar{s}_{i'j'} \bar{s}_{i'j'}']. \nonumber
\end{align}
Finally, we have that the previous expression for the variance can be rewritten as:
\begin{align}
    \text{Var} ( \hat{U}_{N,1} ) &= \left(\frac{12}{N(N-1)}\right)^2 \sum_{i' = 1}^N \sum_{j' \neq i'} \mathbbm{E}[\bar{s}_{i'j'} \bar{s}_{i'j'}'] \\
    &= \left(\frac{12}{N(N-1)}\right)^2 \sum_{i' = 1}^N \sum_{j' \neq i'} \delta_2 = \frac{144}{N(N-1)} \delta_2. \nonumber
\end{align}
Notice that the expression $\mathbbm{E}[\bar{s}_{i'j'} \bar{s}_{i'j'}']$ has the same value for any dyad $(i', j')$ since the random variables $\{X_{i'j'}\}_{i'=1, j' \neq i'}^N$ are identically distributed and so are $\{U_{i'j'}\}_{i'=1, j' \neq i'}$. Finally, by rescaling the statistic:
$$ \text{Var} (\sqrt{N(N-1)} \hat{U}_{N,1}) = 144 \delta_2. $$ \qed

\subsection{Proof of Lemma \ref{lemma:3}}
Denoting $\bar{s}_{i'j',2} = \mathbbm{E}[s_{ijkl} \rvert A_{i'}, A_{j'}, B_{i'}, B_{j'}, U_{i'j'}, U_{j'i'}]$, when $\{i', j'\} \subset \{i,j,k,l\}$. Then, from before we have that:
\begin{align}
    \hat{U}_{N,2} &= \sum_{i'=1}^N \sum_{j' > i'} \frac{1}{{N \choose 4}} \sum_{\mathcal{C} \in \mathcal{C}(\mathcal{N},4)} \mathbbm{E} [s_{ijkl} \rvert A_{i'}, A_{j'}, B_{i'}, B_{j'}, U_{i'j'}, U_{j'i'}] \\
    &= \sum_{i'=1}^N \sum_{j' > i'}  \frac{1}{{N \choose 4}} {N-2 \choose 2} \bar{s}_{i'j',2} \nonumber \\
    &= \sum_{i'=1}^N \sum_{j' > i'} \frac{12}{N(N-1)} \bar{s}_{i'j',2}. \nonumber
\end{align}
To derive the variance of this term, we can first show that, similarly to the case of the first proposed projection, $\bar{s}_{i'j',2}$ and $\bar{s}_{k'l',2}$ are uncorrelated when having at least one different index for instance $i' \neq k'$ (independently of the order). This is a result from the conditional independence of the model, and from the fact that the idiosyncratic errors are i.i.d., as it is given by Assumption \ref{ass:1}. This is analogous to the previous proof for the first variance.

We can further denote this variance as a function of $\Delta_2$. To do so, consider two tetrads ${i',j',k,l}$ and ${i',j',m,p}$ sharing the indices $i'$ and $j'$ in common. Due to the Assumptions \ref{ass:1} - \ref{ass:3}, we have the following result of conditional independence:
$$ \mathbbm{E}[s_{i'j'kl} s_{i'j'mp}' \rvert A_{i'}, A_{j'}, B_{i'}, B_{j'}, U_{i'j'}, U_{j'i'}] = \bar{s}_{i'j',2} \bar{s}_{i'j',2} '. $$
Then, from the previous definition of $\Delta_2$, we have, by law of iterated expectations:
\begin{align}
    \Delta_2 = \text{Cov} (s_{i'j'kl}, s_{i'j'mp}) &= \mathbbm{E}[s_{i'j'kl}s_{i'j'mp}'] - \mathbbm{E}[s_{i'j'kl}] \mathbbm{E}[s_{i'j'mp}]' \\
    &= \mathbbm{E}[ \mathbbm{E}[s_{i'j'kl}s_{i'j'mp}' \rvert A_{i'}, A_{j'}, B_{i'}, B_{j'}, U_{i'j'}, U_{j'i'}]] \nonumber \\
    &= \mathbbm{E} [\bar{s}_{i'j',2} \bar{s}_{i'j',2}'] \nonumber \\
    &= \mathbbm{E} [\bar{s}_{i'j',2}^2 ]. \nonumber
\end{align}
Independently of which indices $\{i', j'\}$ are, the expression $\mathbbm{E} [\bar{s}_{i'j',2}^2 ]$ will always be equal to a constant $\Delta_2$, since from Assumptions \ref{ass:1} and \ref{ass:2}, the random variables $\{X_{i'j'}\}_{i'=1, j' \neq i'}^N$ are identically distributed and so are $\{U_{i'j'}\}_{i'=1, j' \neq i'}$, which is the same argument used previously for the variance of the first projection.

Therefore, the variance of the projection $\hat{U}_{N,2}$ can be rewritten as:
\begin{align}
    \text{Var} (\hat{U}_{N,2}) &= \left( \frac{12}{N(N-1)} \right)^2 \sum_{i'=1}^N \sum_{j' > i'} \text{Var} (\bar{s}_{i'j',2}) \\
    &= \left( \frac{12}{N(N-1)} \right)^2 \sum_{i'=1}^N \sum_{j' > i'} \mathbbm{E} [\bar{s}_{i'j',2}^2 ] \nonumber \\
    &= \frac{144}{(N(N-1))^2} \frac{N(N-1)}{2} \mathbbm{E} [\bar{s}_{i'j',2}^2 ] \nonumber \\
    &= \frac{72}{N(N-1)} \Delta_2. \nonumber
\end{align}
Finally, by rescaling the statistic:
$$ \text{Var} (\sqrt{N(N-1)} \hat{U}_{N,2}) = 72 \Delta_2. $$ \qed

\subsection{Proof of Theorem \ref{theorem:2}}
We demonstrate the proof to the first H\'{a}jek projection, $\hat{U}_{N,1}$. We have that, since $\mathbbm{E}[\hat{U}_{N,1}] = \mathbbm{E}[U_N] = 0$:
\begin{align}
    N(N-1) \mathbbm{E}[(\hat{U}_{N,1} - U_N)^2] &= N(N-1) \mathbb{E} [\hat{U}_{N,1}^2 - 2 \hat{U}_{N,1} U_N + U_N^2] \\
    &= N(N-1) \text{Var} (U_N) + N(N-1) \text{Var} (\hat{U}_{N,1}) \nonumber \\ &- 2 N(N-1)\text{Cov}(U_N, \hat{U}_{N,1}). \nonumber
\end{align}
We know from before that the first term converges in probability to a constant, $144 \delta_2$, as shown in Theorem \ref{theorem:1}. Moreover, from Lemma \ref{lemma:2}, we have that the second term is equal to $144 \delta_2$, therefore, they have the same probability limit, and we can further rewrite:
\begin{align*}
    N(N-1) \mathbbm{E}[(\hat{U}_{N,1} - U_N)^2] &= 2 \times 144 \delta_2 - 2 N(N-1)\text{Cov}(U_N, \hat{U}_{N,1}) + o_p(1).
\end{align*}
Now it only remains to evaluate the covariance term, $- 2 N(N-1)\text{Cov}(U_N, \hat{U}_{N,1})$. To do so, first note that the statistic $U_N$ and the projection $\hat{U}_{N,1}$ are orthogonal, following the arguments of \cite{graham2017econometric}:
\begin{align*}
    &\mathbb{E} \left[ (U_N - \hat{U}_{N,1}) \sum_{i'=1}^N \sum_{j' \neq i'} g_{i'j'} (A_{i'}, B_{j'}, U_{i',j'}) \right] \\
    &= \sum_{i'=1}^N \sum_{j' \neq i'} \mathbb{E} [(U_N - \hat{U}_{N,1})g_{i'j'} (A_{i'}, B_{j'}, U_{i',j'})] \nonumber\\
    &= \sum_{i'=1}^N \sum_{j' \neq i'} \mathbb{E} [\mathbb{E}[U_N - \hat{U}_{N,1} \rvert A_{i'}, B_{j'}, U_{i',j'}] g_{i'j'} (A_{i'}, B_{j'}, U_{i',j'})] \nonumber \\
    &= 0,
\end{align*}
\noindent for $g_{i'j'}(\cdot)$ an arbitrary function of $(A_{i'}, B_{j'}, U_{i',j'})$.

Given this orthogonality result, and given that $\mathbb{E} [\hat{U}_{N,1}] = 0$, we have that:
\begin{align*}
    \text{Cov} (U_N, \hat{U}_{N,1}) &= \mathbbm{E}[\hat{U}_{N,1} U_N] \\
    &= \mathbbm{E} [(U_N-\hat{U}_{N,1}) \hat{U}_{N,1} + \hat{U}_{N,1} \hat{U}_{N,1}] \\
    &= \mathbbm{E}[(U_N - \hat{U}_{N,1}) \hat{U}_{N,1}] + \mathbbm{E}[\hat{U}_{N,1} \hat{U}_{N,1}] \\
    &= \text{Var} (\hat{U}_{N,1}).
\end{align*}
Therefore, the term  $- 2 N(N-1)\text{Cov}(U_N, \hat{U}_{N,1})$ is equal to $-2 \times 144 \delta_2$, which gives us the result:
\begin{align*}
    N(N-1) \mathbbm{E}[(\hat{U}_{N,1} - U_N)^2] &=  o_p(1)
\end{align*}. \qed 

The proof for the second proposed H\'{a}jek projection is analogous.

\subsection{Proof of Proposition \ref{lemma:5}}

The goal is to show that:
$$\left[ \frac{1}{{N \choose 4}} \sum_{\mathcal{C} \in \mathcal{C}(\mathcal{N},4)} \frac{1}{4!} \sum_{\pi \in \mathcal{P}(\mathcal{C},4)} \Tilde{X}_{\pi_1\pi_2\pi_3\pi_4} \Tilde{X}_{\pi_1\pi_2\pi_3\pi_4}' \right] \xrightarrow[]{p} \Gamma = \mathbb{E} [\Tilde{X}_{\pi_1\pi_2\pi_3\pi_4} \Tilde{X}_{\pi_1\pi_2\pi_3\pi_4}']$$
Given that $X_{ij} = f(A_i, B_j)$, and assuming that $f$ is an integrable function, we can use Chebyshev's inequality to prove the above.

The statement above corresponds to, given any $\epsilon > 0$:
$$ \lim_{N \to \infty} \mathbb{P} \left[ \left\rvert \frac{1}{{N \choose 4}} \sum_{\mathcal{C} \in \mathcal{C}(\mathcal{N},4)} \frac{1}{4!} \sum_{\pi \in \mathcal{P}(\mathcal{C},4)} \Tilde{X}_{\pi_1\pi_2\pi_3\pi_4} \Tilde{X}_{\pi_1\pi_2\pi_3\pi_4}' - \mathbb{E} \left[ \Tilde{X}_{\pi_1\pi_2\pi_3\pi_4} \Tilde{X}_{\pi_1\pi_2\pi_3\pi_4}' \right] \right\rvert > \epsilon \right] = 0 $$

By Chebyshev's inequality we have the following:
\begin{align} \label{eq:cheby}
    &\mathbb{P} \left[ \left\rvert \frac{1}{{N \choose 4}} \sum_{\mathcal{C} \in \mathcal{C}(\mathcal{N},4)} \frac{1}{4!} \sum_{\pi \in \mathcal{P}(\mathcal{C},4)} \Tilde{X}_{\pi_1\pi_2\pi_3\pi_4} \Tilde{X}_{\pi_1\pi_2\pi_3\pi_4}' - \mathbb{E} \left[ \Tilde{X}_{\pi_1\pi_2\pi_3\pi_4} \Tilde{X}_{\pi_1\pi_2\pi_3\pi_4}' \right] \right\rvert > \epsilon \right] \nonumber \\
    & \leq \frac{\text{Var} \left(  \frac{1}{{N \choose 4}} \sum_{\mathcal{C} \in \mathcal{C}(\mathcal{N},4)} \frac{1}{4!} \sum_{\pi \in \mathcal{P}(\mathcal{C},4)} \Tilde{X}_{\pi_1\pi_2\pi_3\pi_4} \Tilde{X}_{\pi_1\pi_2\pi_3\pi_4}' \right)}{\epsilon^2} 
\end{align}
Therefore, the first step is to establish the variance term and then taking the limit as $N \xrightarrow[]{} \infty$.

Taking the same first steps as we took when definiting the variance of the statistic $U_N$, we can rewrite:
\begin{align}
    &\text{Var} \left(  \frac{1}{{N \choose 4}} \sum_{\mathcal{C} \in \mathcal{C}(\mathcal{N},4)} \frac{1}{4!} \sum_{\pi \in \mathcal{P}(\mathcal{C},4)} \Tilde{X}_{\pi_1\pi_2\pi_3\pi_4} \Tilde{X}_{\pi_1\pi_2\pi_3\pi_4}' \right) \\
    &= \left( \frac{1}{{N \choose 4}} \right)^2 \left( \frac{1}{4!} \right)^2 \text{Var} \left(\sum_{\mathcal{C} \in \mathcal{C}(\mathcal{N},4)} \sum_{\pi \in \mathcal{P}(\mathcal{C},4)} \Tilde{X}_{\pi_1\pi_2\pi_3\pi_4} \Tilde{X}_{\pi_1\pi_2\pi_3\pi_4}'  \right) \nonumber \\
    &= \left( \frac{1}{{N \choose 4}} \right)^2 \left( \frac{1}{4!} \right)^2 \text{Cov} \left(\sum_{\mathcal{C} \in \mathcal{C}(\mathcal{N},4)} \sum_{\pi \in \mathcal{P}(\mathcal{C},4)} \Tilde{X}_{\pi_1\pi_2\pi_3\pi_4} \Tilde{X}_{\pi_1\pi_2\pi_3\pi_4}', \sum_{\mathcal{C} \in \mathcal{C}(\mathcal{N},4)} \sum_{\pi \in \mathcal{P}(\mathcal{C},4)} \Tilde{X}_{\pi_1\pi_2\pi_3\pi_4} \Tilde{X}_{\pi_1\pi_2\pi_3\pi_4}'  \right) \nonumber
\end{align}
Note that, if we open the cross sums in the covariance term, we have that, in a case where $\{\pi_1, \pi_2, \pi_3, \pi_4\}$ and $\{\pi_1', \pi_2', \pi_3', \pi_4'\}$ share no indices in common, we have that:
$$ \text{Cov} (\Tilde{X}_{\pi_1\pi_2\pi_3\pi_4} \Tilde{X}_{\pi_1\pi_2\pi_3\pi_4}' , \Tilde{X}_{\pi_1'\pi_2'\pi_3'\pi_4'} \Tilde{X}_{\pi_1'\pi_2'\pi_3'\pi_4'}' ) = 0 $$
\noindent since $\{A_i\}_{i=1}^N$ and $\{B_j\}_{j=1}^N$ are mutually independent. 

However, for the cases where we have 1,2,3 or 4 indices in common the same does not hold.

We can start with, for instance, evaluating the covariance when having one index in common, suppose $\pi_1 = \pi_1' = i$. In this case, we have the following:
\begin{align}
    &\text{Cov} (\Tilde{X}_{i\pi_2\pi_3\pi_4} \Tilde{X}_{i\pi_2\pi_3\pi_4}' , \Tilde{X}_{i\pi_2'\pi_3'\pi_4'} \Tilde{X}_{i\pi_2'\pi_3'\pi_4'}' ) \\
    & = \mathbb{E} [\Tilde{X}_{i\pi_2\pi_3\pi_4} \Tilde{X}_{i\pi_2\pi_3\pi_4}' \Tilde{X}_{i\pi_2'\pi_3'\pi_4'} \Tilde{X}_{i\pi_2'\pi_3'\pi_4'}'] - \mathbb{E} [\Tilde{X}_{i\pi_2\pi_3\pi_4} \Tilde{X}_{i\pi_2\pi_3\pi_4}'] \mathbb{E} [\Tilde{X}_{i\pi_2'\pi_3'\pi_4'} \Tilde{X}_{i\pi_2'\pi_3'\pi_4'}'] \nonumber 
\end{align}
The problem here is that we need to evaluate these expressions for all the different possibilities of sharing one element in common in different positions, and then the same for 2, 3 or 4 elements in common. This is necessary, once we would like to show that those covariances are bounded.

If we sum the covariances found for the different possibilities of one element in common, it is the same as checking the covariances for the sum of all the permutations of quadruples having one element in common. As before, since $A_i$ and $B_j$ are i.i.d., those covariances will be constant for any two quadruples sharing one index in common. We can then denote the covariances of quadruples sharing 1, 2, 3 or 4 elements in common by: $\Delta_{X,1}, \Delta_{X,2}, \Delta_{X,3}, \Delta_{X,4}$.

Then, similarly to before, we can check how many combinations of quadruples have how many elements in common arriving at:
\begin{align}
    &\text{Var} \left(  \frac{1}{{N \choose 4}} \sum_{\mathcal{C} \in \mathcal{C}(\mathcal{N},4)} \frac{1}{4!} \sum_{\pi \in \mathcal{P}(\mathcal{C},4)} \Tilde{X}_{\pi_1\pi_2\pi_3\pi_4} \Tilde{X}_{\pi_1\pi_2\pi_3\pi_4}' \right) \\
    &= \frac{1}{3!3!} \frac{(N-4)(N-5)(N-6)}{N(N-1)(N-2)(N-3)} \Delta_{X,1} \nonumber \\
    &+ \frac{1}{2! 2! 2!} \frac{(N-4)(N-5)}{N(N-1)(N-2)(N-3)} \Delta_{X,2} \nonumber \\
    &+ \frac{1}{3!} \frac{(N-4)}{N(N-1)(N-2)(N-3)} \Delta_{X,3} \nonumber \\
    &+ \frac{1}{4!} \frac{1}{N(N-1)(N-2)(N-3)} \Delta_{X,4} \nonumber
\end{align}
Then, if we take the limit as $N \xrightarrow[]{} \infty$ of both sides of expression \ref{eq:cheby}, we arrive at the desired result:
$$ \lim_{N \to \infty} \mathbb{P} \left[ \left\rvert \frac{1}{{N \choose 4}} \sum_{\mathcal{C} \in \mathcal{C}(\mathcal{N},4)} \frac{1}{4!} \sum_{\pi \in \mathcal{P}(\mathcal{C},4)} \Tilde{X}_{\pi_1\pi_2\pi_3\pi_4} \Tilde{X}_{\pi_1\pi_2\pi_3\pi_4}' - \mathbb{E} \left[ \Tilde{X}_{\pi_1\pi_2\pi_3\pi_4} \Tilde{X}_{\pi_1\pi_2\pi_3\pi_4}' \right] \right\rvert > \epsilon \right] = 0 $$ \qed

\subsection{Proof of Theorem \ref{theorem:3}}
To establish the consistency of the estimator, we can first look at:
\begin{align} 
  \text{plim } \hat{\beta}_{1} &=\beta_1 + \left[\text{plim } \frac{1}{{N \choose 4}} \sum_{\mathcal{C} \in \mathcal{C}(\mathcal{N},4)} \frac{1}{4!} \sum_{\pi \in \mathcal{P}(\mathcal{C},4)} \Tilde{X}_{\pi_1\pi_2\pi_3\pi_4} \Tilde{X}_{\pi_1\pi_2\pi_3\pi_4}' \right]^{-1} \times \\  &\left[\text{plim } \frac{1}{{N \choose 4}} \sum_{\mathcal{C} \in \mathcal{C}(\mathcal{N},4)} \frac{1}{4!} \sum_{\pi \in \mathcal{P}(\mathcal{C},4)} \Tilde{X}_{\pi_1\pi_2\pi_3\pi_4} \Tilde{U}_{\pi_1\pi_2\pi_3\pi_4} \right] \nonumber \\
  &= \beta_1 + \left[\text{plim } \frac{1}{{N \choose 4}} \sum_{\mathcal{C} \in \mathcal{C}(\mathcal{N},4)} \frac{1}{4!} \sum_{\pi \in \mathcal{P}(\mathcal{C},4)} \Tilde{X}_{\pi_1\pi_2\pi_3\pi_4} \Tilde{X}_{\pi_1\pi_2\pi_3\pi_4}' \right]^{-1}  \left[\text{plim } U_N \right]. \nonumber
\end{align}
Given the result of Proposition \ref{lemma:5}, it follows from continuos mapping theorem:
\begin{align} 
  \text{plim } \hat{\beta}_{1} &= \beta_1 + \Gamma^{-1}  \Big[\text{plim } U_N \Big]. \nonumber
\end{align}
Moreover, from the result of Lemma \ref{lemma:6} and by Slutsky theorem:
$$ \text{plim } \hat{\beta}_{1} = \beta_1  $$ \qed 
\section{Other relevant calculations}

\subsection{Working out the expression for $\Delta_2$}

Note that, in order to derive the expression for $\Delta_2$, we only need to consider the case of two combinations that contain two indices in common. Given that the $A_i$, $B_j$ and $U_{ij}$ are i.i.d. variables, it does not matter which indices those two combinations exactly have.

Therefore, we will consider for now $\mathcal{C}_1 = {1,2,3,4}$, and $\mathcal{C}_2 = {1,2,5,6}$. 

Remember that the expression for $\Delta_2$ is then given by, in this case:
\begin{align*}
    \Delta_2 &=\text{Cov} \left(\frac{1}{4!} \sum_{\pi(\mathcal{C}_1)} \Tilde{X}_{\pi(\mathcal{C}_1)} \Tilde{U}_{\pi(\mathcal{C}_1)}, \frac{1}{4!} \sum_{\pi(\mathcal{C}_2)} \Tilde{X}_{\pi(\mathcal{C}_2)} \Tilde{U}_{\pi(\mathcal{C}_2)} \right) = \left(\frac{1}{4!}\right)^2 \text{Cov} \left( \sum_{\pi(\mathcal{C}_1)} \Tilde{X}_{\pi(\mathcal{C}_1)} \Tilde{U}_{\pi(\mathcal{C}_1)},\sum_{\pi(\mathcal{C}_2)} \Tilde{X}_{\pi(\mathcal{C}_2)} \Tilde{U}_{\pi(\mathcal{C}_2)} \right) \nonumber
\end{align*}
Therefore, we essentially need to look at the covariances of the cross terms among all the permutations of the first combination with all the permutations of the second combination. The permutations of each set with the respectively constructed variable $\Tilde{U}_{ijkl}$ are:

\begin{table}[H]
    \centering
    \begin{tabular}{rrrrr}
      \hline
        i & j & k & l & $\Tilde{U}_{ijkl}$\\
       \hline
       1 & 2 & 3 & 4 & $(U_{12} - U_{13}) - (U_{42} - U_{43})$ \\
       2 & 1 & 3 & 4 & $(U_{21} - U_{23}) - (U_{41} - U_{43})$ \\ 
       3 & 1 & 2 & 4 & $(U_{31} - U_{32}) - (U_{41} - U_{42})$ \\
       1 & 3 & 2 & 4 & $(U_{13} - U_{12}) - (U_{43} - U_{42})$ \\
       2 & 3 & 1 & 4 & $(U_{23} - U_{21}) - (U_{43} - U_{41})$ \\
       3 & 2 & 1 & 4 & $(U_{32} - U_{31}) - (U_{42} - U_{41})$ \\
       3 & 2 & 4 & 1 & $(U_{32} - U_{34}) - (U_{12} - U_{14})$ \\
       2 & 3 & 4 & 1 & $(U_{23} - U_{24}) - (U_{13} - U_{14})$ \\
       4 & 3 & 2 & 1 & $(U_{43} - U_{42}) - (U_{13} - U_{12})$ \\
       3 & 4 & 2 & 1 & $(U_{34} - U_{32}) - (U_{14} - U_{12})$ \\
       2 & 4 & 3 & 1 & $(U_{24} - U_{23}) - (U_{14} - U_{13})$ \\
       4 & 2 & 3 & 1 & $(U_{42} - U_{43}) - (U_{12} - U_{13})$ \\
       4 & 1 & 3 & 2 & $(U_{41} - U_{43}) - (U_{21} - U_{23})$ \\
       1 & 4 & 3 & 2 & $(U_{14} - U_{13}) - (U_{24} - U_{23})$ \\
       3 & 4 & 1 & 2 & $(U_{34} - U_{31}) - (U_{24} - U_{21})$ \\
       4 & 3 & 1 & 2 & $(U_{43} - U_{41}) - (U_{23} - U_{21})$ \\
       1 & 3 & 4 & 2 & $(U_{13} - U_{14}) - (U_{23} - U_{24})$ \\
       3 & 1 & 4 & 2 & $(U_{31} - U_{34}) - (U_{21} - U_{24})$ \\
       2 & 1 & 4 & 3 & $(U_{21} - U_{24}) - (U_{31} - U_{34})$ \\
       1 & 2 & 4 & 3 & $(U_{12} - U_{14}) - (U_{32} - U_{34})$ \\
       4 & 2 & 1 & 3 & $(U_{42} - U_{41}) - (U_{32} - U_{31})$ \\
       2 & 4 & 1 & 3 & $(U_{24} - U_{21}) - (U_{34} - U_{31})$ \\
       1 & 4 & 2 & 3 & $(U_{14} - U_{12}) - (U_{34} - U_{32})$ \\
       4 & 1 & 2 & 3 & $(U_{41} - U_{42}) - (U_{31} - U_{32})$ \\
    \hline
    \end{tabular}
    \caption{Permutations for $\mathcal{C}_1$}
    \label{tab:perm1}
\end{table} 

\begin{table}[H]
        \centering
        \begin{tabular}{rrrrr}
          \hline
            i & j & k & l & $\Tilde{U}_{ijkl}$\\
           \hline
           1 & 2 & 5 & 6 & $(U_{12} - U_{15}) - (U_{62} - U_{65})$ \\
           2 & 1 & 5 & 6 & $(U_{21} - U_{25}) - (U_{61} - U_{65})$ \\ 
           5 & 1 & 2 & 6 & $(U_{51} - U_{52}) - (U_{61} - U_{62})$ \\
           1 & 5 & 2 & 6 & $(U_{15} - U_{12}) - (U_{65} - U_{62})$ \\
           2 & 5 & 1 & 6 & $(U_{25} - U_{21}) - (U_{65} - U_{61})$ \\
           5 & 2 & 1 & 6 & $(U_{52} - U_{51}) - (U_{62} - U_{61})$ \\
           5 & 2 & 6 & 1 & $(U_{52} - U_{56}) - (U_{12} - U_{16})$ \\
           2 & 5 & 6 & 1 & $(U_{25} - U_{26}) - (U_{15} - U_{16})$ \\
           6 & 5 & 2 & 1 & $(U_{65} - U_{62}) - (U_{15} - U_{12})$ \\
           5 & 6 & 2 & 1 & $(U_{56} - U_{52}) - (U_{16} - U_{12})$ \\
           2 & 6 & 5 & 1 & $(U_{26} - U_{25}) - (U_{16} - U_{15})$ \\
           6 & 2 & 5 & 1 & $(U_{62} - U_{65}) - (U_{12} - U_{15})$ \\
           6 & 1 & 5 & 2 & $(U_{61} - U_{65}) - (U_{21} - U_{25})$ \\
           1 & 6 & 5 & 2 & $(U_{16} - U_{15}) - (U_{26} - U_{25})$ \\
           5 & 6 & 1 & 2 & $(U_{56} - U_{51}) - (U_{26} - U_{21})$ \\
           6 & 5 & 1 & 2 & $(U_{65} - U_{61}) - (U_{25} - U_{21})$ \\
           1 & 5 & 6 & 2 & $(U_{15} - U_{16}) - (U_{25} - U_{26})$ \\
           5 & 1 & 6 & 2 & $(U_{51} - U_{56}) - (U_{21} - U_{26})$ \\
           2 & 1 & 6 & 5 & $(U_{21} - U_{26}) - (U_{51} - U_{56})$ \\
           1 & 2 & 6 & 5 & $(U_{12} - U_{16}) - (U_{52} - U_{56})$ \\
           6 & 2 & 1 & 5 & $(U_{62} - U_{61}) - (U_{52} - U_{51})$ \\
           2 & 6 & 1 & 5 & $(U_{26} - U_{21}) - (U_{56} - U_{51})$ \\
           1 & 6 & 2 & 5 & $(U_{16} - U_{12}) - (U_{56} - U_{52})$ \\
           6 & 1 & 2 & 5 & $(U_{61} - U_{62}) - (U_{51} - U_{52})$ \\
        \hline
        \end{tabular}
        \caption{Permutations for $\mathcal{C}_2$}
        \label{tab:perm2}
\end{table} 

When looking at the cross covariances among the permutations in Table \ref{tab:perm1} and Table \ref{tab:perm2}, the only terms that will be non-zero are the ones that have $U_{12}$ in both tables, or $U_{21}$ in both tables. Therefore, all the remaining permutations can be disregarded.

When looking for instance at the covariance between the first summand in the summation $\sum_{\pi(\mathcal{C}_1)} \Tilde{X}_{\pi(\mathcal{C}_1)} \Tilde{U}_{\pi(\mathcal{C}_1)}$, with the first summand in the summation $\sum_{\pi(\mathcal{C}_2)} \Tilde{X}_{\pi(\mathcal{C}_2)} \Tilde{U}_{\pi(\mathcal{C}_2)}$, we are essentially looking at the covariance between the $\Tilde{X}$'s and $\Tilde{U}$'s generated by the first permutations in both tables, that is:
$$ \text{Cov} (\Tilde{X}_{1234}((U_{12} - U_{13}) - (U_{42} - U_{43})), \Tilde{X}_{1256}((U_{12} - U_{15}) - (U_{62} - U_{65}))) $$
Given that:
\begin{itemize}
    \item $\mathbbm{E}[\Tilde{X}_{1234}((U_{12} - U_{13}) - (U_{42} - U_{43}))] = \mathbbm{E}[\Tilde{X}_{1256}((U_{12} - U_{15}) - (U_{62} - U_{65}))] = 0$
    \item $U_{ij}$ is i.i.d., and independent of $X_{i'j'}$, for any $i,i',j,j'$ 
\end{itemize}
We can further simplify:
\begin{align}
    &\text{Cov} (\Tilde{X}_{1234}((U_{12} - U_{13}) - (U_{42} - U_{43})), \Tilde{X}_{1256}((U_{12} - U_{15}) - (U_{62} - U_{65}))) \\
    &= \mathbbm{E}[\Tilde{X}_{1234}((U_{12} - U_{13}) - (U_{42} - U_{43}))\Tilde{X}_{1256}((U_{12} - U_{15}) - (U_{62} - U_{65}))] \nonumber \\
    &= \mathbbm{E}[\Tilde{X}_{1234} \Tilde{X}_{1256}] \sigma_u^2 \nonumber
\end{align}
We then should do the same calculation for all covariances among the permutations of the first combination with the permutations of the second combination that have either $U_{12}$ or $U_{21}$ in common, and then sum those terms to obtain the final covariance expression for $\Delta_2$.

However, it is possible to further simplify the expression, by noting that since $A_i$'s and $B_j$'s are i.i.d., then we have that, for instance, the following equivalences hold:
$$ \mathbbm{E}[\Tilde{X}_{1234}\Tilde{X}_{1256}] = \mathbbm{E}[\Tilde{X}_{2143}\Tilde{X}_{2156}] $$
$$ \mathbbm{E}[\Tilde{X}_{2143}\Tilde{X}_{6152}] = \mathbbm{E}[\Tilde{X}_{1243}\Tilde{X}_{6251}] $$

That is, the expected value $\mathbbm{E}[\Tilde{X}_{ijkl} \Tilde{X}_{i'j'k'l'}]$ are the same as long as the indices in common to the both $\Tilde{X}$'s are in the same position of the permutation, for instance, if $i=i'$, and $k=k'$.

By applying these equivalences, we get to the following final expression:
\begin{align}
    \Delta_2 = (\frac{1}{4!}\big)^2 \Big[&8 \mathbbm{E}[\Tilde{X}_{1234}\Tilde{X}_{1256}] \sigma_u^2 
    -16 \mathbbm{E}[\Tilde{X}_{1234}\Tilde{X}_{1526}] \sigma_u^2 
    -16 \mathbbm{E}[\Tilde{X}_{1234}\Tilde{X}_{5261}] \sigma_u^2 
    +16 \mathbbm{E}[\Tilde{X}_{1234}\Tilde{X}_{6521}] \sigma_u^2 \nonumber \\
    &+8 \mathbbm{E}[\Tilde{X}_{1324}\Tilde{X}_{1526}] \sigma_u^2 
    +16 \mathbbm{E}[\Tilde{X}_{1324}\Tilde{X}_{5261}] \sigma_u^2 
    -16 \mathbbm{E}[\Tilde{X}_{1324}\Tilde{X}_{6521}] \sigma_u^2
    +8 \mathbbm{E}[\Tilde{X}_{3241}\Tilde{X}_{5261}] \sigma_u^2 \nonumber \\
    &-16 \mathbbm{E}[\Tilde{X}_{3241}\Tilde{X}_{6521}] \sigma_u^2
    +8 \mathbbm{E}[\Tilde{X}_{4321}\Tilde{X}_{6521}] \sigma_u^2 \Big] 
\end{align}

\subsection{Simplifying the first H\'{a}jek projection}

We can further boil down the H\'{a}jek projection by checking when fixing two indices of the permutations to be $i$ and $j$, and letting the other indices to be of any other values, $k$ and $l$, how many summands in $\sum_{\pi \in \mathcal{P}(\mathcal{C},4)}$ are different than zero, and its values:
\begin{itemize}
    \item Term 1:
    $\mathbbm{E}[((X_{ij} - X_{i\pi_3})-(X_{k_4j}-X_{\pi_4\pi_3})) \rvert A_i, B_j] U_{ij} = (X_{ij} - \mathbbm{E}[X_{i\pi_3} \rvert A_i] - \mathbbm{E}[X_{\pi_4j} \rvert B_j] + \mathbbm{E}[X_{\pi_4\pi_3}])U_{ij}$
    \item Term 2:
$    \mathbbm{E}[((X_{ik_3} - X_{ij})-(X_{\pi_4\pi_3}-X_{\pi_4j}))\rvert A_i, B_j] (-U_{ij}) = (\mathbbm{E}[X_{i\pi_3} \rvert A_i] - X_{ij} - \mathbbm{E} [X_{\pi_4\pi_3}] + \mathbbm{E} [X_{\pi_4j} \rvert B_j])(-U_{ij})$
    \item Term 3:
$\mathbbm{E} [((X_{\pi_3j} - X_{\pi_3 \pi_4}) - (X_{ij} - X_{i\pi_4})) \rvert A_i, B_j](-U_{ij})= (\mathbbm{E} [X_{\pi_3j} \rvert B_j] - \mathbbm{E} [X_{\pi_3 \pi_4}] - X_{ij} + \mathbbm{E} [X_{i\pi_4} \rvert A_i]) (-U_{ij}) $
    \item Term 4:
$\mathbbm{E} [((X_{\pi_4\pi_3} - X_{\pi_4j}) - (X_{i\pi_3 - X_{ij}})) \rvert A_i, B_j] U_{ij} = (\mathbbm{E} [X_{\pi_4 \pi_3}] - \mathbbm{E}[X_{\pi_4j} \rvert B_j] - \mathbbm{E} [X_{i\pi_3} \rvert A_i] + X_{ij}) U_{ij} $
    \item Term 5:
$\mathbbm{E} [((X_{\pi_3 \pi_4} - X_{\pi_3j}) - (X_{i\pi_4} - X_{ij})) \rvert A_i, B_j] U_{ij} = (\mathbbm{E} [X_{\pi_3\pi_4}] - \mathbbm{E} [X_{\pi_3j} \rvert B_j] - \mathbbm{E} [X_{i\pi_4} \rvert A_i + X_{ij}) U_{ij} $
    \item Term 6:
$\mathbbm{E} [((X_{\pi_4j} - X_{\pi_4\pi_3}) - (X_{ij} - X_{i\pi_3})) \rvert A_i, B_j] (-U_{ij}) = (\mathbbm{E} [X_{\pi_4j} \rvert B_j] - \mathbbm{E} [X_{\pi_4\pi_3}] - X_{ij} + \mathbbm{E} [X_{i\pi_3} \rvert A_i]) (-U_{ij}) $
    \item Term 7:
$\mathbbm{E} [((X_{ij} - X_{i\pi_4}) - (X_{\pi_3j} - X_{\pi_3 \pi_4})) \rvert A_i, B_j] U_{ij} = (X_{ij} - \mathbbm{E} [X_{i\pi_4} \rvert A_i] - \mathbbm{E} [X_{\pi_3j} \rvert B_j] + \mathbbm{E}[X_{\pi_3\pi_4}]) U_{ij} $
    \item Term 8:
$\mathbbm{E} [((X_{i\pi_4} - X_{ij}) - (X_{\pi_3 \pi_4} - X_{\pi_3j})) \rvert A_i, B_j] (-U_{ij}) = (\mathbbm{E} [X_{i\pi_4} \rvert A_i] - X_{ij} - \mathbbm{E}[X_{\pi_3\pi_4}] + \mathbbm{E} [X_{\pi_3j} \rvert B_j]) (-U_{ij})$
\end{itemize}
Moreover, given that $X_{ij} = f(A_i, B_j)$, where $X_{ij}, A_i, B_j$ are i.i.d., we have that:
$$ \mathbbm{E}[X_{i\pi_3} \rvert A_i] = \mathbbm{E}[X_{i\pi_4} \rvert A_i] = \mathbbm{E}[X_{ij} \rvert A_i] $$
$$ \mathbbm{E}[X_{\pi_4j} \rvert B_j] = \mathbbm{E}[X_{\pi_3j} \rvert B_j] = \mathbbm{E}[X_{ij} \rvert B_j] $$
$$ \mathbbm{E}[X_{\pi_4\pi_3}] =  \mathbbm{E}[X_{\pi_3\pi_4}] =  \mathbbm{E}[X_{ij}] $$
Such that we can further simplify the projection:
\begin{align}
    \hat{U}_N &= \sum_{i=1}^N \sum_{j \neq i} \frac{1}{{N \choose 4}} \frac{1}{4!} {N-2 \choose 2} 8 [(X_{ij} - \mathbbm{E}[X_{ij} \rvert A_i] - \mathbbm{E}[X_{ij} \rvert B_j] + \mathbbm{E}[X_{ij}])U_{ij}] \\
    &= \frac{4}{N(N-1)} \sum_{i=1}^N \sum_{j \neq i}[(X_{ij} - \mathbbm{E}[X_{ij} \rvert A_i] - \mathbbm{E}[X_{ij} \rvert B_j] + \mathbbm{E}[X_{ij}])U_{ij}] \nonumber
\end{align}

\subsection{Simplifying second H\'{a}jek projection}

Then, we sum over the combinations of all tetrads, however, the only ones that will lead to non-zero conditional expectations are the ones that contain the two specific indices $i$ and $j$, and any other two indices $k$ and $l$. There are a total of ${N-2 \choose 2}$ of such combinations. We can then take the conditional expectation of the kernel $s_{ijkl}$ where the indices $\pi_1, \pi_2, \pi_3, \pi_4$ denote the positions of the elements $i,j,k,l$ in a given permutation.

Again, we can further boil down the H\'{a}jek projection by checking when fixing 2 indices of the permutations to be $i$ and $j$, and letting the other indices to be of any other value, how many summands are then different then zero im the sum $\sum_{\pi \in \mathcal{P}(\mathcal{C},4)}$:
\begin{itemize}
    \item Term 1:
$\mathbbm{E}[((X_{ij} - X_{i\pi_3})-(X_{\pi_4j}-X_{\pi_4\pi_3})) \rvert A_i, B_j] U_{ij} = (X_{ij} - \mathbbm{E}[X_{i\pi_3} \rvert A_i] - \mathbbm{E}[X_{\pi_4j} \rvert B_j] + \mathbbm{E}[X_{\pi_4\pi_3}])U_{ij} $
    \item Term 2:
  $\mathbbm{E}[((X_{i\pi_3} - X_{ij})-(X_{\pi_4\pi_3}-X_{\pi_4j}))\rvert A_i, B_j] (-U_{ij}) = (\mathbbm{E}[X_{i\pi_3} \rvert A_i] - X_{ij} - \mathbbm{E} [X_{\pi_4\pi_3}] + \mathbbm{E} [X_{\pi_4j} \rvert B_j])(-U_{ij})$
    \item Term 3:
   $\mathbbm{E} [((X_{\pi_3j} - X_{\pi_3 \pi_4}) - (X_{ij} - X_{i\pi_4})) \rvert A_i, B_j](-U_{ij}) = (\mathbbm{E} [X_{\pi_3j} \rvert B_j] - \mathbbm{E} [X_{\pi_3 \pi_4}] - X_{ij} + \mathbbm{E} [X_{i\pi_4} \rvert A_i]) (-U_{ij}) $
    \item Term 4:
   $\mathbbm{E} [((X_{\pi_4\pi_3} - X_{\pi_4j}) - (X_{i\pi_3 - X_{ij}})) \rvert A_i, B_j] U_{ij} = (\mathbbm{E} [X_{\pi_4 \pi_3}] - \mathbbm{E}[X_{\pi_4j} \rvert B_j] - \mathbbm{E} [X_{i\pi_3} \rvert A_i] + X_{ij}) U_{ij} $
    \item Term 5:
 $\mathbbm{E} [((X_{\pi_3 \pi_4} - X_{\pi_3j}) - (X_{i\pi_4} - X_{ij})) \rvert A_i, B_j] U_{ij} = (\mathbbm{E} [X_{\pi_3\pi_4}] - \mathbbm{E} [X_{\pi_3j} \rvert B_j] - \mathbbm{E} [X_{i\pi_4} \rvert A_i + X_{ij}) U_{ij} $
    \item Term 6:
$\mathbbm{E} [((X_{\pi_4j} - X_{\pi_4\pi_3}) - (X_{ij} - X_{i\pi_3})) \rvert A_i, B_j] (-U_{ij}) = (\mathbbm{E} [X_{\pi_4j} \rvert B_j] - \mathbbm{E} [X_{\pi_4\pi_3}] - X_{ij} + \mathbbm{E} [X_{i\pi_3} \rvert A_i]) (-U_{ij}) $
    \item Term 7:
 $\mathbbm{E} [((X_{ij} - X_{i\pi_4}) - (X_{\pi_3j} - X_{\pi_3 \pi_4})) \rvert A_i, B_j] U_{ij} = (X_{ij} - \mathbbm{E} [X_{i\pi_4} \rvert A_i] - \mathbbm{E} [X_{\pi_3j} \rvert B_j] + \mathbbm{E}[X_{\pi_3\pi_4}]) U_{ij} $
    \item Term 8:
$\mathbbm{E} [((X_{i\pi_4} - X_{ij}) - (X_{\pi_3 \pi_4} - X_{\pi_3j})) \rvert A_i, B_j] (-U_{ij}) = (\mathbbm{E} [X_{i\pi_4} \rvert A_i] - X_{ij} - \mathbbm{E}[X_{\pi_3\pi_4}] + \mathbbm{E} [X_{\pi_3j} \rvert B_j]) (-U_{ij})$
    \item Term 9:
$\mathbbm{E}[((X_{ji} - X_{j\pi_3})-(X_{\pi_4i}-X_{\pi_4\pi_3})) \rvert A_j, B_i] U_{ji} = (X_{ji} - \mathbbm{E}[X_{j\pi_3} \rvert A_j] - \mathbbm{E}[X_{\pi_4i} \rvert B_i] + \mathbbm{E}[X_{\pi_4\pi_3}])U_{ji} $
    \item Term 10:
  $\mathbbm{E}[((X_{j\pi_3} - X_{ji})-(X_{\pi_4\pi_3}-X_{\pi_4i}))\rvert A_j, B_i] (-U_{ji}) = (\mathbbm{E}[X_{j\pi_3} \rvert A_j] - X_{ji} - \mathbbm{E} [X_{\pi_4\pi_3}] + \mathbbm{E} [X_{\pi_4i} \rvert B_i])(-U_{ji}) $ 
    \item Term 11:
    $\mathbbm{E} [((X_{\pi_3i} - X_{\pi_3 \pi_4}) - (X_{ji} - X_{j\pi_4})) \rvert A_j, B_i](-U_{ji}) = (\mathbbm{E} [X_{\pi_3i} \rvert B_i] - \mathbbm{E} [X_{\pi_3 \pi_4}] - X_{ji} + \mathbbm{E} [X_{j\pi_4} \rvert A_j]) (-U_{ji}) $
    \item Term 12:
  $\mathbbm{E} [((X_{\pi_4\pi_3} - X_{\pi_4i}) - (X_{j\pi_3 - X_{ji}})) \rvert A_j, B_i] U_{ji} = (\mathbbm{E} [X_{\pi_4 \pi_3}] - \mathbbm{E}[X_{\pi_4i} \rvert B_i] - \mathbbm{E} [X_{j\pi_3} \rvert A_j] + X_{ji}) U_{ji} $
    \item Term 13:
$\mathbbm{E} [((X_{\pi_3 \pi_4} - X_{\pi_3i}) - (X_{j\pi_4} - X_{ji})) \rvert A_j, B_i] U_{ji} = (\mathbbm{E} [X_{\pi_3\pi_4}] - \mathbbm{E} [X_{\pi_3i} \rvert B_i] - \mathbbm{E} [X_{j\pi_4} \rvert A_j + X_{ji}) U_{ji} $
    \item Term 14:
  $\mathbbm{E} [((X_{\pi_4i} - X_{\pi_4\pi_3}) - (X_{ji} - X_{j\pi_3})) \rvert A_j, B_i] (-U_{ji}) = (\mathbbm{E} [X_{\pi_4i} \rvert B_i] - \mathbbm{E} [X_{\pi_4\pi_3}] - X_{ji} + \mathbbm{E} [X_{j\pi_3} \rvert A_j]) (-U_{ji}) $
    \item Term 15:
   $\mathbbm{E} [((X_{ji} - X_{j\pi_4}) - (X_{\pi_3i} - X_{\pi_3 \pi_4})) \rvert A_j, B_i] U_{ji} = (X_{ji} - \mathbbm{E} [X_{j\pi_4} \rvert A_j] - \mathbbm{E} [X_{\pi_3i} \rvert B_i] + \mathbbm{E}[X_{\pi_3\pi_4}]) U_{ji} $
    \item Term 16:
  $\mathbbm{E} [((X_{j\pi_4} - X_{ji}) - (X_{\pi_3 \pi_4} - X_{\pi_3i})) \rvert A_j, B_i] (-U_{ji}) = (\mathbbm{E} [X_{j\pi_4} \rvert A_j] - X_{ji} - \mathbbm{E}[X_{\pi_3\pi_4}] + \mathbbm{E} [X_{\pi_3i} \rvert B_i]) (-U_{ji})$
\end{itemize}

Again, given that $X_{ij} = f(A_i, B_j)$, where $X_{ij}, A_i, B_j$ are i.i.d., we have that:
$$ \mathbbm{E}[X_{i\pi_3} \rvert A_i] = \mathbbm{E}[X_{i\pi_4} \rvert A_i] = \mathbbm{E}[X_{ij} \rvert A_i] $$
$$ \mathbbm{E}[X_{\pi_4j} \rvert B_j] = \mathbbm{E}[X_{\pi_3j} \rvert B_j] = \mathbbm{E}[X_{ij} \rvert B_j] $$
$$ \mathbbm{E}[X_{\pi_4\pi_3}] =  \mathbbm{E}[X_{\pi_3\pi_4}] =  \mathbbm{E}[X_{ij}] $$
Also, given that $X_{ji} = f(A_j, B_i)$, where $X_{ji}, A_j, B_i$ are i.i.d., we have also that:
$$ \mathbbm{E}[X_{j\pi_3} \rvert A_j] = \mathbbm{E}[X_{j\pi_4} \rvert A_j] = \mathbbm{E}[X_{ji} \rvert A_j] $$
$$ \mathbbm{E}[X_{\pi_4i} \rvert B_i] = \mathbbm{E}[X_{\pi_3i} \rvert B_i] = \mathbbm{E}[X_{ji} \rvert B_i] $$
$$ \mathbbm{E}[X_{\pi_4\pi_3}] =  \mathbbm{E}[X_{\pi_3\pi_4}] =  \mathbbm{E}[X_{ji}] $$

\section{Figures}

\begin{figure}[H] 
    \label{ fig7} 
    \begin{minipage}[b]{0.5\linewidth}
      \centering
      \includegraphics[width=0.8\linewidth]{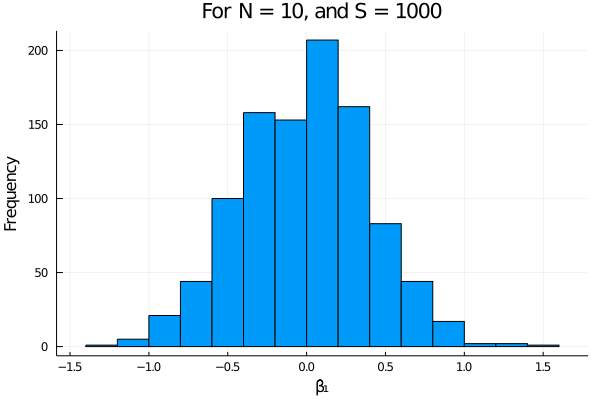} 
    \end{minipage}%%
    \begin{minipage}[b]{0.5\linewidth}
      \centering
      \includegraphics[width=0.8\linewidth]{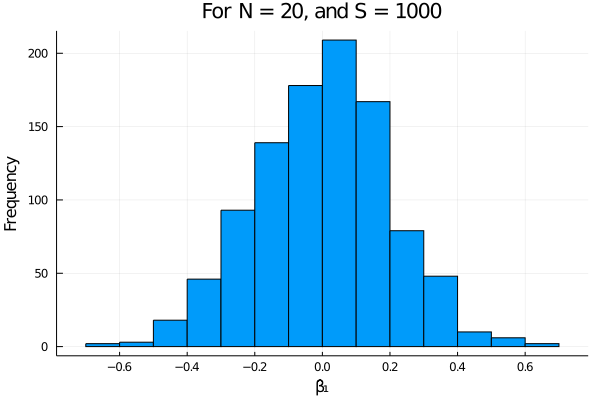} 
    \end{minipage} 
    \begin{minipage}[b]{0.5\linewidth}
      \centering
      \includegraphics[width=0.8\linewidth]{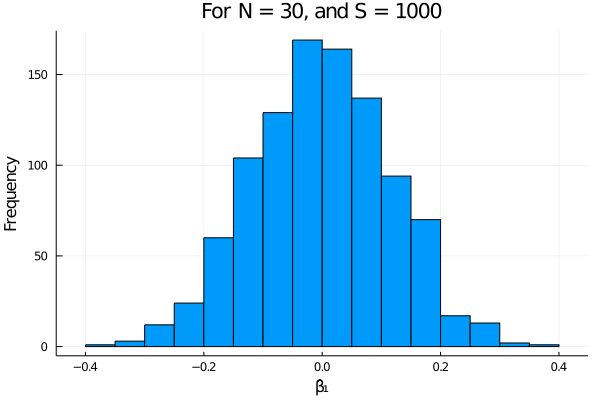} 
    \end{minipage}%% 
    \begin{minipage}[b]{0.5\linewidth}
      \centering
      \includegraphics[width=0.8\linewidth]{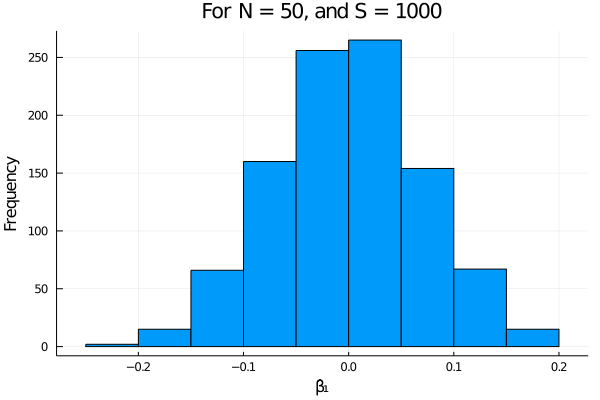} 
    \end{minipage} 
    \caption{Histograms of $\hat{\beta}_1$ for design 1 and $S=1000$} 
  \end{figure}

  \begin{figure}[H] 
    \label{ fig7} 
    \begin{minipage}[b]{0.5\linewidth}
      \centering
      \includegraphics[width=0.8\linewidth]{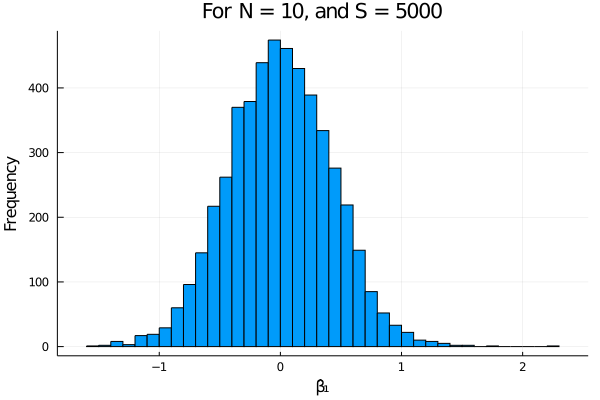} 
    \end{minipage}%%
    \begin{minipage}[b]{0.5\linewidth}
      \centering
      \includegraphics[width=0.8\linewidth]{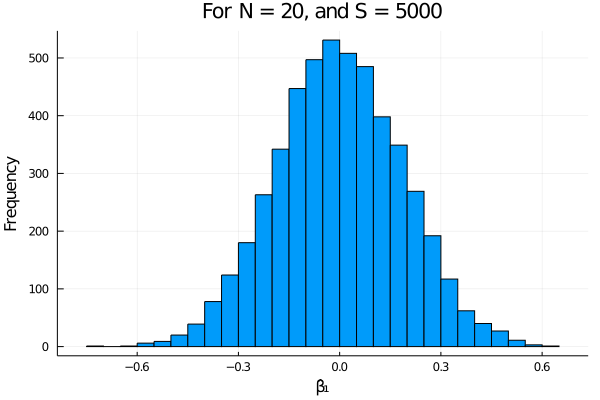} 
    \end{minipage} 
    \begin{minipage}[b]{0.5\linewidth}
      \centering
      \includegraphics[width=0.8\linewidth]{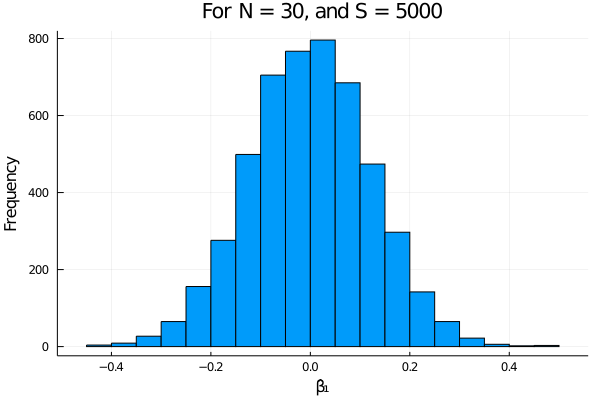} 
    \end{minipage}%% 
    \begin{minipage}[b]{0.5\linewidth}
      \centering
      \includegraphics[width=0.8\linewidth]{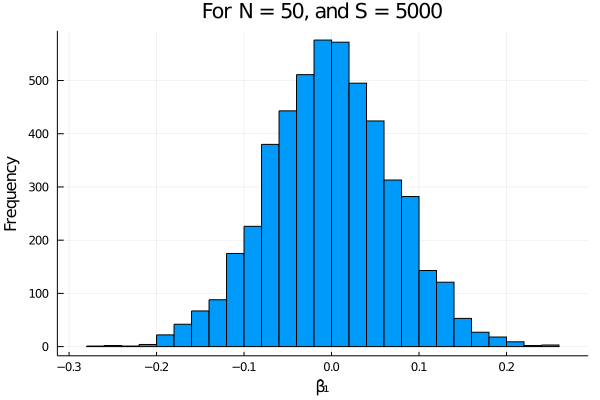} 
    \end{minipage} 
    \caption{Histograms of $\hat{\beta}_1$ for design 1 and $S=5000$} 
  \end{figure}

  \begin{figure}[H] 
    \label{ fig7} 
    \begin{minipage}[b]{0.5\linewidth}
      \centering
      \includegraphics[width=0.8\linewidth]{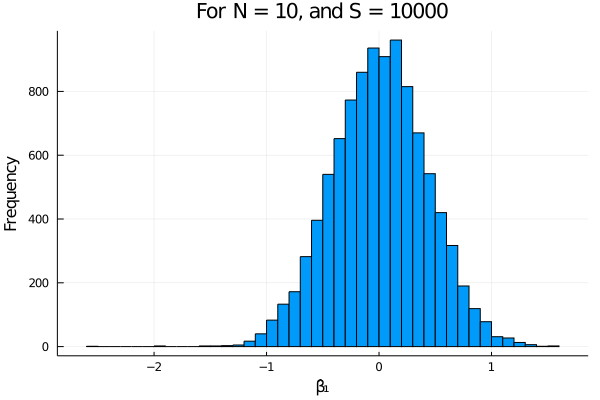} 
    \end{minipage}%%
    \begin{minipage}[b]{0.5\linewidth}
      \centering
      \includegraphics[width=0.8\linewidth]{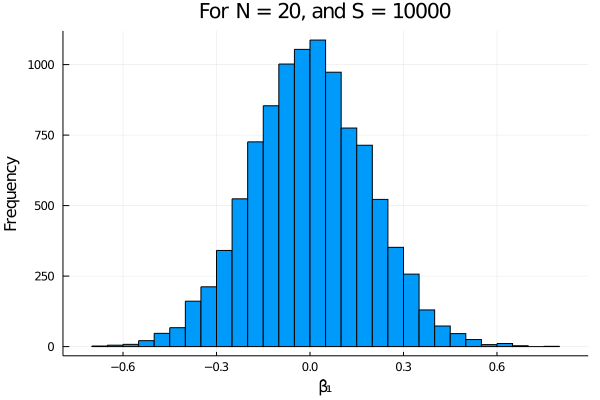} 
    \end{minipage} 
    \begin{minipage}[b]{0.5\linewidth}
      \centering
      \includegraphics[width=0.8\linewidth]{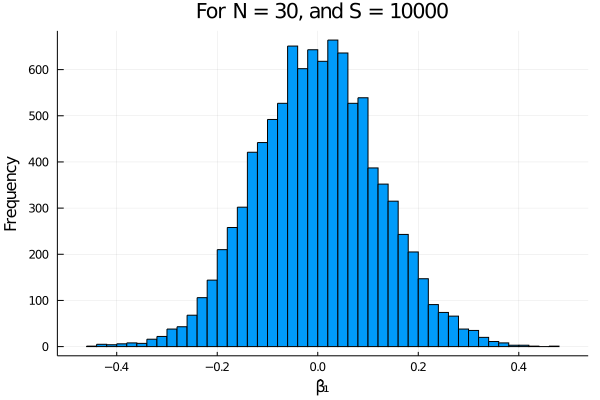} 
    \end{minipage}%% 
    \begin{minipage}[b]{0.5\linewidth}
      \centering
      \includegraphics[width=0.8\linewidth]{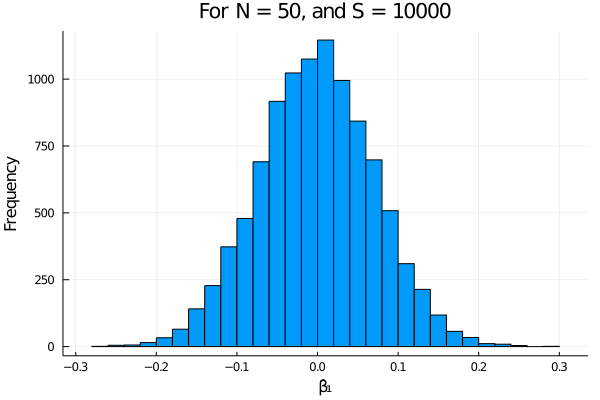} 
    \end{minipage} 
    \caption{Histograms of $\hat{\beta}_1$ for design 1 and $S=10000$} 
  \end{figure}

  \begin{figure}[H] 
    \label{ fig7} 
    \begin{minipage}[b]{0.5\linewidth}
      \centering
      \includegraphics[width=0.8\linewidth]{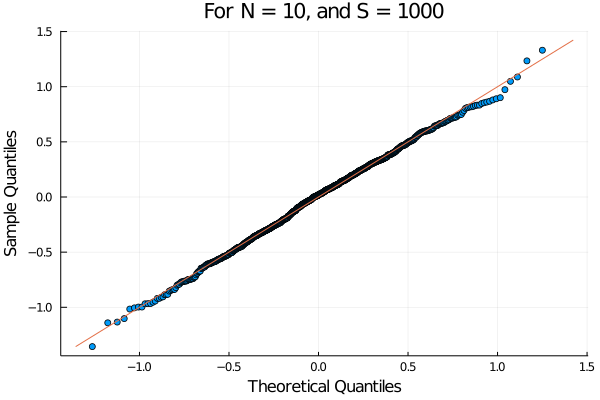} 
    \end{minipage}%%
    \begin{minipage}[b]{0.5\linewidth}
      \centering
      \includegraphics[width=0.8\linewidth]{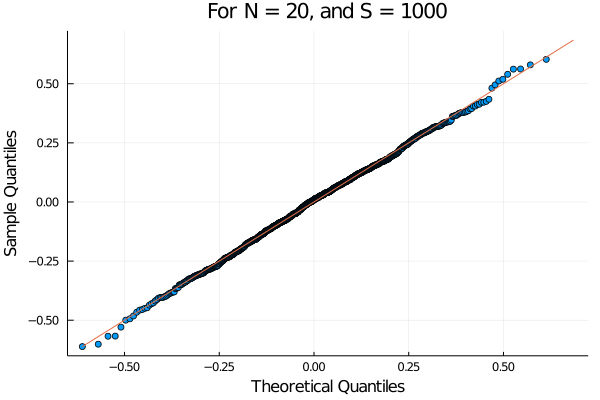} 
    \end{minipage} 
    \begin{minipage}[b]{0.5\linewidth}
      \centering
      \includegraphics[width=0.8\linewidth]{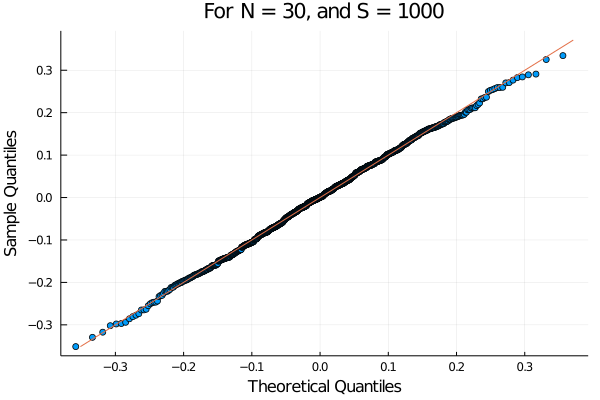} 
    \end{minipage}%% 
    \begin{minipage}[b]{0.5\linewidth}
      \centering
      \includegraphics[width=0.8\linewidth]{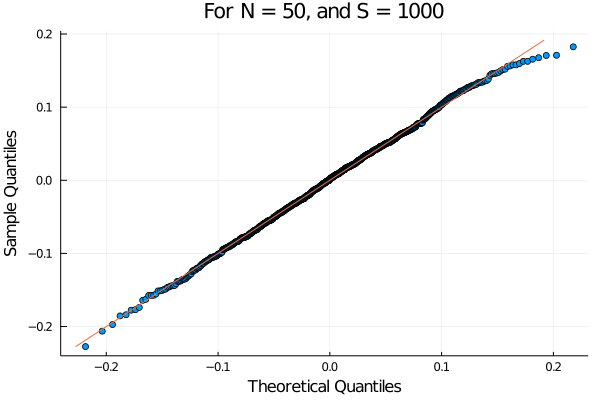} 
    \end{minipage} 
    \caption{QQ-plot of $\hat{\beta}_1$ for design 1 and $S=1000$} 
  \end{figure}

  \begin{figure}[H] 
    \label{ fig7} 
    \begin{minipage}[b]{0.5\linewidth}
      \centering
      \includegraphics[width=0.8\linewidth]{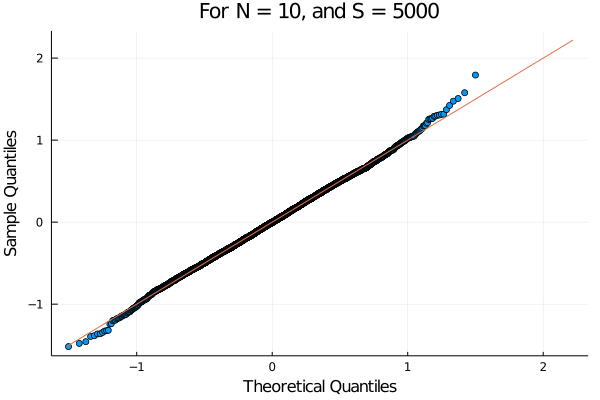} 
    \end{minipage}%%
    \begin{minipage}[b]{0.5\linewidth}
      \centering
      \includegraphics[width=0.8\linewidth]{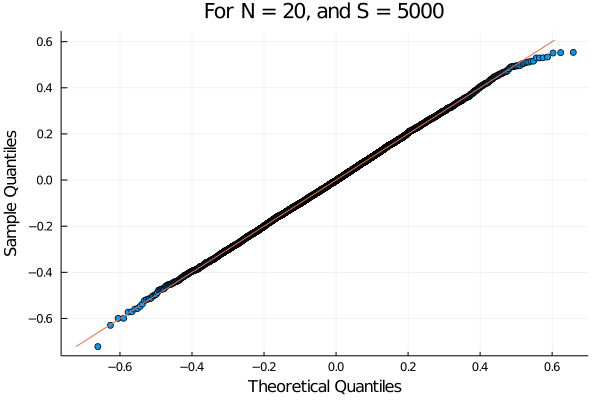} 
    \end{minipage} 
    \begin{minipage}[b]{0.5\linewidth}
      \centering
      \includegraphics[width=0.8\linewidth]{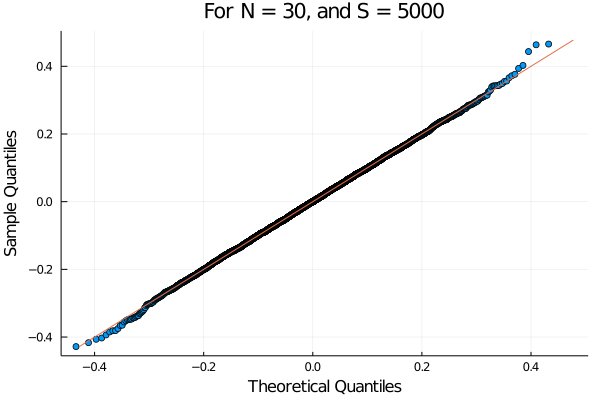} 
    \end{minipage}%% 
    \begin{minipage}[b]{0.5\linewidth}
      \centering
      \includegraphics[width=0.8\linewidth]{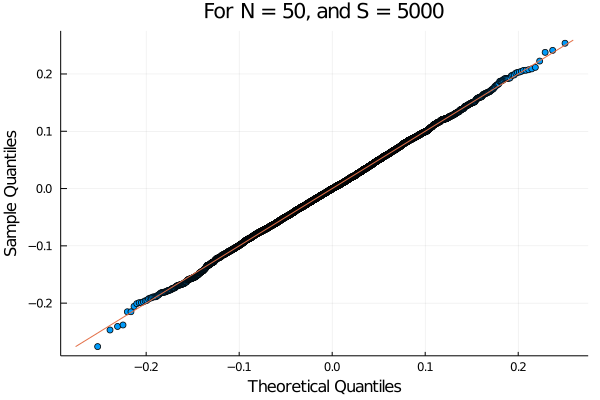} 
    \end{minipage} 
    \caption{QQ-plot of $\hat{\beta}_1$ for design 1 and $S=5000$} 
  \end{figure}

  \begin{figure}[H] 
    \label{ fig7} 
    \begin{minipage}[b]{0.5\linewidth}
      \centering
      \includegraphics[width=0.8\linewidth]{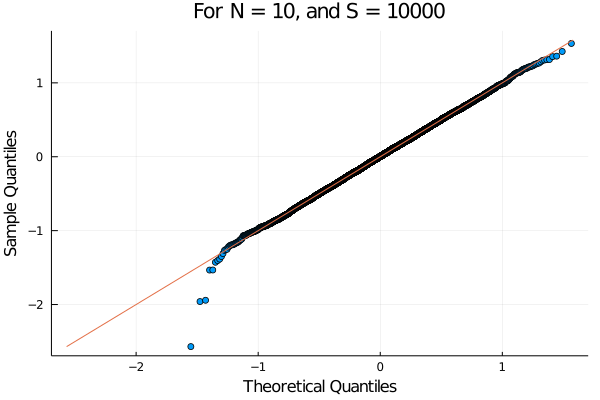} 
    \end{minipage}%%
    \begin{minipage}[b]{0.5\linewidth}
      \centering
      \includegraphics[width=0.8\linewidth]{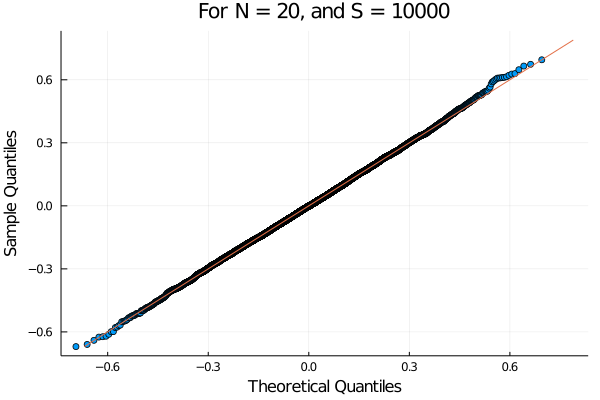} 
    \end{minipage} 
    \begin{minipage}[b]{0.5\linewidth}
      \centering
      \includegraphics[width=0.8\linewidth]{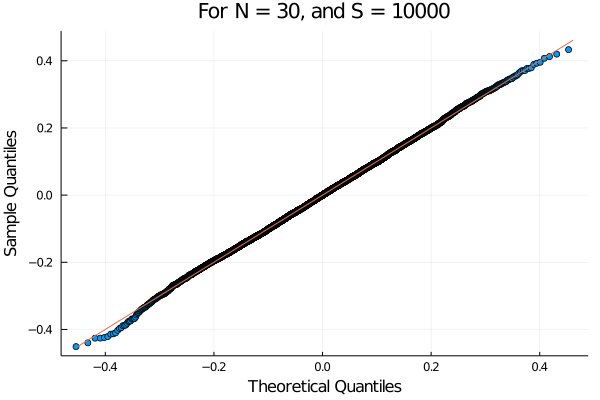} 
    \end{minipage}%% 
    \begin{minipage}[b]{0.5\linewidth}
      \centering
      \includegraphics[width=0.8\linewidth]{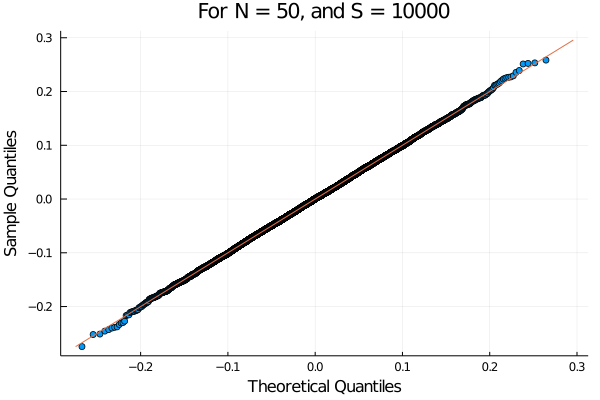} 
    \end{minipage} 
    \caption{QQ-plot of $\hat{\beta}_1$ for design 1 and $S=10000$} 
  \end{figure}

  \begin{figure}[H] 
    \label{ fig7} 
    \begin{minipage}[b]{0.5\linewidth}
      \centering
      \includegraphics[width=0.8\linewidth]{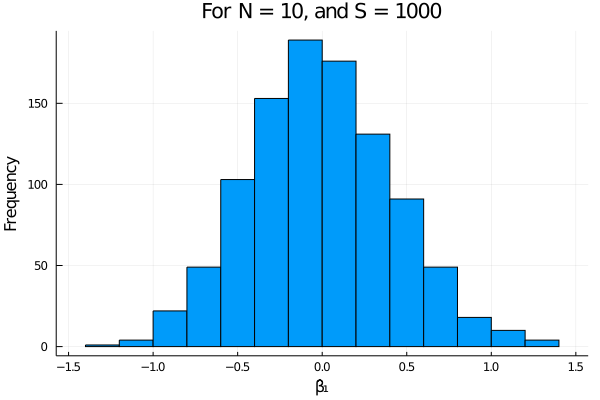} 
    \end{minipage}%%
    \begin{minipage}[b]{0.5\linewidth}
      \centering
      \includegraphics[width=0.8\linewidth]{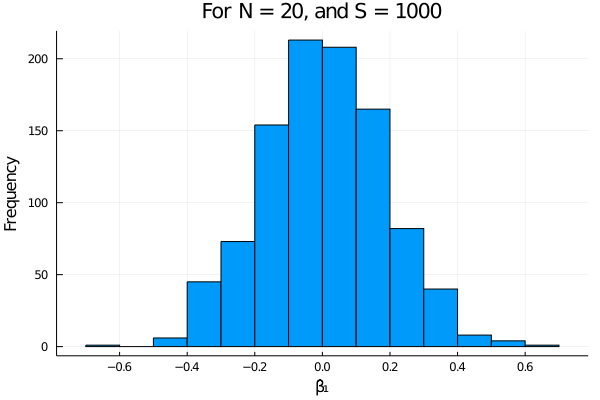} 
    \end{minipage} 
    \begin{minipage}[b]{0.5\linewidth}
      \centering
      \includegraphics[width=0.8\linewidth]{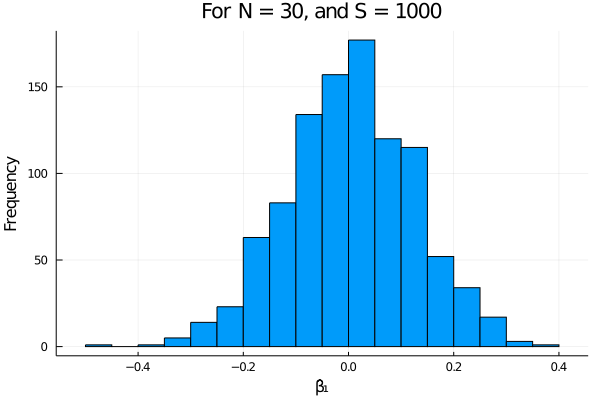} 
    \end{minipage}%% 
    \begin{minipage}[b]{0.5\linewidth}
      \centering
      \includegraphics[width=0.8\linewidth]{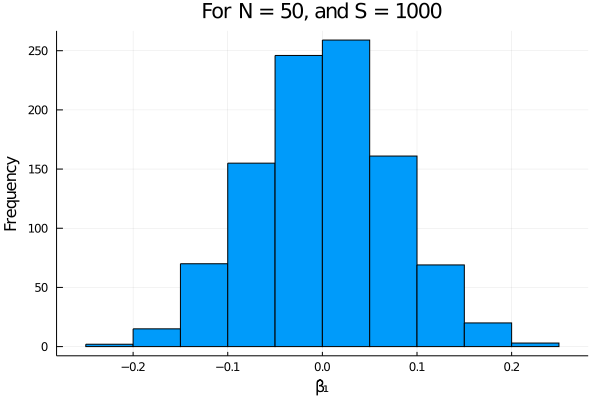} 
    \end{minipage} 
    \caption{Histograms of $\hat{\beta}_1$ for design 3 and $S=1000$} 
  \end{figure}

  \begin{figure}[H] 
    \label{ fig7} 
    \begin{minipage}[b]{0.5\linewidth}
      \centering
      \includegraphics[width=0.8\linewidth]{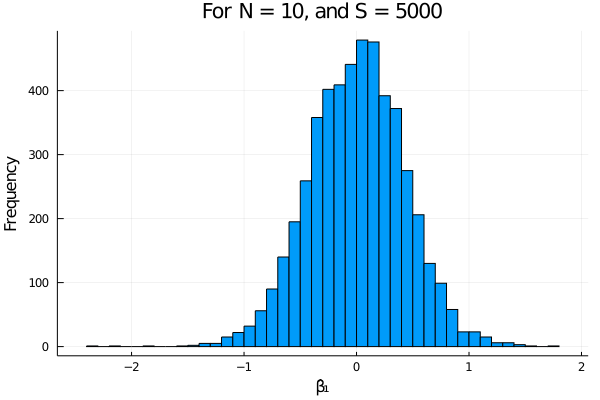} 
    \end{minipage}%%
    \begin{minipage}[b]{0.5\linewidth}
      \centering
      \includegraphics[width=0.8\linewidth]{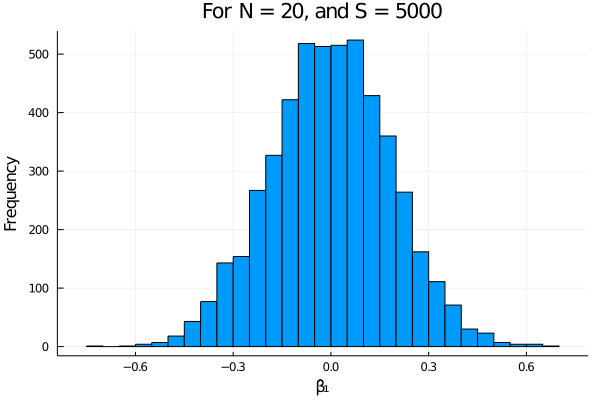} 
    \end{minipage} 
    \begin{minipage}[b]{0.5\linewidth}
      \centering
      \includegraphics[width=0.8\linewidth]{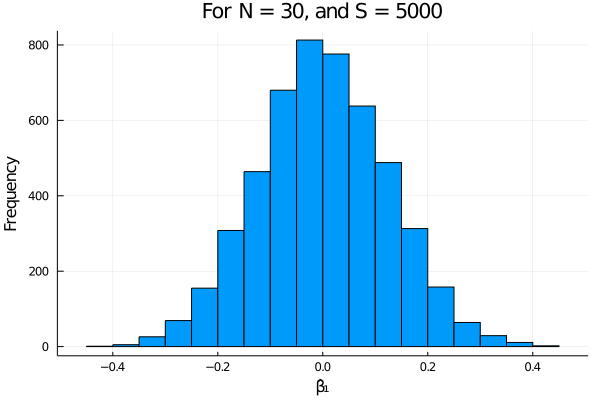} 
    \end{minipage}%% 
    \begin{minipage}[b]{0.5\linewidth}
      \centering
      \includegraphics[width=0.8\linewidth]{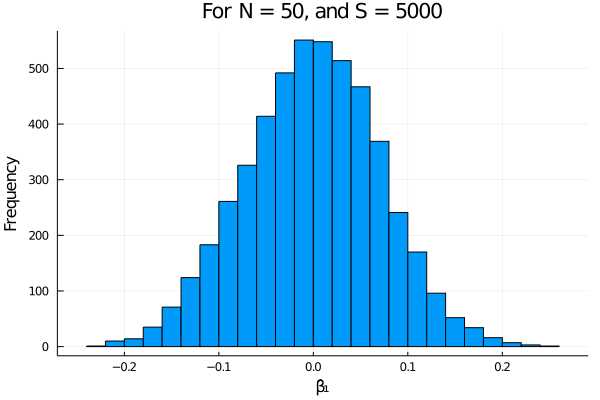} 
    \end{minipage} 
    \caption{Histograms of $\hat{\beta}_1$ for design 3 and $S=5000$} 
  \end{figure}

  \begin{figure}[H] 
    \label{ fig7} 
    \begin{minipage}[b]{0.5\linewidth}
      \centering
      \includegraphics[width=0.8\linewidth]{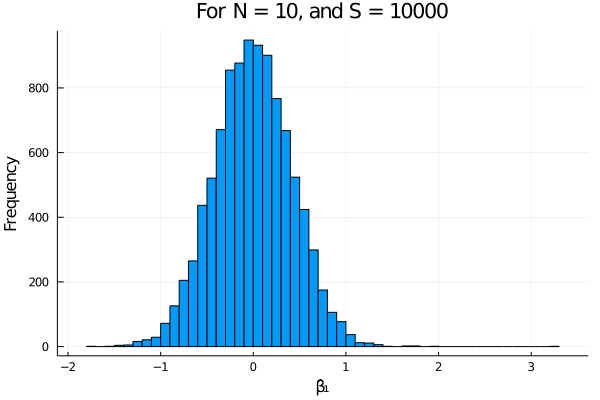} 
    \end{minipage}%%
    \begin{minipage}[b]{0.5\linewidth}
      \centering
      \includegraphics[width=0.8\linewidth]{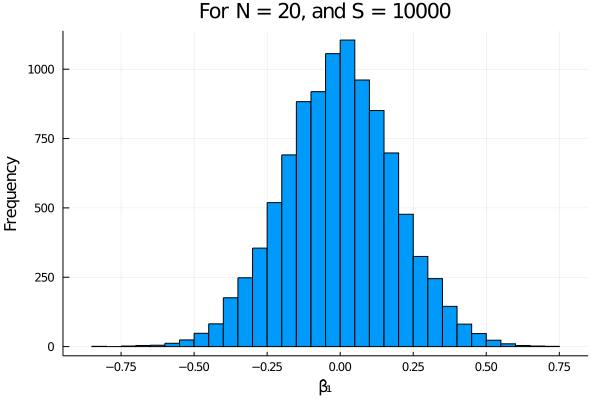} 
    \end{minipage} 
    \begin{minipage}[b]{0.5\linewidth}
      \centering
      \includegraphics[width=0.8\linewidth]{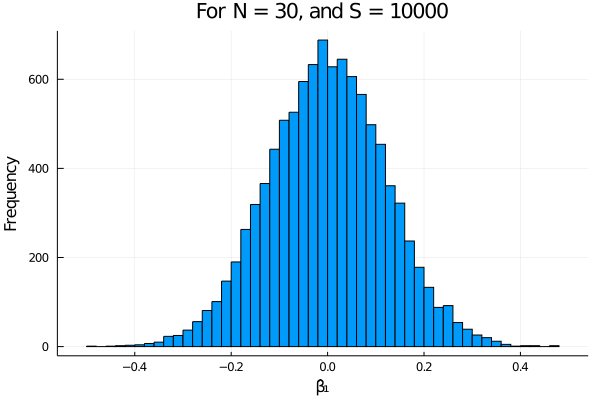} 
    \end{minipage}%% 
    \begin{minipage}[b]{0.5\linewidth}
      \centering
      \includegraphics[width=0.8\linewidth]{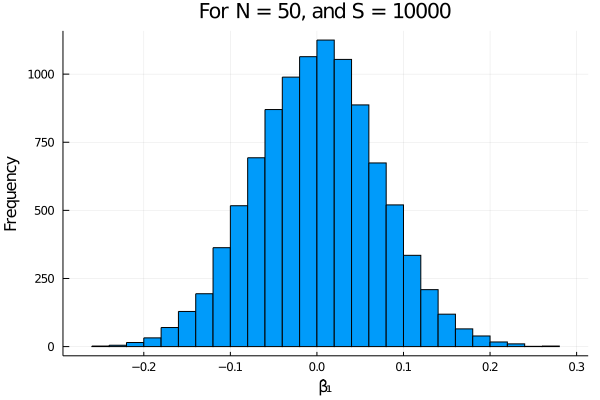} 
    \end{minipage} 
    \caption{Histograms of $\hat{\beta}_1$ for design 3 and $S=10000$} 
  \end{figure}

  \begin{figure}[H] 
    \label{ fig7} 
    \begin{minipage}[b]{0.5\linewidth}
      \centering
      \includegraphics[width=0.8\linewidth]{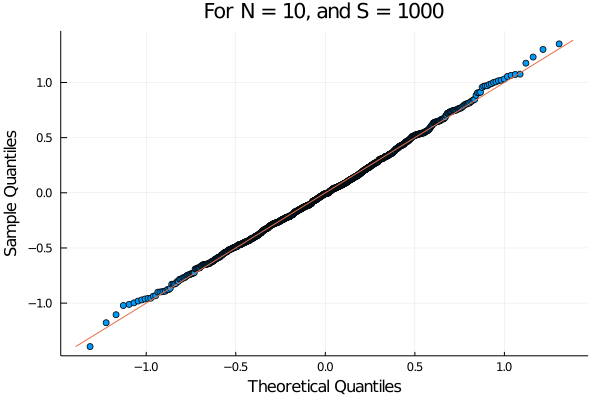} 
    \end{minipage}%%
    \begin{minipage}[b]{0.5\linewidth}
      \centering
      \includegraphics[width=0.8\linewidth]{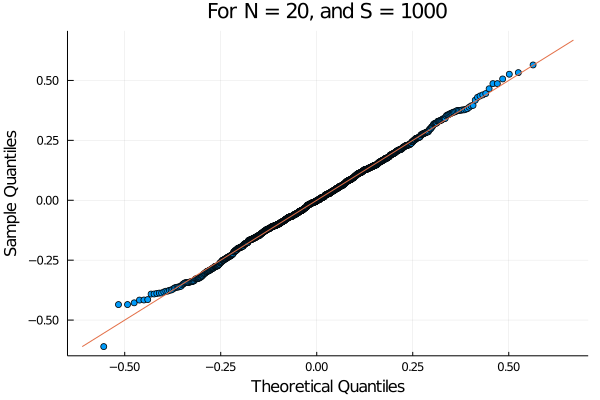} 
    \end{minipage} 
    \begin{minipage}[b]{0.5\linewidth}
      \centering
      \includegraphics[width=0.8\linewidth]{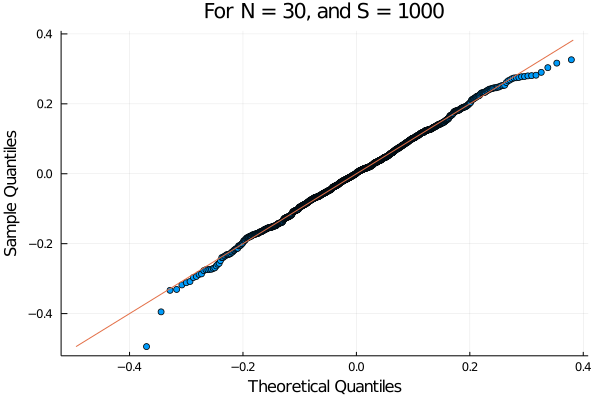} 
    \end{minipage}%% 
    \begin{minipage}[b]{0.5\linewidth}
      \centering
      \includegraphics[width=0.8\linewidth]{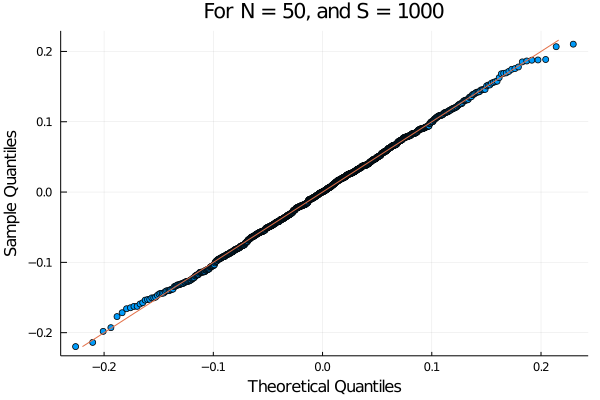} 
    \end{minipage} 
    \caption{QQ-plot of $\hat{\beta}_1$ for design 3 and $S=1000$} 
  \end{figure}

  \begin{figure}[H] 
    \label{ fig7} 
    \begin{minipage}[b]{0.5\linewidth}
      \centering
      \includegraphics[width=0.8\linewidth]{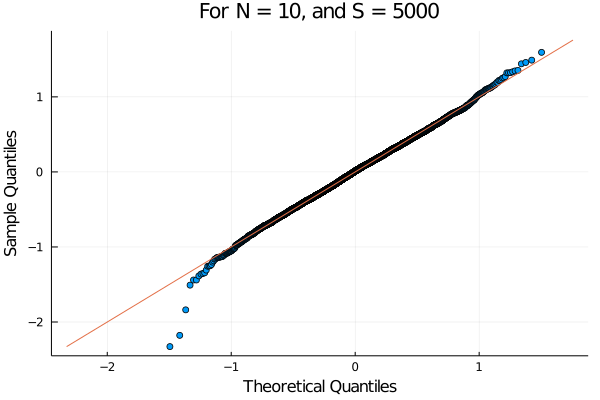} 
    \end{minipage}%%
    \begin{minipage}[b]{0.5\linewidth}
      \centering
      \includegraphics[width=0.8\linewidth]{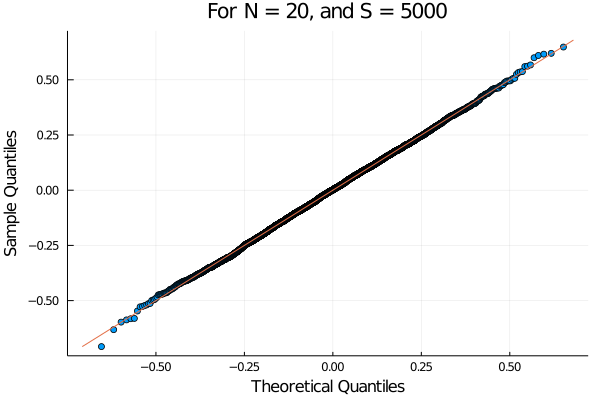} 
    \end{minipage} 
    \begin{minipage}[b]{0.5\linewidth}
      \centering
      \includegraphics[width=0.8\linewidth]{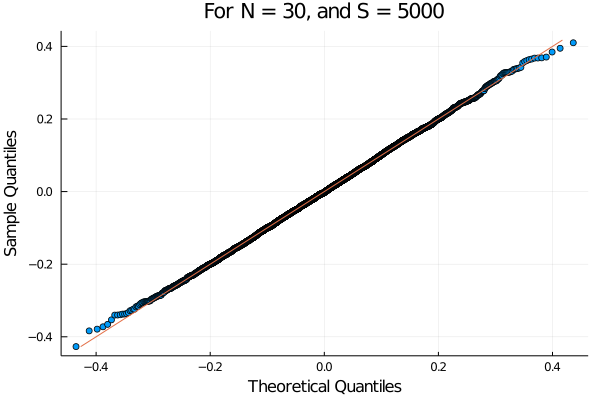} 
    \end{minipage}%% 
    \begin{minipage}[b]{0.5\linewidth}
      \centering
      \includegraphics[width=0.8\linewidth]{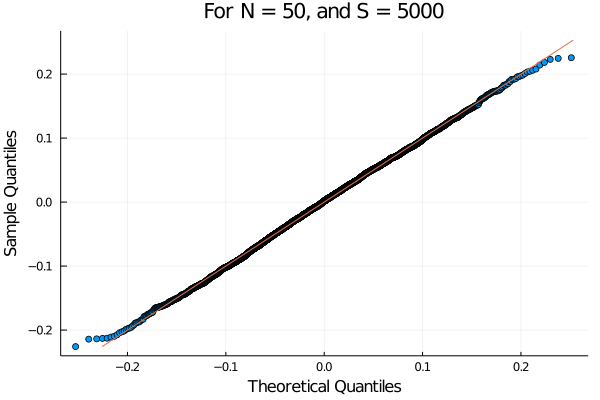} 
    \end{minipage} 
    \caption{QQ-plot of $\hat{\beta}_1$ for design 3 and $S=5000$} 
  \end{figure}

  \begin{figure}[H] 
    \label{ fig7} 
    \begin{minipage}[b]{0.5\linewidth}
      \centering
      \includegraphics[width=0.8\linewidth]{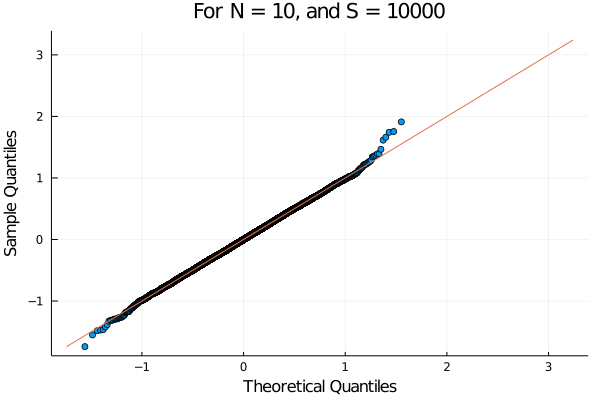} 
    \end{minipage}%%
    \begin{minipage}[b]{0.5\linewidth}
      \centering
      \includegraphics[width=0.8\linewidth]{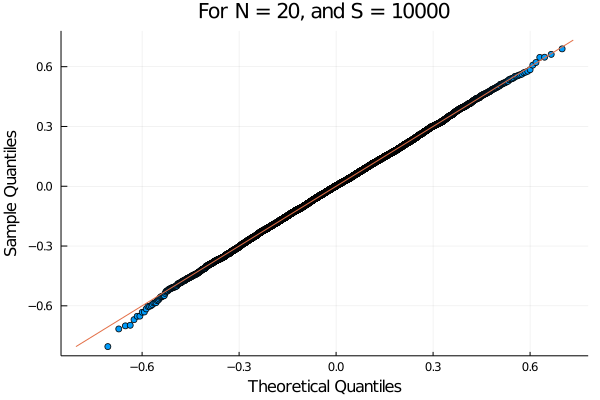} 
    \end{minipage} 
    \begin{minipage}[b]{0.5\linewidth}
      \centering
      \includegraphics[width=0.8\linewidth]{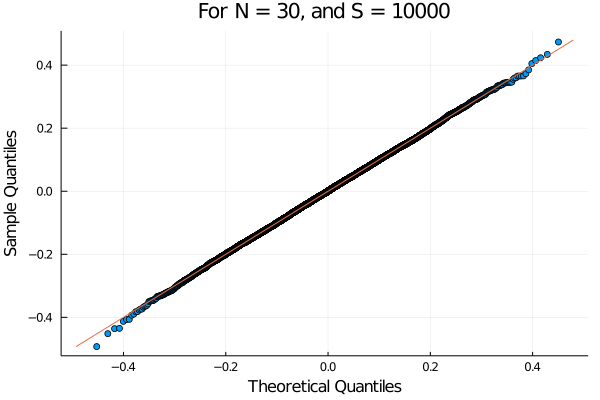} 
    \end{minipage}%% 
    \begin{minipage}[b]{0.5\linewidth}
      \centering
      \includegraphics[width=0.8\linewidth]{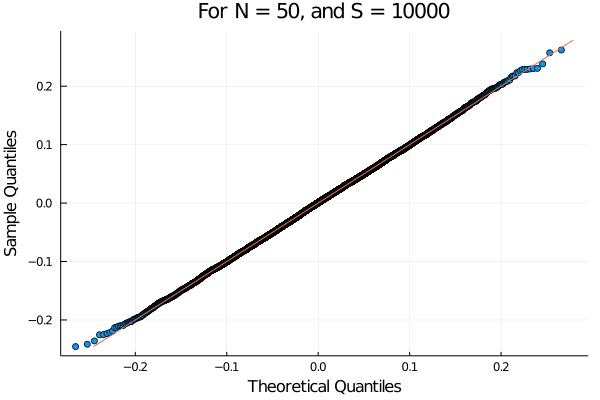} 
    \end{minipage} 
    \caption{QQ-plot of $\hat{\beta}_1$ for design 3 and $S=10000$} 
  \end{figure}

\end{document}